\documentclass[onecolumn,showpacs,preprintnumbers]{revtex4}

\usepackage{amsfonts}
\usepackage{amsmath}
\usepackage{makeidx}
\usepackage{amssymb}
\usepackage{amsmath}
\usepackage{graphicx}
\usepackage{dcolumn}
\usepackage{bm}
\usepackage{hyperref}
\usepackage{longtable}
\usepackage{lineno,hyperref}
\begin{document}

\title{Spatiotemporal engineering of matter-wave
solitons in Bose-Einstein condensates}

\author{Emmanuel Kengne$^{1}$}
\thanks{Corresponding author: ekengne6@zjnu.edu.cn}
\author{ Wu-Ming Liu$^{2}$ and Boris A. Malomed$^{3,4}$}
\address{$^{1}$ School of Physics and Electronic Information Engineering,
Zhejiang Normal University, Jinhua 321004, China}
\address{$^{2}$ Laboratory of Condensed Matter Theory and Materials
Computation, Institute of Physics, Chinese Academy of Sciences,
No. 8 South-Three Street, ZhongGuanCun, Beijing 100190, China}
\address{$^{3}$ Department of Physical Electronics, School of Electrical Engineering,
Faculty of Engineering, and Center for Light-Matter Interaction,
Tel Aviv University, P.O.B. 39040, Ramat Aviv, Tel Aviv, Israel}
\address{$^{4}$ Instituto de Alta Investigaci\'{o}n, Universidad de Tarapac\'{a},
Casilla 7D, Arica, Chile}

\begin{abstract}
Since the realization of Bose-Einstein condensates (BECs) trapped
in optical potentials, intensive experimental and theoretical
investigations have been carried out for bright and dark
matter-wave solitons, coherent structures, modulational
instability (MI), and nonlinear excitation of BEC matter waves,
making them objects of fundamental interest in the vast realm of
nonlinear physics and soft condensed-matter physics. Many of these
states have their counterparts in optics, as concerns the
nonlinear propagation of localized and extended light modes in the
spatial, temporal, and spatiotemporal domains. Ubiquitous models,
which are relevant to the description of diverse nonlinear media
in one, two, and three dimensions (1D, 2D, and 3D), are provided
by the nonlinear Schr\"{o}dinger (NLS), alias Gross-Pitaevskii
(GP), equations. In many settings, nontrivial solitons and
coherent structures, which do not exist or are unstable in free
space, can be created and/or stabilized by means of various
\textit{management} techniques, which are represented by NLS and
GP equations with coefficients in front of linear or nonlinear
terms which are functions of time and/or coordinates. Well-known
examples are \textit{dispersion management} in nonlinear fiber
optics, and \textit{nonlinearity management} in 1D, 2D, and 3D
BEC. Developing this direction of research in various settings,
efficient schemes of the spatiotemporal modulation of coefficients
in the NLS/GP equations have been designed to \textit{engineer}
desirable robust nonlinear modes. This direction and related ones
are the main topic of the present review. In particular, a broad
and important theme is the creation and control of 1D matter-wave
solitons in BEC by means of combination of the temporal or spatial
modulation of the nonlinearity strength (which may be imposed by
means of the Feshbach resonance induced by variable magnetic
fields) and a time-dependent trapping potential. An essential
ramification of this topic is analytical and numerical analysis of
MI of continuous-wave (constant-amplitude) states, and control of
the nonlinear development of MI. Another physically important
topic is stabilization of 2D solitons against the critical
collapse, driven by the cubic self-attraction, with the help of
temporarily periodic nonlinearity management, which makes the sign
of the nonlinearity periodically flipping. In addition to that,
the review also includes some topics that do not directly include
spatiotemporal modulation, but address physically important
phenomena which demonstrate similar soliton dynamics. These are
soliton motion in binary BEC, three-component solitons in spinor
BEC, and dynamics of two-component 1D solitons under the action of
spin-orbit coupling.\bigskip

\textbf{Highlights}
\begin{itemize}

\item A review is focused on 1D, 2D, and 3D matter-wave solitons
in Bose-Einstein condensates under the action of spatiotemporally
modulated cubic nonlinearity and time-dependent trapping
potentials

\item Most essential problems under the consideration is the shape
and stability of solitons and other coherent structures, including
stabilization against the critical collapse

\item Both analytical results (exact and approximate ones) and
systematically produced numerical findings are summarized

\item The modulational instability in these models and its
nonlinear development is addressed in detail

\item Stability and motion of multi-component solitons in binary
and spinor (triple) solitons is considered
\end{itemize}

\textbf{Keywords} Nonlinear Schr\"{o}dinger equations;
Gross-Pitaevskii equations; dispersion management; nonlinearity
management; critical collapse; Feshbach-resonance technique;
spin-orbit coupling; spinor Bose--Einstein condensate

\end{abstract}

\maketitle

\textbf{Acronyms}:

1D one-dimensional

2D two-dimensional

3D three-dimensional

BdG Bogoliubov - de Gennes (equations for small perturbations)

BEC Bose-Einstein condensate

CW continuous-wave (solution, alias plane wave)

DM dispersion management

FM ferromagnetic

FR Feshbach resonance

GP Gross-Pitaevskii (equation)

HO harmonic oscillator

MI modulational instability

NLS nonlinear Schr\"{o}dinger (equation)

NM nonlinearity management

ODE ordinary differential equation

PIT phase-imprinting technique

SOC spin-orbit coupling

SPM self-phase-modulation (nonlinear self-interaction)

TS Townes' soliton

VA variational approximation

XPM cross-phase-modulation (nonlinear interaction between two
components)

\newpage
\tableofcontents
\newpage

\section{Introduction}

The nonlinear Schr\"{o}dinger (NLS) equations are basic models in a broad
spectrum of disciplines in physics and engineering. Among these models, most
thoroughly elaborated ones belong to the realms of nonlinear optics and
Bose-Einstein condensates (BECs) of ultracold atoms. In this latter context,
the NLS equation is often called the Gross-Pitaevskii (GP) equation.
Commonly known fundamental solutions of the NLS equations are solitons
(self-trapped localized states) \cite{Peyrard,Yang}. Many species of
solitons have been predicted and observed in optics \cite{HK,0.1} and in BEC
\cite{3.30,Abdullaev,Luca}. In particular, many results have been produced
in the framework of the \textit{dynamical management of solitons}, under the
action of time-dependent factors, which are represented by time-dependent
terms in the respective NLS or GP equations \cite{2.1}, that may also
include linear losses and compensating gain. A well-known example is the
\textit{dispersion management} (DM) in fiber optics \cite{2.1,Turitsyn},
with the group-velocity-dispersion coefficient periodically alternating
between positive and negative values along the propagation distance (which
is the evolution variable in the guided-wave propagation theory \cite{HK}).
This setting helps to stabilize DM solitons against various perturbations,
such as random noise and collisions between solitons in multi-channel
systems \cite{Nijhof,coll1,coll2,0.2}. Furthermore, the DM scheme was
predicted to help stabilize 2D \cite{Michal} and 3D \cite{Wagner}
spatiotemporal solitons (\textquotedblleft light bullets").

Another important variant of the NLS/GP models with the dynamical modulation
include \textit{nonlinearity management} (NM), i.e., a time-varying
coefficient in front of the cubic term. This possibility was first
elaborated in terms of the optics model, based on the two-dimensional (2D)
spatial-domain NLS equation for the beam propagation in bulk media. Assuming
that the medium was composed of alternating elements with self-focusing and
defocusing Kerr (cubic) nonlinearity, it was demonstrated \cite{4.7} that
this version of NM stabilizes oscillating 2D solitons against the \textit{%
critical collapse} (blowup), which makes all static 2D solitons (called
Townes' soliton (TS) \cite{Townes} in this case) unstable in the uniform
medium with the cubic self-focusing \cite{Fibich}.

The concept of NM in optics, which ultimately aims to create nontrivial
spatiotemporal light patterns, such as \textquotedblleft bullets", is
naturally related to the rapidly developing area of experimental and
theoretical studies aimed at the creation and various applications of
\textit{structured light} \cite%
{str-light0,str-light1,str-light3,str-light2,Forbes,str-light-roadmap,str-light-book}%
. In most cases, the structure in light beams is induced by means of linear
techniques (see, e.g., Ref. \cite{linear-shaping} and references therein).
Frequently, the so created light patterns take the form of a
\textquotedblleft flying focus", with the focal point featuring controlled
spatiotemporal motion \cite{flying-focus}. Because nonlinearity is the
underlying topic of the present article, it is relevant to mention that,
very recently, a more powerful method was elaborated, which makes it
possible to build optical fields with an embedded \textquotedblleft
self-flying-focus", by means of inherent nonlinearity of the medium \cite%
{self-flying}. Another experimentally implemented technique, which is very
relevant in the context of models considered below in the present article,
makes it possible to create arbitrarily structured averaged optical patterns
\textquotedblleft painted" by a rapidly moving laser beam \cite{paint}.

In the GP equation, the cubic term accounts for effects of inter-atomic
collisions in BEC, treated in the mean-field approximation, with the
coefficient in front of it proportional to the \textit{s}-wave collision
scattering length \cite{Pit}. In the experiment, this parameter can be
controlled by means of the Feshbach-resonance (FR) effect, i.e., a change of
the scattering length under the action of magnetic field, which creates
short-lived quasi-bound states in the course of the inter-atomic collision
\cite{Feshbach,FR-review}. It was demonstrated that the FR may be used to
very accurately adjust the nonlinearity strength in BEC of potassium \cite%
{4.15} and lithium \cite{4.12,0.7} atoms, as well as in binary BEC\
mixtures, such as one formed by $^{23}$Na and $^{87}$Rb atoms \cite%
{FR-mixture}. The FR can also be imposed optically, by appropriate laser
illumination \cite{FR-optical,0.5,0.4,FR-Killian}, as well as by microwave
\cite{FR-micro} and electrostatic \cite{FR-electro} fields. Then, NM for BEC
may be realized using time-dependent and/or nonuniform control fields \cite%
{2.1,0.1,0.3,0.4,0.5,Pelster,Barcelona,05feb}. In particular, the FR
technique makes it possible to quickly reverse (in time) the interaction
sign from repulsion to attraction (positive to negative scattering length),
which gives rise, via the onset of the collapse, to abrupt shrinkage of the
condensate, followed by a burst of emitted atoms and the formation of a
stable residual condensate \cite{Weiman}. A similar method relies on the use
of \textit{quench}, i.e., sudden change of the self-attraction strength in
BEC, which makes it possible to transform a quasi-1D fundamental soliton
into an oscillatory state in the form of a breather \cite{breather}. A still
more spectacular outcome is \textquotedblleft bosonic fireworks" in the form
of violent emission of matter-wave jets from the original condensate \cite%
{fireworks1,fireworks2}. In other experiments, the time-periodic modulation
of the nonlinearity strength, applied to the quasi-1D (cigar-shaped) BEC,
initiates a transition to a granulated state in the condensate \cite%
{granulation}.

In the case of the attractive interaction, matter-wave solitons may be
readily formed in an effectively 1D condensate \cite{3.30}. However, in the
2D geometry the attraction results in the critical collapse of the
condensate, if the number of atoms (the norm of the wave field, in terms of
the GP equation) exceeds a critical value. The FR technique was predicted to
be quite useful in these settings: similar to the above-mentioned results
for the 2D spatial solitons in the optical bulk waveguide with the periodic
alternation of the self-focusing and defocusing \cite{4.7}, 2D matter-wave
solitons may be readily stabilized against the critical collapse by the
time-periodic variation of the nonlinearity coefficient between negative and
positive values. This approach was elaborated in detail in various forms
\cite{3.1,Ueda,VPG,Itin}. Unlike the fundamental 2D solitons, the NM scheme
cannot stabilize 2D solitons with embedded vorticity against the splitting
instability (to which the vortex solitons are most vulnerable \cite{PhysicaD}%
), and it cannot stabilize 3D solitons against the \textit{supercritical
collapse} either (in the latter case, the critical norm is zero, i.e., any
value of the norm may lead to the collapse). However, 3D solitons can be
stabilized if the NM is combined with a quasi-1D spatially periodic
potential (optical lattice) applied to the condensate \cite{Michal2}. A
generalization of the FR technique for controlling the strength and sign of
the interaction between atoms in BEC and, thus, the coefficient in front of
the cubic term in the corresponding GP equation, is the application of a
control field containing time-periodic (ac) and constant (dc) components.
Another essential ramification is the use of FR in BEC mixtures, to adjust
the relation between strengths of self- and cross-component interactions
(alias self- and cross-phase modulation, SPM and XPM, in terms of nonlinear
optics \cite{0.1}), which helps to produce robust multi-component states in
the experiment \cite{0.6,Dimitri,mixed-droplet}.

A natural generalization of the NM scenarios for BEC\ is to combine the
variable coefficient in front of the nonlinear term in the GP equation,
imposed by the temporal and/or spatial modulation of the scattering length,
with a temporarily modulated potential trapping the condensate. In
particular, a specially designed relation between the spatiotemporal
dependence of nonlinear and linear factors in the GP equation makes it
possible to produce integrable versions of the equation in 1D \cite%
{WMLiu-PRL,Zhong,Zhong4,2.7,Lakshmanan,LuLi,WMLiu,Cardoso,add8}. A similar
approach was developed for constructing solvable 3D models, which admit
factorization of the respective 3D equation into a product of relatively
simple 1D equations, which admit exact solutions (in particular, solitons)
\cite{Zhong2,Zhong3,Belic,Zhong3D,Zhong3D-2}. In fact, the integrability of
such specially designed (\textit{engineered}) models is not a fundamentally
new mathematical finding, because they may be transformed, by means of
tricky but explicit transformations of the wave function (or several wave
functions, in the case of multi-component systems), spatial coordinates, and
the temporal variable, into the classical integrable 1D NLS equation with
constant coefficients (or the Manakov's \cite{4.14} integrable system of the
NLS equations) \cite{Suslov,cubic-quartic}. Accordingly, a great variety of
integrable and nearly integrable models can be generated by means of \textit{%
inverse engineering}, applying generic transformations of the same type to
the underlying integrable equation(s) \cite{Stepa}. Although this approach
to the expansion of the set of solvable models seems somewhat artificial, it
is meaningful, and quite useful in many cases, as it has a potential to
predict tractable nontrivial configurations in BEC and, to a lesser degree,
in optics (it is not easy to apply flexible modulation of local nonlinearity
to optical media, where there is no direct counterpart of FR).

In this review article, we summarize basic results (chiefly, theoretical
ones) for the dynamics of BEC and optical fields trapped in time-dependent
potentials, combined with the variable (time-modulated) nonlinearity. The
respective models are based on one-, two-, and three-component nonautonomous
GP/NLS equations. It is relevant to stress that all the results included in
the review have already been published, although some of them are quite
recent. In particular, an essential direction in these studies, elaborated
in many theoretical works, is the design of specific nontrivial models which
can be explicitly transformed into an integrable form, thus making it
possible to use the huge stock of exact solutions, available in the original
models. The analytically found exact solutions are, in many cases, confirmed
by numerical simulations. It is necessary to stress that, while the exact
solutions are stable in the framework of the integrable models, they may be
subject to the structural instability, as a small deviation of the physical
model from the integrable form may lead to the loss of stability of the
original exact solutions, or even make them nonexistent, in the rigorous
mathematical sense. In such cases, the former solutions will suffer slow
decay; however, it may often happen that the decay time (or propagation
distance, in optics) will be essentially larger than the actual extension of
plausible experiments, hence the approximate solutions still predict
physically relevant states.

The rest of the article is composed of sections which are related by
gradually increasing complexity of models presented in them. Section 2
addresses the basic model, in which the approach outlined above leads to the
construction and management of nonautonomous solitons of 1D cubic
self-defocusing NLS equations with spatiotemporally modulated coefficients,
that may be transformed into the classical integrable NLS equation. In this
setting, matter-wave soliton solutions are constructed in an analytical
form, and it is shown that the instability of those solitons, if any, may be
delayed or completely eliminated by varying the nonlinearity's strength in
time. In Section 3, we continue the presentation by considering engineered
nonautonomous matter-wave solitons in BEC with spatially modulated local
nonlinearity and a time-dependent harmonic-oscillator (HO) potential. The
modulational instability (MI) in that setting is considered too. In Section
4, we proceed to more general nonintegrable models, which are treated by
means of the semi-analytical variational approximation (VA) and direct
numerical simulations. In this case, we address the dynamics of 2D and 3D
condensates with the nonlinearity strength containing constant and
harmonically varying parts, which can be implemented with the help of ac
magnetic field tuned to FR. In particular, the spatially uniform temporal
modulation of the nonlinearity may readily play the role of an effective
trap that confines the condensate, and sometimes enforces its collapse.
Section 5 deals with dynamics of a binary (two-component) condensate in an
expulsive time-varying HO potential with the time-varying attractive
interaction. In this case, the condensates in the expulsive time-modulated
HO potential may sustain the stability, while their counterparts in the
time-independent potential rapidly decay. In Section 6, we address the
dynamics of matter-wave solitons of coupled GP equations for binary BEC.
Strictly speaking, this model does not include spatiotemporal engineering of
solitons, but the results are closely related to those produced by means of
the engineering. In particular, the results reported in section 6 show that,
in the absence of the XPM\ interaction, the solutions maintain properties of
one-component condensates, such as MI, while in the presence of the
interaction between the components, the solutions exhibit different
properties, such as restriction of MI and soliton splitting. Section 7 deals
with the soliton states in a model based on a set of three coupled GP
equations modeling the dynamics of the spinor BEC, with atomic spin $F=1$.
Both nonintegrable and integrable versions of the system are considered, and
exact soliton solutions are demonstrated. The stability of the solutions was
checked, in most cases, by direct simulations and, in some cases, it was
investigated in a more rigorous form, based on linearized Bogoliubov - de
Gennes (BdG) equations for small perturbations. In Section 8, we introduce a
new physical setting, considering the motion of bright and dark matter-wave
solitons in 1D BEC in the presence of spin-orbit coupling (SOC). We
demonstrate that the spin dynamics of the SOC solitons is governed by a
nonlinear Bloch equation and affects the orbital motion of the solitons,
leading to SOC effects in the dynamics of macroscopic quantum objects.
Similar to what is mentioned about Section 6, the settings addressed in
Sections 7 and 8 do not directly include ingredients of the spatiotemporal
engineering. Nevertheless, the models addressed in these sections, as well
as methods applied to them and obtained results, are quite similar to those
produced by the engineering. In particular, the macroscopic SOC
phenomenology is explained by the fact that an effective time-periodic force
produced by rotation of the soliton's (pseudo-) spin plays the role of
temporal management which affects motion of the same soliton. Each Section 2
through 8 contains formulation of the underlying model, summary of
analytical and numerical results, and a conclusion focused in the topic
under the consideration, with references to original papers in which those
results have been reported. Finally, Section 9 concludes the review and
mentions perspectives for further work in this area.

The presentation has a rather technical form, as the topic surveyed in this
article is inherently technical. Nevertheless, an effort is made, in each
section, to highlight the main findings, on top of technical details. The
sections include an outline of physical realizations of the considered
theoretical models in BEC (and in some cases, in nonlinear optics).
Characteristic physical parameters of the respective settings, such as
atomic species and particular atomic states, relevant numbers of atoms, the
scattering and transverse-confinement lengths, etc., are given too.

\section{Management of matter-wave solitons in BEC with time modulation of
the scattering length and trapping potential}

The present section deals with nonautonomous solitons produced by 1D
self-defocusing GP equations with the nonlinearity coefficient subject to
spatiotemporal modulation and a time-dependent trapping potential. The
respective model is the basic one in the framework of the topic considered
in this review. We start by presenting stable higher-order modes\ in the
form of arrays of dark solitons nested in a finite-width background. Then,
we show that the dynamics of those modes from this class which are not
stable can be efficiently controlled (attenuated or completely suppressed)
by varying the nonlinearity strength in time.

Results collected in this section are based on original works \cite%
{2.12,2.13,2.6,2.7} and \cite{2.5a}.

\subsection{The model and stationary solitons}

\subsubsection{Formulation of the model}

The setting addressed in this section is based on the 1D NLS/GP equation,
written in the normalized form:%
\begin{equation}
i\frac{\partial \psi }{\partial \xi }=-\frac{\partial ^{2}\psi }{\partial
x^{2}}+g(\xi ,x)\left\vert \psi \right\vert ^{2}\psi +V(\xi ,x)\psi ,
\label{eq_2.1}
\end{equation}%
where $\psi $ is the mean-field wave function of the quasi-1D (cigar-shaped)
BEC if $\xi $ stands for time, $x$ is the axial coordinate, $V\left( \xi
,x\right) $ the time-dependent axial potential, and the strength of the
cubic nonlinearity, $g$, is proportional to the interatomic \textit{s}-wave
scattering length. The effective 1D equation is derived from the full 3D GP
equation if the cigar-shaped configuration is confined, in transverse plane $%
\left( y,z\right) $, by the strong transverse HO potential with frequency $%
\omega _{\perp }$ and the respective confinement radius $a_{\perp }=\sqrt{%
\hslash /(m\omega _{\bot })}$, where $m$ is the atomic mass. To this end,
the 3D mean-field wave function is approximately factorized into the product
of its axial (1D) counterpart, in which time $\xi $ and coordinate $x$ are
measured in units of $2/\omega _{\bot }$ and $a_{\bot }$, and the HO
ground-state wave function in the transverse plane \cite{Salasnich}:%
\begin{equation}
\Psi (\mathbf{r},\xi )=\frac{1}{\sqrt{2\pi a_{\mathrm{B}}a_{\bot }}}\exp
\left( -i\omega _{\bot }\xi -\frac{y^{2}+z^{2}}{2a_{\bot }^{2}}\right) \psi
\left( \omega _{\bot }\xi ,\frac{x}{a_{\bot }}\right) ,  \label{Psi}
\end{equation}%
where $a_{\mathrm{B}}$ is the Bohr radius. It is relevant to mention that,
for very dense BEC, the 3D $\rightarrow $ 1D dimension reduction may lead to
more complex effective 1D equations, with nonpolynomial nonlinearities \cite%
{Salasnich,Delgado}. Another caveat is that the reduction to the 1D GP
equation is relevant for $a_{\perp }$ taking values in the $\mathrm{\mu }$m
range, which is the usual experimental situation. If the
transverse-confinement radius is reduced much deeper, to make it comparable
to the effective atomic size, the quasi-1D BEC transforms into a different
quantum state, viz., the Tonks-Girardeau gas \cite{TG}.

As outlined above, the spatiotemporal modulation may be imposed on the
nonlinearity coefficient $g(\xi ,x)$ by the FR technique, controlled by
variable magnetic field \cite{2.1,2.2,Zhang,granulation} or by an
appropriately shaped optical field \cite{submicroOL}. Positive and negative
values of $g$ correspond, respectively, to the repulsive (alias
self-defocusing) and attractive (focusing) contact interactions. In the
latter case, the nonlinear self-attraction drives the onset of MI in the
condensate \cite{2.3,RandyMI,CanberraMI}. Note that the spatiotemporal
modulation may create a sign-changing pattern of $g$, with alternating
self-focusing and defocusing regions \cite{Barcelona}.

If $\xi \ $is the propagation distance, Eq. (\ref{eq_2.1}) governs the
propagation of an optical beam in a planar waveguide along the $\xi $
direction, the external potential being, in this case, generated by a
modulation of the local refractive index \cite{2.4}, while various forms of $%
g(\xi ,x)$ may be realized by means of accordingly designed distributions of
nonlinearity-enhancing dopants \cite{2.5}.

\subsubsection{Stationary soliton states for $V=V(x)$ and $g=g(x)$:}

First, we address stationary solutions of Eq. (\ref{eq_2.1}) with both $g$
and $V$ being functions of coordinate $x$ only. Obviously, in the general
case the respective stationary GP equation is not analytically solvable in
the general form. However, there is a method to select specific forms of $%
g(x)$ and $V(x)$ which admit exact solutions \cite{2.6,2.7,2.5a}. To this
end, full soliton solutions to Eq. (\ref{eq_2.1}) were looked for in Ref.
\cite{2.5a} as
\begin{equation}
\psi (\xi ,x)=\rho (x)u(X)\exp \left( -i\mu \xi \right) ,  \label{eq_2.2}
\end{equation}%
where $\mu $ is a real chemical potential (in the application to optics, $%
-\mu $ is the propagation constant). Further, a new coordinate is identified
as
\begin{equation}
X(x)\equiv \int_{-\infty }^{x}ds/\rho ^{2}(s),  \label{X}
\end{equation}%
with real function%
\begin{equation}
\rho (x)=\sqrt{\alpha \varphi _{1}^{2}+2\beta \varphi _{1}\varphi
_{2}+\gamma \varphi _{2}^{2}},  \label{eq_2.3}
\end{equation}%
where $\varphi _{1,2}(x)$ are two linearly independent solutions of the
ordinary differential equation (ODE)
\begin{equation}
d^{2}\varphi /dx^{2}+\left[ \mu -V(x)\right] \varphi =0,  \label{eq_2.4a}
\end{equation}%
$\alpha ,\beta >0$ and $\gamma $ being real constants satisfying constraint
\begin{equation}
\alpha \gamma -\beta ^{2}>0  \label{abg}
\end{equation}%
(these conditions imply that $\rho (x)$ remains real and never vanishes).
Inserting expression (\ref{eq_2.2}) under conditions (\ref{eq_2.3}) and (\ref%
{eq_2.4a}) and defining the nonlinearity coefficient as%
\begin{equation}
g(x)=\frac{F(u)}{\rho ^{6}(x)},  \label{eq_2.4}
\end{equation}%
where $F(u)$ is another real function, leads to ODE%
\begin{equation}
Eu=\frac{d^{2}u}{dX^{2}}+F(u)u^{3}\text{, \ \ }E=\left( \alpha \gamma -\beta
^{2}\right) W^{2},  \label{eq_2.5}
\end{equation}%
where $W=W\left[ \varphi _{1},\varphi _{2}\right] =\varphi _{1}d\varphi
_{2}/dx-\varphi _{2}d\varphi _{1}/dx$ is the constant Wronskian of the pair
of solutions $\varphi _{1,2}(x)$. Because condition (\ref{abg}) guarantees
that $\rho (x)$ does not vanish at all $x$, the coefficient function $g$
defined by Eq. (\ref{eq_2.4}) does not have singularities. Moreover,
condition (\ref{abg}) also secures $E>0$ in Eq. (\ref{eq_2.5}).

Finally, looking for exact soliton solutions to the underlying equation (\ref%
{eq_2.1}) in the form given by Eqs. (\ref{eq_2.2})-(\ref{eq_2.3}) amounts to
finding exact solutions of ODE (\ref{eq_2.5}), which may be obtained with
various choices of function $F(u)$. As it follows from Eq. (\ref{eq_2.4}), $%
F(u)$ is positive or negative in the whole spatial domain for the NLS
equation whose nonlinearity is, respectively, defocusing (positive) or
focusing (negative). In particular, the choice of $F(u)=g_{3}+g_{5}u^{2}$,
casts Eq. (\ref{eq_2.5}) into the following solvable cubic-quintic ODE: $%
d^{2}u/dX^{2}+g_{3}u^{3}+g_{5}u^{5}+\left( \beta ^{2}-\alpha \gamma \right)
W^{2}u=0,$ where the cubic and quintic coefficients $g_{3}$ and $g_{5}$ with
opposite signs, $g_{3}g_{5}<0$, correspond to the \textit{competing
nonlinearities}.

In what follows, we consider exact solutions of Eq. (\ref{eq_2.5}) for a
more sophisticated choice,
\begin{equation}
F(u)=\frac{Eu-\sin \left[ Eu\right] }{u^{3}},  \label{eq_2.6}
\end{equation}%
which makes Eq. (\ref{eq_2.5}) tantamount to the stationary sine-Gordon
equation, $d^{2}u/dX^{2}+\sin \left( Eu\right) =0$ \cite{2.5a}. Under
condition $E>0$, it represents motion of a pendulum, i.e., either
oscillations, with $u$ being a periodic function of $X$, or the rotation,
with $U$ linearly growing, on the average. Relevant solutions of Eq. (\ref%
{eq_2.1}) correspond to periodic solutions for $u(X)$. Thus, one takes%
\begin{equation}
u(X)=\frac{2}{E}\arcsin \left[ q~\mathrm{sn}\left( \sqrt{E}X,q\right) \right]
,  \label{eq_2.7}
\end{equation}%
where $\mathrm{sn}$ is the Jacobi's elliptic sine with modulus $q$, the
respective period of $u(X)$, as given by Eq. (\ref{eq_2.7}), being $4K(q)/%
\sqrt{E}$, where $K(q)$ is the complete elliptic integral of the first kind.

Inserting expressions (\ref{eq_2.6}) and (\ref{eq_2.7}) in Eq. (\ref{eq_2.4}%
), one finds that the nonlinearity coefficient supporting this exact
solution has the defocusing sign, i.e., $g(x)>0$. Further, an external axial
potential $V(x)$ is needed to confine the solution. The most physically
relevant confining potential is the HO one, $V(x)=\omega _{0}x^{2}$. In this
case, solutions $\varphi _{1}(x)$ and $\varphi _{2}(x)$ of ODE (\ref{eq_2.4a}%
) can be expressed in terms of the Whittaker and Watson functions $M$ and $W$
\cite{2.8}, and the Gamma function, $\Gamma $, as follows:
\begin{gather}
\varphi _{1}(x)=\frac{\text{\textrm{sign}}\left( x\right) }{\omega _{0}^{3/4}%
\sqrt{\left\vert x\right\vert }}M(x)\text{, \ \ }  \notag \\
\varphi _{2}(x)=\frac{1}{\sqrt[4]{\omega _{0}}\sqrt{\pi \left\vert
x\right\vert }}\left[ \Gamma \left( \frac{3\omega _{0}-\mu }{4\omega _{0}}%
\right) W(x)+2\sqrt{\pi }M(x)\right] ,  \label{varphi}
\end{gather}%
provided that
\begin{equation}
\mu \neq \widetilde{\mu }^{(\ell )}\equiv \left( 2\ell +1\right) \omega
_{0},\ell =0,1,2,\cdots ,  \label{mu}
\end{equation}%
with $\widetilde{\mu }^{(\ell )}$ being the $\ell $-th HO\ energy
eigenvalue. Because $\varphi _{1}(0)=d\varphi _{2}/dx|_{x=0}=0$ and $%
d\varphi _{1}/dx|_{x=0}=\varphi _{2}(0)=1$, the respective Wronskian is $%
W=-1 $.

\begin{figure}[tbp]
\centerline{\includegraphics[scale=.95]{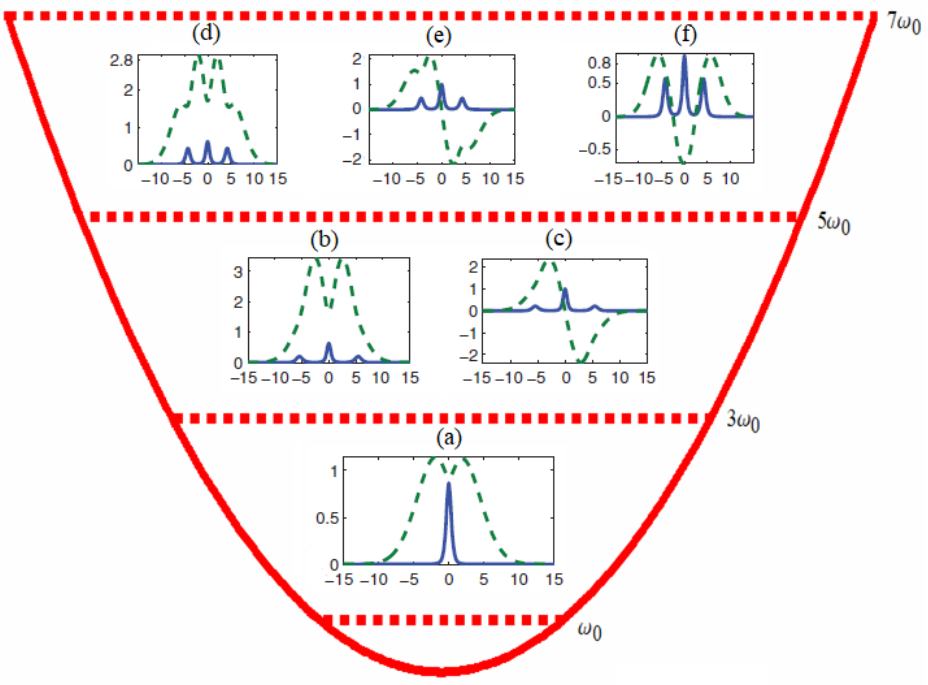}}
\caption{(Color online) The relation between the number of exact soliton
solutions of Eq. (\protect\ref{eq_2.1}), given by Eqs. (\protect\ref{eq_2.2}%
), (\protect\ref{X}), (\protect\ref{eq_2.3}), (\protect\ref{eq_2.4}), and (%
\protect\ref{eq_2.7}), and the discrete energy levels of the
harmonic-oscillator potential (the underlying parabola). (a) Modulation
profiles of the defocusing nonlinearity $g(x)$ (the solid line), and the
shape of the exact soliton (the dashed line), for $\protect\mu =0.2$, $n=1$,
and $q=0.7458$. The horizontal axis is coordinate $x$. Other plots display
the same for (b) $\protect\mu =0.4$, $n=1$, $q=0.9984$, (c) $\protect\mu %
=0.4 $, $n=2$, $q=0.8586$, (d) $\protect\mu =0.6$, $n=1$, $q=0.9998$, (e) $%
\protect\mu =0.6$, $n=2$, $q=0.9530$, and (f) $\protect\mu =0.6$, $n=3,$ $%
q=0.5933$. Other parameters are $\protect\alpha =\sqrt[3]{6}$, $\protect%
\beta =0$, $\protect\gamma =1$, and $\protect\omega _{0}=0.1$. Source:
Reproduced from Ref. \protect\cite{2.5a}.}
\label{fig01}
\end{figure}

Here we focus on the basic case of symmetric profiles of the nonlinearity
modulation and the corresponding solitons by setting $\beta =0$ in Eq. (\ref%
{eq_2.3}). To produce a typical example, one can set $\omega _{0}=0.1$, $%
\alpha =6^{1/3}$, and $\gamma =1$, which makes $g(x)\sim 1$ close to the
midpoint, $x=0$. Then, for given $\mu $, $\rho (x)$ and $X(x)$ can be
calculated. To meet the zero boundary conditions at $x=\pm \infty $, which
implies the localization of the solution, elliptic modulus $q$ must satisfy
a constraint, $2nK(q)=\sqrt{E}R$, where $n$ is a positive integer and $%
R\equiv \int_{-\infty }^{+\infty }ds/\rho ^{2}(s)$. It follows from
condition $K(q)>\pi /2$ that $n$ is bounded from above, $n<\sqrt{E}R/\pi $,
hence there is only a finite number of the exact solutions for given $\mu $.
Integer $n$ is the order of the soliton mode, which features $n-1$ density
nodes. Further analysis demonstrates that $\sqrt{E}R/\pi $ increases
monotonically as $\mu $ increases, while other parameters are fixed, and $n$
takes values up to $\ell +1$ when $\widetilde{\mu }^{(\ell )}<\mu <%
\widetilde{\mu }^{(\ell +1)}$ (see Eq. (\ref{mu})), there being no solutions
at $\mu <\widetilde{\mu }^{(0)}$, see Fig. \ref{fig01}. An explanation for
this result is that the defocusing nonlinearity pushes the energy levels up
relative to the HO spectrum. Thus, the fundamental solitons ($n=1$) exist at
$\mu >\widetilde{\mu }^{(0)}$, first-order excited solitons ($n=2$) exist at
$\mu >\widetilde{\mu }^{(1)}$, and so on. This conclusion coincides with
findings reported in Ref. \cite{2.9} for the spatially homogeneous
defocusing nonlinearity. A similar result holds for gap solitons in
self-defocusing media: only first $n$ families of the solitons exist when $%
\mu $ lies in the $n$-$\mathrm{th}$ optical-lattice-induced bandgap \cite%
{2.10}.

\begin{figure}[tbp]
\centerline{\includegraphics[scale=.9]{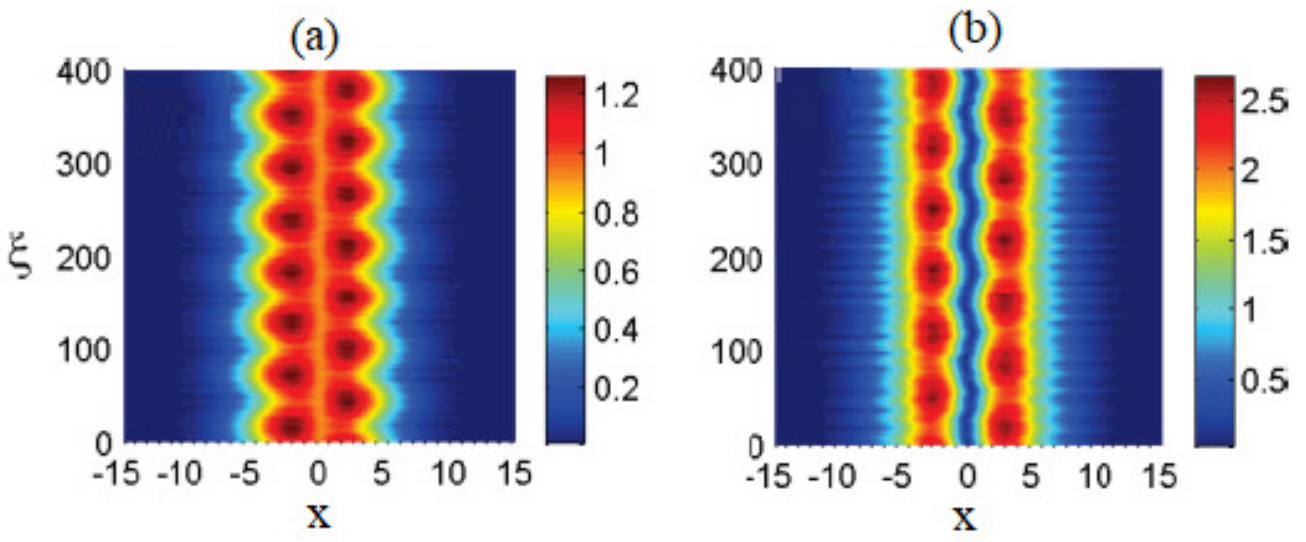}}
\caption{(Color online) (a) Oscillations of the soliton from Fig. \protect
\ref{fig01}(a) which was left-kicked with velocity $0.1$, the corresponding
initial condition for Eq. (\protect\ref{eq_2.1}) being $\protect\rho %
(x)u[X(x)]\exp \left( -0.05ix\right) $. (b) Oscillations of the soliton from
Fig. \protect\ref{fig01}(c), which was left-kicked with initial velocity $%
0.05$. Reproduced from Ref. \protect\cite{2.5a}. }
\label{fig02}
\end{figure}

In accordance with the above analysis, we show in Fig. \ref{fig01}(a) that
only one exact soliton exists when $\mu =0.2$. The corresponding
single-humped nonlinearity-modulation profile is localized near $x=0$, and
the exact soliton has a small dip, because of the self-repulsive sign of the
nonlinearity. Stability of the exact soliton solutions was checked by means
of direct simulations of the perturbed evolution in the framework of Eq. (%
\ref{eq_2.1}). In particular, if kicked with initial velocity $0.1$, the
soliton remains stable, featuring periodic oscillations in the trapping, as
shown in Fig. \ref{fig02}(a).

\begin{figure}[tbp]
\centerline{\includegraphics[scale=.9]{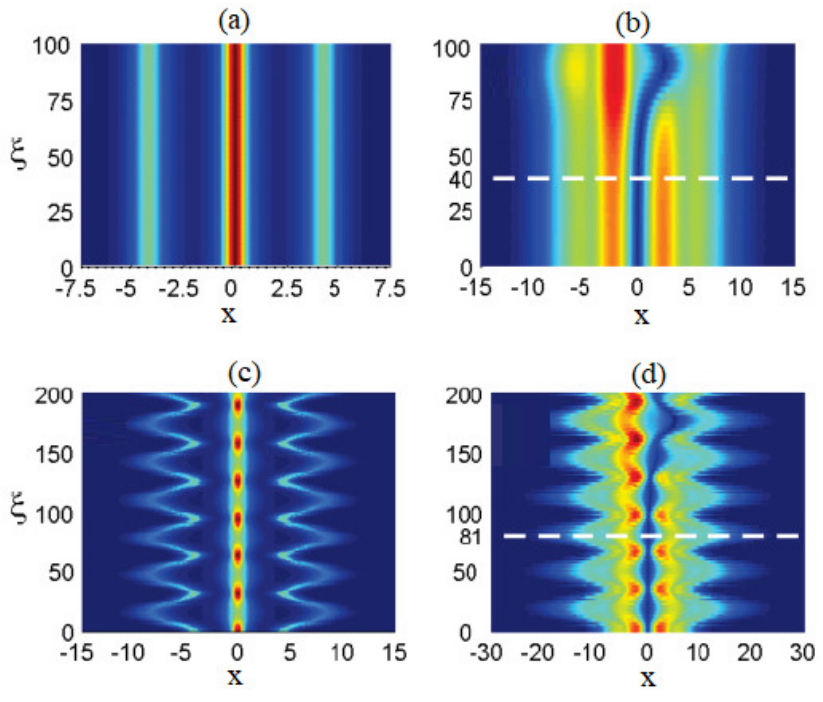}}
\caption{(Color online) (a) The profile of the spatially modulated
nonlinearity and (b) unstable evolution of the soliton from Fig. \protect\ref%
{fig01}(e). (c) The spatiotemporal nonlinearity modulation and (d) the
unstable evolution of the soliton, for the same initial conditions as in
panel (b), but with the nonlinearity taken as per Eq. (\protect\ref{eq_2.9}%
), and with half the trapping frequency, $\protect\omega =\protect\omega %
_{0}/2$. Reproduced from Ref. \protect\cite{2.5a}. }
\label{fig03}
\end{figure}

\begin{figure}[tbp]
\centerline{\includegraphics[scale=.81]{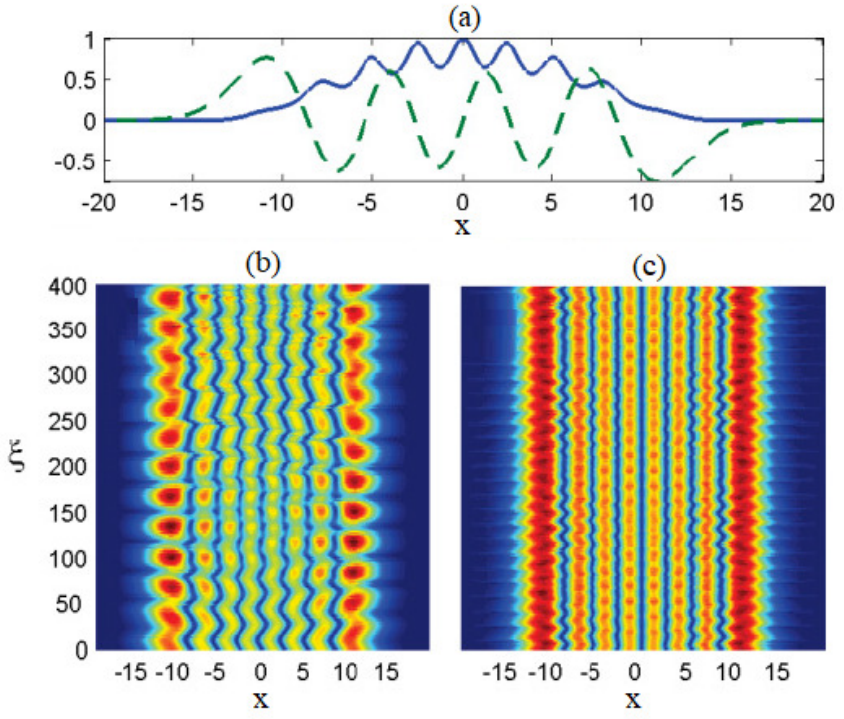}}
\caption{(Color online) (a) The modulation profile of the defocusing
nonlinearity and the corresponding exact soliton (solid and dashed lines,
respectively) for $\protect\mu =1.6$, $n=8$, and $q=0.4752$. (b,c) The
quasi-stable evolution of the soliton. The input for Eq. (\protect\ref%
{eq_2.1}), used to generate panel (b), is $\protect\rho (x)u[X(x)]\exp
\left( -0.05ix\right) $, while panel (c) was created by input $\protect\rho %
(x)u[X(x)]\exp \left( -0.05i\left\vert x\right\vert \right) $. Reproduced
from Ref. \protect\cite{2.5a} }
\label{fig04}
\end{figure}

With the help of the numerical integration in imaginary time, it was found
that the fundamental soliton in the present model represents the ground
state, which explains its robustness. As the solution parameter $\mu $
increases, more and more exact solitons appear, while the nonlinearity
profile develops several peaks. In particular, it is seen from Figs. \ref%
{fig01}(b) and \ref{fig01}(c) that two exact solitons exist in the case of $%
\mu =0.4$. Figure \ref{fig02}(b) demonstrates that, besides the fundamental
soliton, the first excited-state solution, which looks like a dark soliton
nested in a finite-width background, is also stable. However, for $\mu =0.6$%
, the exact first and second excited-state solitons are unstable, see Fig. %
\ref{fig03}(b) for the former one. While it was not easy to analyze the
stability of all the solitons corresponding to the excited states, it was
possible for larger values of $\mu $ to generate arrays of nested dark
solitons which maintain their stability---in particular, if small spatially
homogeneous (Fig. \ref{fig04}(b)) or inhomogeneous (Fig. \ref{fig04}(c))
kicks are applied to them. This finding suggests a new approach to
constructing soliton chains in the form of the Newton's cradle \cite{cradle}%
, and creating \textit{supersolitons}, i.e., robust localized excitations
running through a soliton chain \cite{supersol} in systems described by the
scalar NLS equation. In BEC trapped in a shallow HO potential, chains of
dark solitons supporting the propagation of supersoliton modes, can be
readily generated in the experiment by means of the phase-imprinting
technique (PIT) \cite{imprint,2.11,2.5a,Hamburg}.

Note that, for a fixed chemical potential $\mu $, there is one-to-one
correspondence between the exact solution and nonlinearity profile $g(x)$.
On the other hand, for fixed $g(x)$, one can always find other numerical
soliton solutions, by varying $\mu $ and using the relaxation method. It was
checked that, if the exact soliton is stable, its counterparts numerically
found for the same spatially modulated nonlinearity coefficient $g(x)$ are
also stable, provided that the chemical potential $\mu $ does not vary too
much \cite{2.5a}. Further, it was checked that the numerically constructed
solitons remain stable too when the nonlinearity profiles were taken
somewhat different from the special form defined by (\ref{eq_2.4}).
Therefore, higher-order solitons are physically meaningful objects.

\subsection{Stability control of nonautonomous solitons}

Next, following Ref. \cite{2.12}, we address the model based on Eq. (\ref%
{eq_2.1}) in which $V$ and $g$ are functions of both $\xi $ and $x$. With
specially designed coefficient functions $g(\xi ,x)$ and $V\left( \xi
,x\right) $, exact soliton solutions can be constructed by dint of a
self-similar transformation which reduces the original equation to the
standard integrable NLS equation \cite{2.13}. The transformation is
introduced as%
\begin{equation}
\psi (\xi ,x)=\frac{1}{\sqrt{w(\xi )}}\Phi (T,Y)\exp \left( -\frac{1}{4w}%
\frac{dw}{d\xi }x^{2}\right) ,  \label{eq_2.10}
\end{equation}%
where
\begin{equation}
T\equiv \int_{0}^{\xi }ds/w^{2}(s),~Y\equiv x/w(\xi ).  \label{TY}
\end{equation}%
Inserting expression (\ref{eq_2.10}) in Eq. (\ref{eq_2.1}) and choosing $w$
as a solution of ODE%
\begin{equation}
\frac{d^{2}w}{d\xi ^{2}}+4\omega (\xi )w-4\frac{\omega _{0}^{2}}{w^{3}(\xi )}%
=0  \label{eq_2.10a}
\end{equation}%
yields%
\begin{equation}
i\frac{\partial \Phi }{\partial T}=-\frac{\partial ^{2}\Phi }{\partial Y^{2}}%
+\omega _{0}^{2}Y^{2}\Phi +g_{2}(Y)\left\vert \Phi \right\vert ^{2}\Phi ,
\label{eq_2.8}
\end{equation}%
if the cubic nonlinearity parameter $g$ and external potential $V$ are
chosen as%
\begin{equation}
g(\xi ,x)=g_{1}(\xi )g_{2}(Y),\text{ \ \ }V(\xi ,x)=\omega (\xi )x^{2},\text{
\ with }g_{1}(\xi )=1/w(\xi )\text{.}  \label{eq_2.9}
\end{equation}%
As seen from Eq. (\ref{eq_2.9}), $w(\xi )$ is the width of the nonlinearity
modulation, and determines the soliton's width.

To produce exact solutions of Eq. (\ref{eq_2.8}), the exact stationary
soliton solutions displayed in Fig. \ref{fig01} are used, replacing $Y$ by $%
x $, $g_{2}$ by $g$, $T$ by $\xi $ , and $\Phi $ by $\psi $. Moreover, given
the functional form of $\omega (\xi )$, such as, e.g., $a_{0}+a_{1}\cos \xi $%
, equation (\ref{eq_2.10a}) for $w(\xi )$ can be solved. Then, one can
construct exact nonstationary soliton solutions of Eq. (\ref{eq_2.1}) as per
Eq. (\ref{eq_2.10}).

The stability of the nonstationary solitons is related to that of the
corresponding stationary solution and the functional form of $w(\xi )$. If
the stationary soliton of Eq. (\ref{eq_2.8}) is stable, the respective
nonstationary soliton of Eq. (\ref{eq_2.1}), produced by the transformation
of the stable stationary one, is stable too. However, if the stationary
soliton is unstable, with the instability setting in, say, at $T=T_{0}$, the
instability of the nonstationary soliton may be delayed or even suppressed,
choosing an appropriate form of $w(\xi )$. Namely, if $T$, defined as per
Eq. (\ref{TY}), remains smaller than $T_{0}$, the instability is completely
prevented, because the system does not have enough time (propagation
distance) to reach the instability threshold. Otherwise, the onset of the
instability is delayed.

As a typical example, we display the exact soliton corresponding to Fig. \ref%
{fig01}(e). Without the temporal nonlinearity modulation, its instability
starts at $\xi \approx 40$ [see Fig. \ref{fig03}(b)]. If, at $\xi =0$, the
trapping frequency is abruptly reduced by half, and after that the
nonlinearity is varied as per Eq. (\ref{eq_2.9}) with $w(\xi )=\sqrt{1+3\sin
^{2}\left( \omega _{0}\xi \right) }$ (see Fig. \ref{fig03}(c)), one finds
that the soliton first experiences exact self-similar evolution in
accordance with Eq. (\ref{eq_2.10}), and then it starts to develop
instability at $\xi \approx 81$, determined by condition $40=\int_{0}^{\xi
}ds/w^{2}(s)$, see Fig. 3(d). Clearly, the instability onset is delayed. It
can be delayed further if one decreases the trapping frequency more, varying
the nonlinearity accordingly. It is also relevant to mention that
self-similar dynamical regimes find other important realizations in the
mean-field dynamics, such as the collapse (blowup) regime \cite%
{self-sim,Fibich}.

\subsection{Conclusion of the section}

The subject of the section is to demonstrate well-known results which play
the basic role in the topic of the present review, as they are produced by
the basic model. These results produce exact soliton solutions to the NLS
equation with the coefficient in front of the cubic nonlinearity and
external potential subject to specially designed spatiotemporally
modulation, that allows one to explicitly transform the equation into the
classical integrable NLS equation. The number of solitons is determined by
the value of the chemical potential and discrete energy levels of the
trapping HO potential. The existence of stable higher-order modes, built as
arrays of dark solitons embedded in the finite-width background, is
demonstrated too. Finally, it is shown how one can control instability of
nonstationary solitons, by choosing the temporal modulation which delays or
completely eliminates the onset of the instability.

\section{Engineering nonautonomous solitons in Bose-Einstein condensates
with a spatially modulated scattering length}

As a characteristic example of BEC models with a spatially modulated
scattering length, diverse versions of which have been theoretically
elaborated in many works, see a review in Ref. \cite{Barcelona}, we here
consider, first, the corresponding cubic GP equation, and then the
application of PIT to obtain a nonautonomous cubic derivative NLS equation,
which includes a time-dependent HO potential. PIT is a relatively new tool
used for wave-function engineering in BEC. It may be extended to control the
wave function by means of absorption provided by proximity to a resonance
with frequencies of external laser illumination. The action of PIT onto BEC
amounts to modifying the phase pattern in the mean-field wave function---for
example, by exposing the condensate to the action of pulsed, off-resonant
laser light with a specially designed intensity pattern. As a result, atoms
experience the action of a spatially varying light-induced potential, and
thus acquire the corresponding phase. The main advantage of the application
of PIT for BECs is that it conserves the total number of atoms.

Results collected in this section are based on original works \cite{3.39},
\cite{3.35}, \cite{3.33}, and \cite{3.33b}. Some methods presented in the
section refer to a recently published book \cite{3.5}.

\subsection{Introduction to the section}

In the derivative NLS equation with the spatiotemporal modulation of the
contact nonlinear term and time-variable potential,%
\begin{equation}
i\,\frac{\partial \phi (x,t)}{\partial t}+\left[ \frac{\partial ^{2}}{%
\partial x^{2}}+\widetilde{g}(x,t)\left\vert \phi (x,t)\right\vert ^{2}+%
\widetilde{V}(x,t)\right] \phi (x,t)+i\widetilde{\beta }\frac{\partial
\left( |\phi |^{2}\phi \right) }{\partial x}=0,  \label{eq_3.1}
\end{equation}%
the derivative cubic term represents the delayed nonlinear response of the
system. Well-known in plasma physics, Eq. (\ref{eq_3.1}) models the
propagation of finite-amplitude Alfv\'{e}n waves in directions nearly
parallel to the external magnetic field in a plasma with the gas pressure
much smaller than the magnetic pressure (low-$\beta $ plasma) \cite{3.26}.
Other physical realizations of the derivative NLS equation are provided by
convection in binary fluids \cite{3.34} and propagation of signals in
electric transmission lines \cite{3.35}. Furthermore, an equation of type (%
\ref{eq_3.1}) governs the behavior of large-amplitude magnetohydrodynamic
waves propagating in an arbitrary direction with respect to the magnetic
field in high-$\beta $ plasmas \cite{3.27}, see also review \cite{Kamchatnov}%
. In nonlinear optics, the propagation equation for very short pulses the
local Kerr nonlinearity has to be supplemented by the derivative \textit{%
self-steepening term}, which accounts for the nonlinear dispersion of the
optical material \cite{3.28,4.2,YangShen,6.3}. If $\widetilde{g}=\mathrm{%
const}$ and the external potential is absent, i.e., $\widetilde{V}=0$, Eq. (%
\ref{eq_3.1}) reduces to the-well known integrable derivative cubic NLS
equation, which can be derived from the usual NLS equation by means of the $%
U(1)$ gauge transformation \cite{KaupNewell,3.29}. Two basic questions
arise, as concerns equations of this type, in the context of this review:
(i) How should one introduce a GP model describing the impact of the cubic
derivative nonlinearity on the condensates? (ii) How does the derivative
cubic term in the GP equation affect the MI in BEC? The main aim of the
present section is to address these questions. To this end, we first
demonstrate derivation of the extended NLS equation of type (\ref{eq_3.1}),
in the case when PIT \cite{3.33,3.33b} is applied to the setting modeled by
GP equation with the spatiotemporally modulated contact--nonlinearity
coefficient and time-dependent HO trapping potential.

\subsection{The cubic inhomogeneous NLS equation}

As mentioned above, FRs may be widely used to control the nonlinearity of
matter waves, by modulating the scattering length of inter-atomic collisions
temporarily, spatially, or spatiotemporally, which leads to generation of
many novel nonlinear phenomena \cite{3.30,3.31,3.32}. In particular, as
outlined in the Introduction, it has been predicted that time-dependent
modulation of the scattering length can be used to stabilize attractive 2D
BECs against the critical collapse \cite{3.1}, and to create robust
matter-wave breathers in 1D BECs \cite{3.2}. It has been found too that
atomic matter waves exhibit novel features under the action of a spatially
varying scattering length, i.e., with a spatially varying mean-field
nonlinearity \cite{3.3,3.4,HSBM,Barcelona}.

Here, the starting point is the GP equation (\ref{eq_2.1}), written as
\begin{equation}
i\,\frac{\partial u(x,t)}{\partial t}+\left[ \frac{\partial ^{2}}{\partial
x^{2}}+g(x,t)\left\vert u(x,t)\right\vert ^{2}+V(x,t)\right] u(x,t)=0,
\label{eq_3.01}
\end{equation}%
with the time-dependent HO potential,
\begin{equation}
V(x,t)=-\alpha (t)x^{2}.  \label{eq_3.02}
\end{equation}%
The strength of the HO trap $\alpha (t)$ may be negative or positive,
corresponding to the confining or expulsive potential, respectively. In most
experiments, factor $\alpha (t)$ is typically fixed to a constant value, but
adiabatic changes in the strength of the trap are experimentally feasible
too.

In the case of the cigar-shaped BEC, the aforementioned self-consistent
reduction of the 3D GP equation to the 1D form with the external potential
can be provided by means of a multiple-scale expansion \cite{3.5} which
exploits a small parameter $\delta ^{2}=(Na_{s}/a_{0})\alpha \ll 1$, where $%
a_{s}$ is the $s$-wave scattering length. Parameter $\delta $ evaluates the
relative strength of the two-body interactions as compared to the kinetic
energy of the atoms. In the quasi-1D case, where the dynamics along the
cigar's axis is of primary interest, the same small parameter $\delta $
defines the ratio of the tight transverse confinement to a characteristic
scale along the longitudinal axis, as $a_{\bot }/a_{0}\sim \delta \sqrt{%
\alpha }$. For example, for BEC composed of $N=10^{4}$ of $^{23}\mathrm{Na}$
atoms (with $a_{s}\approx 2.75$ $\mathrm{nm}$) the characteristic lengths
are $a_{0}=300$ $\mathrm{\mu m}$ and $a_{\bot }=10$ $\mathrm{\mu m}$, which
means $\alpha =0.11$ and $\delta ^{2}\simeq 0.01$. \

Recall that the rescaled 1D mean-field wave function $u(x,t)$ of the
condensate appearing in Eq.~(\ref{eq_3.01}) is connected to the underlying
3D order parameter, $\Psi (\mathbf{r},t)$, by relation
\begin{equation}
\Psi (\mathbf{r},t)=\frac{\delta }{a_{\bot }\sqrt{a_{s}}}\exp (-i\omega
_{\bot }t)\exp \left( -\frac{\mathbf{r}^{2}}{2a_{\bot }^{2}}\right) u\left(
\frac{\delta x}{a_{\bot }},\frac{1}{2}\delta ^{2}\omega _{\bot }\,t\right) ,
\label{eq_3.03}
\end{equation}%
where $\mathbf{r=(}y,z\mathbf{)}$ are coordinates in the transverse plane,
and $\omega _{\bot }$ is the confining HO frequency in this plane. Potential
$V(x,t)$ appearing in Eq. (\ref{eq_3.01}) is measured in units of $\hslash
^{2}a_{\bot }^{2}/8m$. Under the above conditions, the sign of the cubic
nonlinearity coefficient $g(x,t)\ $is opposite to that of $a_{s}$, i.e., $%
g(x,t)$ is positive and negative for the focusing or defocusing
nonlinearity, respectively. In this section, the spatiotemporal modulation
of the interaction coefficient is introduced in the spatially-linear form:%
\begin{equation}
g(x,t)=x\widetilde{g}(t)+\widetilde{g}_{0}(t),  \label{eq_3.03a}
\end{equation}%
where $\widetilde{g}(t)$ and $\widetilde{g}_{0}(t)$\ are real functions of
time.

In the framework of inhomogeneous NLS equation (\ref{eq_3.01}) with the
time-varying HO potential (\ref{eq_3.02}) and the focusing sign of the
nonlinearity, MI was investigated in Ref. \cite{2.3}, and recently
experimentally demonstrated in \cite{RandyMI}. The present section,
following Refs. \cite{3.33} and \cite{3.33b}, addresses MI in the framework
of a modified version of the GP equation (\ref{eq_3.01}) with potential (\ref%
{eq_3.02}) and two cubic terms, including the derivative one. In particular,
the linear-stability analysis yields an analytical expression for the MI
gain of the BEC state.

\subsubsection{Derivation of the cubic derivative inhomogeneous NLS}

To introduce the inhomogeneous cubic derivative NLS equation, one applies
PIT to the mean-field wave function $u(x,t)$ governed by the usual NLS, to
generate a new wave function $\psi (x,t)$ \cite{3.33b}:
\begin{subequations}
\begin{eqnarray}
u(x,t) &=&\psi (x,t)\exp \left[ -i\theta (x,t)\right] ,  \label{eq_3.04a} \\
\partial \theta /\partial x &=&-3\beta (t)\left\vert \psi \right\vert ^{2},
\label{eq_3.04b} \\
\partial \theta /\partial t &=&i\beta (t)\left( \psi \partial \psi ^{\ast
}/\partial x-\psi ^{\ast }\partial \psi /\partial x\right) +9\beta
^{2}(t)\left\vert \psi \right\vert ^{4}.  \label{eq_3.04c}
\end{eqnarray}%
In Eqs. (\ref{eq_3.04a})--(\ref{eq_3.04c}), $\theta (x,t)$ is the imprinted
real phase, which is constructed according to Eqs. (\ref{eq_3.04b}) and (\ref%
{eq_3.04c}), with a time-dependent real coefficient $\beta (t)$. PIT in this
form can be realized experimentally by instantaneously exposing BEC,
governed by the GP equation, to the action of a properly designed optical
field \cite{3.33,3.33b}, with coefficient $\beta (t)$ representing the
phase-imprint strength.

Inserting ansatz (\ref{eq_3.04a}) in Eq.~(\ref{eq_3.01}), after
straightforward manipulations one arrives at the derivative NLS equation:
\end{subequations}
\begin{equation}
i\,\frac{\partial \psi }{\partial t}+\left[ \frac{\partial ^{2}}{\partial
x^{2}}+V(x,t)+g(x,t)\left\vert \psi \right\vert ^{2}\right] \psi +4i\beta
(t)\,\frac{\partial (|\psi |^{2}\psi )}{\partial x}=0,  \label{eq_3.05}
\end{equation}%
see further technical details in Ref. \cite{3.33b}. Note that the
transformation defined by Eqs. (\ref{eq_3.04a})-(\ref{eq_3.04c}) conserves
the norm of the wave function, i.e., it does not affect the number $N$ of
atoms of the condensate.

Thus far, the PIT scheme which would exactly lead to the setting described
by Eq. (\ref{eq_3.05}) was not realized experimentally in BEC. However, a
similar experimental result, \textit{viz}., creation of quasi-1D dark
solitons by means of sufficiently strong PIT, was reported recently \cite%
{Hamburg}

\subsubsection{Modulational instability in the cubic derivative NLS equation
with constant coefficients and without external potential}

Before addressing MI for the cubic derivative NLS (\ref{eq_3.05}) in its
full form, it is relevant, following Refs. \cite{Kamchatnov} and \cite{3.35}%
, to recapitulate results for MI in the equation with constant coefficients
and without an external potential:%
\begin{equation}
i\,\frac{\partial \psi }{\partial t}+\left( \frac{\partial ^{2}}{\partial
x^{2}}+g_{0}\left\vert \psi \right\vert ^{2}\right) \psi +4i\beta \,\frac{%
\partial \left( |\psi |^{2}\psi \right) }{\partial x}=0.  \label{eq_3.06}
\end{equation}%
For arbitrary real constants $\phi _{0}$ and $q$, a continuous-wave (CW),
i.e., constant-amplitude, solution of Eq.~(\ref{eq_3.06}) is
\begin{equation}
\psi (x,t)=\phi _{0}\exp \left\{ iqx-i[q^{2}+\left( 4\beta q-g_{0}\right)
\phi _{0}^{2}]t\right\}  \label{eq_3.07}
\end{equation}%
Following the usual scheme of the MI analysis \cite{6.3}, one perturbs
solution (\ref{eq_3.07}) as follows:
\begin{equation}
\psi (x,t)=[\phi _{0}+\varepsilon (x,t)]\exp \left( iqx-i[q^{2}+\left(
4\beta q-g_{0}\right) \phi _{0}^{2}]t\right) ,  \label{eq_3.08}
\end{equation}%
where $\varepsilon (x,t)$ is a small complex perturbation, satisfying $%
\left\vert \varepsilon (x,t)\right\vert \ll \left\vert \phi _{0}\right\vert $%
. Substituting ansatz (\ref{eq_3.08}) in Eq.~(\ref{eq_3.06}), one derives
the BdG equation:%
\begin{equation}
i\,\frac{\partial \varepsilon }{\partial t}+\frac{\partial ^{2}\varepsilon }{%
\partial x^{2}}+2i\left( q+4\beta \phi _{0}^{2}\right) \frac{\partial
\varepsilon }{\partial x}+4i\beta \phi _{0}^{2}\,\frac{\partial \varepsilon
^{\ast }}{\partial x}+\phi _{0}^{2}\left( g_{0}-4\beta q\right) \left(
\varepsilon +\varepsilon ^{\ast }\right) =0.  \label{eq_3.09}
\end{equation}%
Looking for eigenmodes of the perturbation as%
\begin{equation}
\varepsilon =U_{1}\exp \left[ i\left( Qx-\Omega t\right) \right]
+U_{2}^{\ast }\exp \left[ -i\left( Qx-\Omega ^{\ast }t\right) \right]
\label{eq_3.010}
\end{equation}%
leads to the dispersion relation connecting wave number $Q$ and frequency $%
\Omega $ of the perturbation:
\begin{equation}
\left[ \Omega -2Q\left( q+4\beta \phi _{0}^{2}\right) \right] ^{2}=Q^{2}%
\left[ \phi _{0}^{2}\left( 16\beta ^{2}\phi _{0}^{2}+8q\beta -2g_{0}\right)
+Q^{2}\right] .  \label{eq_3.011}
\end{equation}%
As one can see from Eq. (\ref{eq_3.011}), the MI occurs when the
perturbation wave numbers $Q$ satisfies condition
\begin{equation}
Q^{2}<2\phi _{0}^{2}\left( g_{0}-8\beta ^{2}\phi _{0}^{2}-4q\beta \right) ,
\label{eq_3.012}
\end{equation}%
which may hold for both focusing and defocusing signs of th nonlinearities.
In fact, condition (\ref{eq_3.012}) is valid for values of $\beta $
belonging to interval%
\begin{equation}
D_{\beta }\equiv \left( -2q-2\sqrt{q^{2}+2g_{0}\phi _{0}^{2}}\,\right)
/8\phi _{0}^{2}<\beta <\left( -2q+2\sqrt{q^{2}+2g_{0}\phi _{0}^{2}}\,\right)
/8\phi _{0}^{2}.  \label{D}
\end{equation}%
This condition implies
\begin{equation}
q^{2}+2g_{0}\phi _{0}^{2}>0,  \label{necessary}
\end{equation}%
which is identically satisfied for the focusing nonlinearity ($g_{0}=+1)$;
in the case of the defocusing nonlinearity, ($g_{0}=-1$), inequality (\ref%
{necessary}) holds only for $q$ and $\phi _{0}$ that satisfy constraint $%
q^{2}-2\phi _{0}^{2}>0$. Thus, Eq. (\ref{necessary}) is the necessary
condition for MI of the CW solutions of the cubic derivative NLS equation
with constant coefficients in the free space. At $\beta =0$, MI in the
framework of the usual cubic NLS equation is the classical result by
Benjamin and Feir \cite{BF}.

\begin{figure}[tbp]
\centerline{\includegraphics[scale=.88]{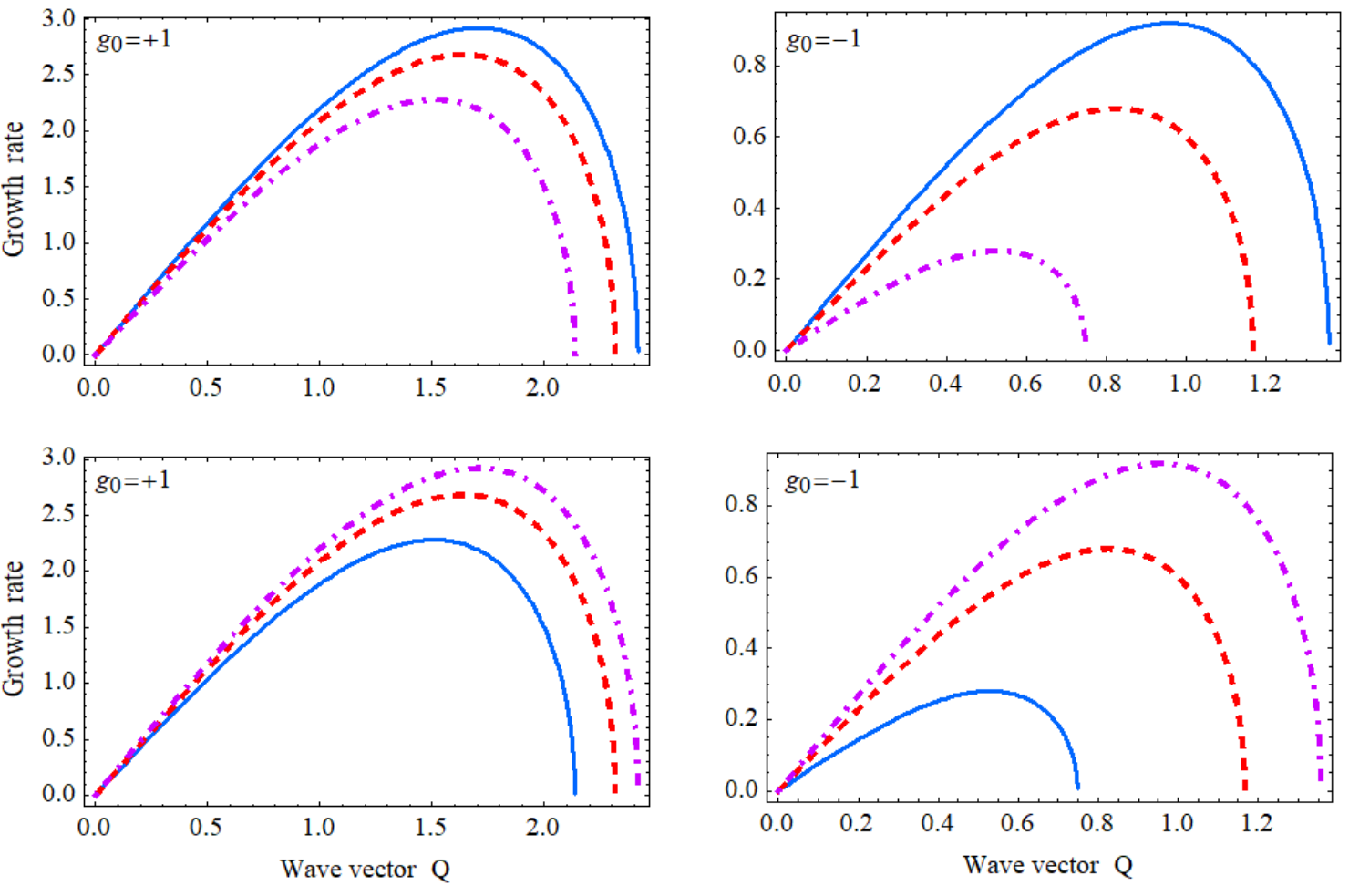}}
\caption{(Color online) The MI gain as produced by Eq.~(\protect\ref%
{eq_3.013}) for three values of the imprint parameter $\protect\beta $, with
$\protect\phi _{0}=1$ and $q=2$. The top panel includes three values of $%
\protect\beta >\protect\beta _{c}=-0.5$: $\protect\beta =-0.4$, $0.3$, and $%
0.2$ (the blue, red, and green lines, respectively). The bottom panel
includes three values of $\protect\beta <\protect\beta _{c}=-0.5$: $\protect%
\beta =-0.8$, $0.7$, and $0.6$ (the blue, red, and green lines,
respectively). The left and right panels correspond to the
focusing and defocusing nonlinearities, respectively. The figure
reproduces known results for MI produced by the derivative NLS
equation with constant coefficients \protect\cite{Kamchatnov}.}
\label{fig05}
\end{figure}

Under condition (\ref{eq_3.012}), the growth rate (gain) of the MI for the
derivative NLS equation in the free space is%
\begin{equation}
\left\vert \text{\textrm{Im}}\Omega \right\vert =\left\vert Q\right\vert
\sqrt{\left( 2g_{0}-16\beta ^{2}\phi _{0}^{2}-8q\beta \right) \phi
_{0}^{2}-Q^{2}}.  \label{eq_3.013}
\end{equation}%
It is evident that function $B(\beta )=2g_{0}-16\beta ^{2}\phi
_{0}^{2}-8q\beta $\ reaches its maximum at the critical point $\beta
_{c}=-q/\left( 4\phi _{0}^{2}\right) $, which does not depend on $g_{0}$.
The presence of the imprint parameter $\beta $ significantly modifies the
instability domain and brings new effects. In particular, it allows MI in
the case of the defocusing nonlinearity ($g_{0}=-1$). In Fig. \ref{fig05} we
plot the MI gain, defined as per Eq.~(\ref{eq_3.013}), for different values
of $\beta $ with $\phi _{0}=1$ and $q=2$. According to this figure, there
are two scenarios, depending on whether $\beta $, belonging to interval (\ref%
{D}), is above or below the critical value ($\beta \geq \beta _{c}$ or $%
\beta \leq \beta _{c}$). In the top panel of Fig. \ref{fig05}, which
corresponds to $\beta \geq \beta _{c}$, the gain decreases with $\left\vert
\beta \right\vert $, while in the bottom panel, with $\beta \leq \beta _{c}$%
, the gain increases when $\left\vert \beta \right\vert $\ decreases. Thus,
the imprint parameter $\beta $, when taken above $\beta _{c}$, softens MI,
and, on the other hand, MI enhances when $\beta $ falls below $\beta _{c}$.
Comparing the left panel of Fig. \ref{fig05} (for the focusing nonlinearity)
with the right one (the defocusing nonlinearity), it appears that, quite
naturally, MI is stronger in the case of the focusing nonlinearity.

\subsubsection{Modulational instability in the inhomogeneous cubic
derivative NLS equation with the HO potential}

To examine MI in the general case of Eq. (\ref{eq_3.05}), Ref. \cite{3.33b}
made use of the modified lens transformation (LT),%
\begin{equation}
\psi (x,t)=\frac{1}{\ell (t)}\phi (X,T)\exp [if(t)x^{2}+\eta (t)],
\label{eq_3.014}
\end{equation}%
where $T=T(t)$, $\ell (t)$, $\eta (t)$, and $f(t)$ are real functions of
time, and $X(x,t)=x\ell ^{-1}(t)$. Originally, LT was introduced by Talanov
\cite{Talanov} as an invariant transformation for the 2D NLS equation, which
changes the scale of the coordinates and adds the radial chirp (a phase term
quadratic in the radial coordinate) to the wave function. The action of LT
on the wave function is similar to the result of the ray propagation of the
field governed by geometric optics. The significance of LT is stressed by
its compatibility with adiabatic variation of the wave function governed by
the cubic NLS equation, and with the dynamics driven by the critical
collapse in 2D \cite{Gadi,Fibich}. Furthermore, LT applies as well to the 2D
NLS equation including the isotropic HO potential \cite{Gadi}.

To preserve the scaling of Eq. (\ref{eq_3.05}) we set%
\begin{equation}
\frac{dT}{dt}=\frac{1}{\ell ^{2}(t)}.  \label{eq_3.015}
\end{equation}%
Demanding that
\begin{subequations}
\begin{gather}
\frac{df}{dt}+4f^{2}+\alpha =0,  \label{eq_3.016} \\
4f(t)-\frac{1}{\ell }\frac{d\ell }{dt}=0,  \label{eq_3.017} \\
\frac{d\eta }{dt}+2f(t)-\frac{1}{\ell }\frac{d\ell }{dt}=0,
\label{eq_3.017a} \\
\beta (t)=\widetilde{g}(t)f^{-1}(t)/8,  \label{eq_3.018}
\end{gather}%
and inserting ansatz (\ref{eq_3.014}) in Eq. (\ref{eq_3.05}) leads to
\end{subequations}
\begin{equation}
i\frac{\partial \phi }{\partial T}+\frac{\partial ^{2}\phi }{\partial X^{2}}%
+\lambda _{0}(T)\left\vert \phi \right\vert ^{2}\phi +i\lambda (T)\frac{%
\partial (|\phi |^{2}\phi )}{\partial X}=0,  \label{eq_3.020}
\end{equation}%
where real functions of time are%
\begin{equation}
\lambda _{0}(T)=\widetilde{g}_{0}(t)\exp [2\eta (t)],\quad \lambda (T)=4\,%
\frac{\beta (t)}{\ell (t)}\exp [2\eta (t)].  \label{eq_3.021}
\end{equation}%
Thus, the invariance of the inhomogeneous cubic derivative NLS equation with
respect to LT is maintained.

Solving Eqs.~(\ref{eq_3.017}), (\ref{eq_3.017a}) and (\ref{eq_3.015}) in
terms of $f(t)$ yields
\begin{subequations}
\begin{eqnarray}
\ell (t) &=&\ell (0)\exp \biggl(4\int_{0}^{t}f(\upsilon )\,d\upsilon \biggr),
\label{eq_3.019} \\
\eta (t) &=&2\int_{0}^{t}f(\upsilon )\,d\upsilon +\eta (0),
\label{eq_3.019a} \\
T(t) &=&\ell ^{-2}(0)\int_{0}^{t}\exp \biggl(-8\int_{0}^{s}f(\upsilon
)\,d\upsilon \biggr)ds+T(0).  \label{eq_3.019b}
\end{eqnarray}%
Thus, the problem of finding time-dependent parameters $\ell (t)$, $\eta (t)$%
, and $T(t)$ is reduced to solving the Riccati equation~(\ref{eq_3.016}).

According to ansatz (\ref{eq_3.014}) $\eta (t)$ affects the total number of
atoms (the norm of the wave function), $N=\int_{\mathrm{R}}\left\vert \psi
\right\vert ^{2}dx=\ell ^{-1}(t)\exp [\eta (t)]\int_{\mathrm{R}}\left\vert
\phi \right\vert ^{2}dX$, if $\ell ^{-1}(t)\exp [\eta (t)]\neq \text{%
constant.}$ According to Eq. (\ref{eq_3.019a}), $\exp [\eta (t)]=\exp
[2\int_{0}^{t}f(\upsilon )\,d\upsilon +\eta (0)]$, which means that $N$
exponentially grows if $f(t)$, the solution of the Riccati equation (\ref%
{eq_3.016}), is positive, and exponentially decays if $f(t)$ is negative. In
other words, positive $f(t)$ represents the feeding of atoms into the
condensate, while negative $f(t)$ implies loss of atoms. While $f(t)$
depends on the strength of the HO trap, $\alpha (t)$, its sign is
independent of the sign of $\alpha (t)$. For example, if $\alpha =-\alpha
_{0}^{4}$ is a negative constant (which corresponds to the confining HO
potential), then $f=\pm \alpha _{0}^{2}/2$ are two particular solutions of
the Riccati equation (\ref{eq_3.016}).

To investigate MI for the derivative\ NLS equation (\ref{eq_3.020}) with
variable coefficients, Ref. \cite{3.33b} introduced perturbed solutions as
\end{subequations}
\begin{equation}
\phi =\left[ \phi _{0}+\varepsilon (X,T)\right] \exp \biggl[%
-iQX-i\int_{0}^{T}\Omega (\upsilon )\,d\upsilon \biggr],  \label{mi01}
\end{equation}%
where $\Omega (T)$ is a real time-dependent function representing the
nonlinear frequency shift, $\phi _{0}$ is a real constant, $Q$ is the wave
number of the carrier, and $\varepsilon (X,T)$ is a small perturbation.
Substituting ansatz (\ref{mi01}) into Eq.~(\ref{eq_3.020}), linearizing it
with respect to the small perturbation, and substituting
\begin{equation}
\Omega (T)=Q^{2}-\lambda _{0}(t)\phi _{0}^{2}-Q\phi _{0}^{2}\lambda (t),
\label{mi02}
\end{equation}%
one obtains
\begin{equation}
i\,\frac{\partial \varepsilon }{\partial T}+\frac{\partial ^{2}\varepsilon }{%
\partial X^{2}}+2i\left( \lambda \phi _{0}^{2}-Q\right) \,\frac{\partial
\varepsilon }{\partial X}+i\lambda \phi _{0}^{2}\frac{\partial \varepsilon
^{\ast }}{\partial X}+\phi _{0}^{2}\left( \lambda _{0}+\lambda Q\right)
\left( \varepsilon +\varepsilon ^{\ast }\right) ,  \label{mi03}
\end{equation}%
where $\ast $ stands for the complex conjugation. Solutions to Eq. (\ref%
{mi03}) are sought for as

\begin{equation}
\varepsilon =U_{1}\exp \biggl(iKX-i\int_{0}^{T}\omega (\upsilon )\,d\upsilon %
\biggr)+U_{2}^{\ast }\exp \biggl(-iKX+i\int_{0}^{T}\omega ^{\ast }(\upsilon
)\,d\upsilon \biggr),  \label{mi04}
\end{equation}%
where $KX-\int_{0}^{T}\omega (\upsilon )\,d\upsilon $ is the modulation
phase in which $K$ and $\omega $ are the wavenumber and the complex
frequency of the perturbation, $U_{1,2}$ being complex amplitudes. MI sets
in if frequency $\omega $ has a nonzero imaginary part. Inserting
expression~(\ref{mi04}) into Eq.~(\ref{mi03}) yields the time-dependent
dispersion relation,%
\begin{equation}
\left[ \omega -2K\left( \lambda \phi _{0}^{2}-Q\right) \right] ^{2}-K^{2}%
\left[ \allowbreak K^{2}+\phi _{0}^{2}\left( \lambda ^{2}\phi
_{0}^{2}-\allowbreak 2Q\lambda -2\lambda _{0}\right) \right] =0.
\label{mi05}
\end{equation}%
For $\omega $ to have a nonzero imaginary part, it is necessary and
sufficient to have%
\begin{equation}
\allowbreak K^{2}+\phi _{0}^{2}\left( \lambda ^{2}\phi _{0}^{2}-\allowbreak
2Q\lambda -2\lambda _{0}\right) <0.  \label{mi06}
\end{equation}%
Inequality (\ref{mi06}) is the MI criterion for the cubic derivative NLS
equation (\ref{eq_3.020}). If the criterion holds, the local MI gain is
given by
\begin{equation}
\left\vert \text{Im}\,\omega (t)\right\vert =\left\vert K\right\vert \sqrt{%
\phi _{0}^{2}\left[ 2Q\lambda (t)+2\lambda _{0}(t)-\lambda ^{2}(t)\phi
_{0}^{2}\right] -K^{2}}.  \label{mi07}
\end{equation}

A particularly simple and interesting case is one with constant $\lambda
_{0} $ and $\lambda $. Then it follows from Eqs. (\ref{eq_3.015})-(\ref%
{eq_3.018}) that $\beta \ $is constant, while $\alpha (t)$ and $\widetilde{g}%
_{0}(t)$ satisfy the nonlinear second-order ODE,%
\begin{equation}
\widetilde{g}_{0}\frac{d^{2}\widetilde{g}_{0}}{dt^{2}}-2\left( \frac{d%
\widetilde{g}_{0}}{dt}\right) ^{2}-4\alpha \widetilde{g}_{0}^{2}=0.
\label{mi08}
\end{equation}%
Thus, in the special case of constant $\lambda _{0}$, $\lambda $ and $\beta $%
, the problem of finding $\ell (t)$, $\eta (t)$, and $T(t)$ amounts to
solving Eq.~(\ref{mi08}). The simplest possibility is to solve it for $%
\alpha (t)$ if $\widetilde{g}_{0}(t)$ is known. For instance, $\widetilde{g}%
_{0}(t)=\widetilde{a}_{0}\exp \left( \widetilde{\lambda }t\right) $ produces
constant $\alpha =-\widetilde{\lambda }^{2}/4.$

Following Ref. \cite{2.3}, one of the most interesting cases in the setting
with the HO potential is the one with
\begin{equation}
\alpha (t)=A(t+t^{\ast })^{-2},  \label{mi09}
\end{equation}%
for real constants $A$ and $t^{\ast }$ in Eq. (\ref{eq_3.02}), which
determine, respectively, the strength of the potential and its width at $t=0$%
. Inserting expression~(\ref{mi09}) in Eq.~(\ref{mi08}) yields%
\begin{equation}
\widetilde{g}_{0}(t)=\frac{4\lambda _{0}\beta }{\lambda \ell (0)}\left(
\frac{t+t^{\ast }}{t^{\ast }}\right) ^{m},\text{ \ }m=\frac{-1\pm \sqrt{1-16A%
}}{2}.  \label{mi010}
\end{equation}%
For $\widetilde{g}_{0}(t)$ to be a real function of time $t$, strength $A$
of the magnetic trap $\alpha (t)$, defined by Eq. (\ref{mi09}), must satisfy
condition $A<1/16$, which allows one to investigate MI for both the
confining and expulsive potentials ($A<0$ and $A>0$, respectively). Note
that $t^{\ast }<0$ describes BEC in a shrinking trap, while $t^{\ast }>$%
\thinspace $0$ corresponds to a broadening condensate. Inserting
expressions~(\ref{mi010}) in the system of Eqs. (\ref{eq_3.016})-(\ref%
{eq_3.018}) determines all time-dependent parameters; in particular, $%
T(t)=t^{\ast }\left( 2m+1\right) ^{-1}\ell ^{-2}(0)\left[ (t/t^{\ast
}+1)^{2m+1}-1\right] $. To secure the variation of $T$ from zero to
infinity, it is necessary to take $m=(-1+\sqrt{1-16A}\,)/2$ and $m=(-1-\sqrt{%
1-16A})/2$ in the cases of the broadening and shrinking trap, respectively.
In the latter case, we focus on $t$ varying from $0$ to $-t^{\ast }$, to
provide the variation of $T$ from $0$ to $+\infty $.

\begin{figure}[tbp]
\centerline{\includegraphics[scale=.89]{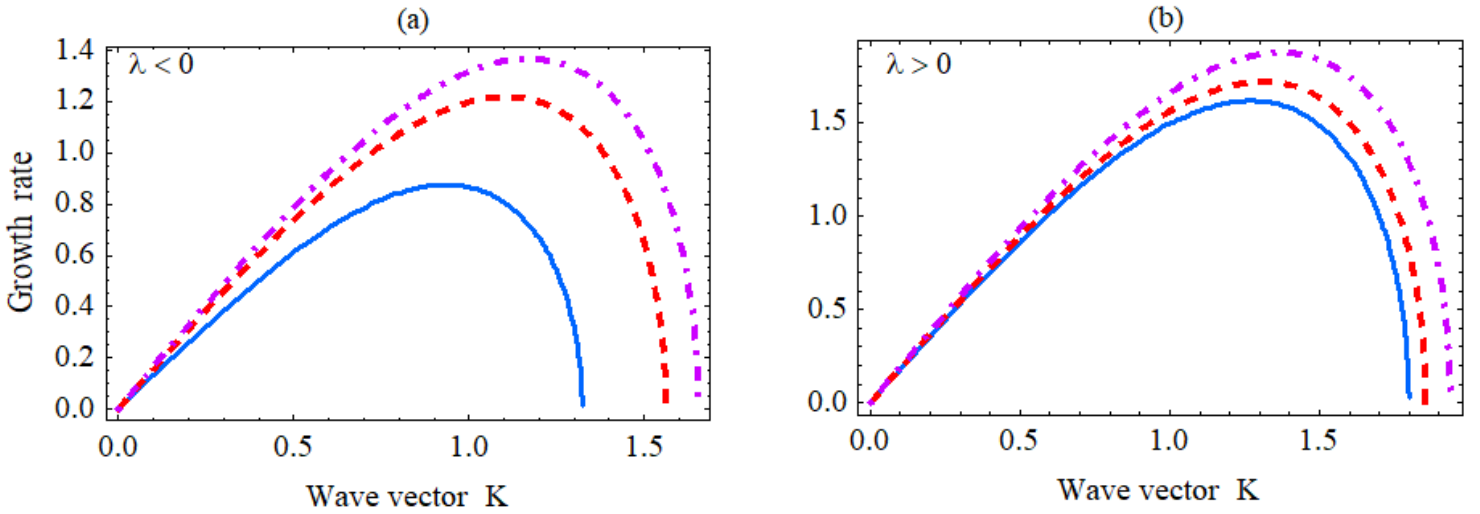}}
\caption{(Color online) The MI gain produced by Eq.~(\protect\ref{eq_3.0gain}%
) for three positive and three negative values of the imprint parameter $%
\protect\lambda (\protect\beta )$. (a): $\protect\lambda (\protect\beta %
)=-0.5$ (the solid line), $\protect\lambda (\protect\beta )=-0.25$ (the
dashed line), and $\protect\lambda (\protect\beta )=-0.125$ (the
dotted-dashed line); (b): $\protect\lambda (\protect\beta )=0.125$ (the
solid line), $\protect\lambda (\protect\beta )=0.25$ (the dashed line), and $%
\protect\lambda (\protect\beta )=0.50$ (the dotted-dashed line). Other
parameters are $Q=1$, $\protect\phi _{0}=1$, and $\protect\lambda _{0}=3/2$.
The results are reproduced from Ref. \protect\cite{3.33b}.}
\label{fig06}
\end{figure}

In the case of constant $\lambda _{0}$ and $\lambda $, the MI gain is
time-independent, but it depends on the phase-imprint parameter $\beta $:
\begin{equation}
\left\vert \text{Im}\,\omega (\beta )\right\vert =\left\vert K\right\vert
\sqrt{\phi _{0}^{2}\left[ -\lambda ^{2}(\beta )\phi _{0}^{2}+2Q\lambda
(\beta )+2\lambda _{0}\right] -K^{2}}.  \label{eq_3.0gain}
\end{equation}%
It is evident that the variation of the gain, controlled by $\beta $, may
significantly modify the instability domain and bring new effects. In fact,
to different values of $\beta $ there correspond different instability
diagrams, depending on whether $\lambda (\beta )$ is positive or negative.
Negative $\lambda (\beta )$ softens the instability, while $\lambda (\beta
)>0$ enhances it. This behavior is shown in Fig. \ref{fig06}, which displays
the MI gain provided by Eq.~(\ref{eq_3.0gain}) as a function of perturbation
wavenumber $K$, for three values of $\lambda (\beta )<0$ (plot (a)), and
three values of $\lambda (\beta )>0$ (plot (b)). In Fig. \ref{fig06}(a)
corresponding to $\lambda (\beta )<0$, it is easy to see that the gain
decreases with the growth of parameter $\lambda $, while in Fig. \ref{fig06}%
(b), with $\lambda (\beta )>0$, the gain increases with the growth of $%
\lambda $.

The results for constant $\lambda _{0}$ and $\lambda $ can be summarized as
follows. For the occurrence of MI of CW solutions $\phi =\phi _{0}\exp
\left( -iQX-i\left[ Q^{2}-\lambda _{0}\phi _{0}^{2}-Q\phi _{0}^{2}\lambda %
\right] T\right) $, it is necessary and sufficient for wavenumber $K$ of the
modulation perturbation to satisfy the MI criterion (\ref{mi06}). Moreover,
for given $\phi _{0}$, $Q$, and $\lambda _{0}$, the imprint parameter $\beta
$ should be chosen such that $\lambda ^{2}(\beta )\phi _{0}^{2}-2Q\lambda
(\beta )-2\lambda _{0}<0.$

Next, it is relevant to look at the case when at least either $\lambda _{0}$
or $\lambda $ is not constant. As in the previous case, if the HO potential
is considered with the same time dependence of the strength as in Eq. (\ref%
{mi09}) with $A<1/16$, one can find a particular solution of Riccati
equation~(\ref{eq_3.016}) in the form of
\begin{equation}
f(t)=B(t+t^{\ast })^{-1},\quad B=\frac{1\pm \sqrt{1-16A}}{8}.  \label{mi011}
\end{equation}%
With expression (\ref{mi011}), the corresponding solution of Eq.~(\ref%
{eq_3.016}) reads
\begin{equation}
f(t)=\frac{1-4B+CB\left( 1-8B\right) (t+t^{\ast })^{8B-1}}{C\left(
1-8B\right) (t+t^{\ast })^{8B}+4(t+t^{\ast })},  \label{mi012}
\end{equation}%
where $C=\left[ 1-4B-4f(0)t^{\ast }\right] /\left[ f(0)\left( 1-8B\right)
(t^{\ast })^{8B}-B\left( 1-8B\right) (t^{\ast })^{8B-1}\right] $. Using Eqs.
(\ref{mi09}) and (\ref{mi011}), one obtain the following time dependence of
the parameters, cf. Eqs. (\ref{eq_3.016})-(\ref{eq_3.018}):
\begin{subequations}
\begin{eqnarray}
\ell (t) &=&\ell (0)\left( \frac{t+t^{\ast }}{t^{\ast }}\right) ^{4B},
\label{mig} \\
\eta (t) &=&2B\ln \left\vert \frac{t+t^{\ast }}{t^{\ast }}\right\vert +\eta
(0),  \label{mih} \\
\beta (t) &=&\frac{\widetilde{g}(t)(t+t^{\ast })}{8B},  \label{mii} \\
T(t) &=&\frac{1}{\ell (0)\left( 1-4B\right) }\left[ \left( t+t^{\ast
}\right) \left( \frac{t^{\ast }}{t+t^{\ast }}\right) ^{4B}-t^{\ast }\right] .
\label{mij}
\end{eqnarray}%
It follows from Eqs. (\ref{mig})-(\ref{mij}) that $\beta (t)$ is no longer a
free parameter, as it depends of $\widetilde{g}(t)$. In the case of $t^{\ast
}>0$ (broadening condensate), it is reasonable to take $B<1/4$; a proper
choice of $\ell (0)$ then ensures the variation of $T(t)$ from $0$ to $%
+\infty $. For BECs in the shrinking trap ($t^{\ast }<0$), the appropriate
choice of $\ell (0)$ and $B>1/4$ demonstrates that the variation of $t$ in
interval $0\leq t<t^{\ast }$ corresponds to $0<T(t)<+\infty $.

In the case when, at least, either $\lambda _{0}$ or $\lambda $ is not
constant, the MI gain given by Eq. (\ref{mi07}) is time-dependent. In this
situation, the variation of the gain, related to the sign of $t^{\ast }$
(recall that $t^{\ast }<0$ or $t^{\ast }>0$ determines, respectively, the
shrinking or expanding trap), may significantly affect the MI domain and
introduce new effects. The instability is enhanced or attenuated by $t^{\ast
}<0$ or $t^{\ast }>0$, respectively, as shown in Fig. \ref{fig07}. The
figure displays the MI gain produced by Eq.~(\ref{mi07}), as a function of
perturbation wavenumber $K$, for three negative and three positive values of
$t^{\ast }$, in plots \ref{fig07}(a) and (b), respectively. In Fig. \ref%
{fig07}(a), corresponding to the shrinking trap ($t^{\ast }<0$), one easily
sees that the gain indeed increases with $t^{\ast }$, while in Fig. \ref%
{fig07}(b), obtained for the broadening trap ($t^{\ast }>0$), the gain
decreases as $t^{\ast }$ increases. The plots in this figure are produced
with $\widetilde{g}_{0}(t)=2\exp \left( 0.5t\right) $ and $\widetilde{g}%
(t)=2\exp \left( -t\right) $.
\begin{figure}[tbp]
\centerline{\includegraphics[scale=.95]{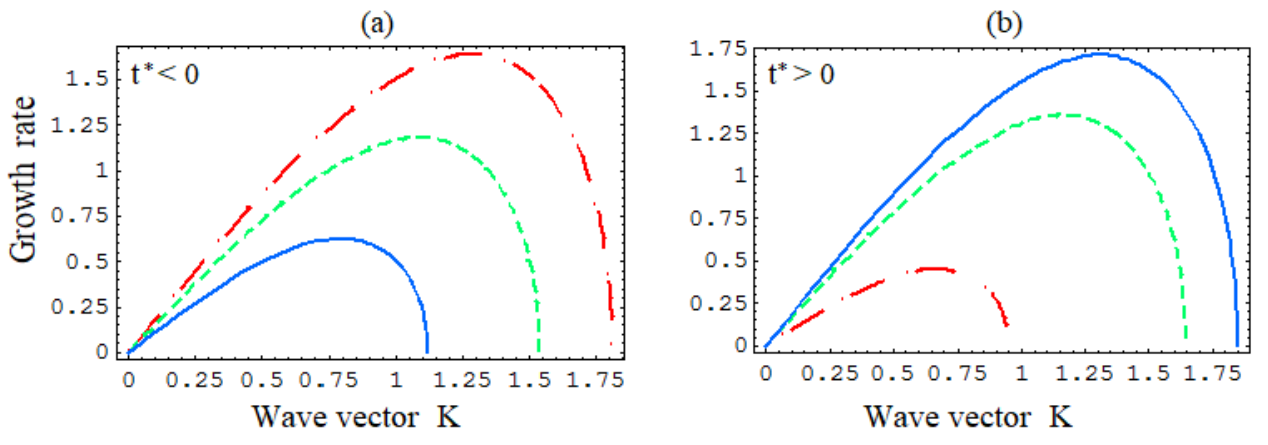}}
\caption{(Color online) The MI gain produced by Eq.~( \protect\ref{mi07})
for three positive and three negative values of $t^{\ast }$ at time $t=0$.
(a): The case of the shrinking trap with $t^{\ast }=-0.6$, $-0.4$, and $-0.2$
(the solid, dashed, and dotted-dashed lines, respectively). (b) The case of
the broadening trap with $t^{\ast }=0.1$, $0.2$, $0.4$ (the solid, dashed,
and dotted-dashed lines, respectively). Other parameters are $Q=1$, $%
\protect\phi _{0}=1$, $A=-1$, $\ell (0)=1$, $\protect\eta (0)=0$, $%
\widetilde{g}_{0}(t)=2\exp \left( 0.5t\right) $, $\widetilde{g}(t)=2\exp
\left( -t\right) $, $B=(1+\protect\sqrt{17}\,)/8$ for plots (a), and $B=(1-%
\protect\sqrt{17}\,)/8$ for (b). The results are reproduced from Ref.
\protect\cite{3.33b}.}
\label{fig07}
\end{figure}

The analysis makes it clear that the simplest and most interesting case in
the setting with the time-dependent HO potential is the one with the
inverse-square time dependence of the trap strength, as defined by Eq.~(\ref%
{mi09}) with $A<1/16$. In this case, the modified LT demonstrates the
equivalence of the setting to the cubic derivative NLS equation. In this
case, the coefficients of the cubic derivative NLS equation are either
constant or time dependent, suggesting that frequencies of eigenmodes of the
modulational perturbations are either constant or effectively time-dependent.

\subsection{Matter-wave solitons of the cubic inhomogeneous NLS equation~(%
\protect\ref{eq_3.01}) with the spatiotemporal HO potential (\protect\ref%
{eq_3.02})}

As said above, the condition of the transformability to the cubic derivative
NLS with constant coefficients, where the MI analysis has been performed in
the complete form, makes it most relevant to consider in Eq. (\ref{eq_3.01})
with the spatiotemporal potential (\ref{eq_3.02}), taken with $\alpha
(t)=A(t+t^{\ast })^{-2}$ and $A<1/16$. The present subsection, following
Ref. \cite{3.33b}, addresses this case in the analytical form under the
assumption that $\lambda _{0}$ and $\lambda $ in Eq.~(\ref{eq_3.021}) are
constant.

The starting point is the CW solution of Eq.~(\ref{eq_3.020}),
\end{subequations}
\begin{equation}
\phi (X,T)=\phi _{0}\exp \left[ -ik_{0}X+i\left( \lambda _{0}\phi
_{0}^{2}+k_{0}\lambda \phi _{0}^{2}-k_{0}^{2}\right) T\right] .  \label{CW}
\end{equation}%
Writing a solution of Eq.~(\ref{eq_3.020}) in the Madelung form,
\begin{equation}
\phi (X,T)=R(X,T)\exp \left[ i\Phi (X,T)\right] ,  \label{eq_3.022}
\end{equation}%
one arrives at the following system of equations for the real amplitude and
phase:
\begin{equation}
\left\{
\begin{array}{r}
-R\partial \Phi /\partial T+\partial ^{2}R/\partial X^{2}-R\left( \partial
\Phi /\partial X\right) ^{2}-\lambda R^{3}\partial \Phi /\partial X+\lambda
_{0}R^{3}=0, \\
\partial R/\partial T+2\left( \partial \Phi /\partial X\right) \left(
\partial R/\partial X\right) +R\partial ^{2}\Phi /\partial X^{2}+3\lambda
R^{2}\partial R/\partial X=0.%
\end{array}%
\right.   \label{eq_3.023}
\end{equation}%
Further, the CW solution (\ref{CW}) suggests to look for a solution to Eqs. (%
\ref{eq_3.023}) in the following traveling-wave form:%
\begin{equation}
\left\{
\begin{array}{l}
R(X,T)=\phi _{0}+\rho (z=X-\upsilon T), \\
\Phi (X,T)=-k_{0}z+\varphi (z)+\left( \lambda _{0}\phi _{0}^{2}+\lambda
k_{0}\phi _{0}^{2}-k_{0}^{2}-k_{0}\upsilon \right) T,%
\end{array}%
\right.   \label{eq_3.024}
\end{equation}%
with arbitrary velocity $\upsilon $. Inserting ansatz (\ref{eq_3.024}) in
the system of equations (\ref{eq_3.023}), and integrating the second
equation yields%
\begin{equation}
\frac{d\varphi }{dz}=\frac{C_{0}}{\left( \phi _{0}+\rho \right) ^{2}}+\frac{%
\left( \upsilon +2k_{0}\right) }{2}-\frac{3\lambda }{4}\left( \phi _{0}+\rho
\right) ^{2},  \label{eq_3.025}
\end{equation}%
where $C_{0}$ is a constant of integration. Inserting this expression for $%
d\varphi /dz$ into the first equation yields %
\begin{equation}
\left( \frac{d\zeta }{dz}\right) ^{2}=\widetilde{\alpha }\zeta ^{4}+4%
\widetilde{\beta }\zeta ^{3}+6\widetilde{\gamma }\zeta ^{2}+4\widetilde{%
\delta }\zeta +\widetilde{\epsilon }\equiv f(\zeta ),  \label{eq_3.026}
\end{equation}%
where%

\begin{equation}
\left.
\begin{array}{l}
\zeta (z)=\left( \phi _{0}+\rho (z)\right) ^{2},\text{ }\widetilde{\alpha }%
=-\lambda ^{2},\text{ }\widetilde{\beta }=\left( \lambda \upsilon
-2\lambda
_{0}\right) /4,\text{ }\widetilde{\epsilon }=-4C_{0}^{2}, \\
\widetilde{\gamma }=\left[ 4\lambda _{0}\phi _{0}^{2}+2\lambda
\left(
2k_{0}\phi _{0}^{2}-C_{0}\right) -4\upsilon k_{0}-\upsilon ^{2}-4k_{0}^{2}%
\right] /6,%
\end{array}%
\right.   \label{eq_3.028}
\end{equation}%
$\widetilde{\delta }$ and $C_{0}$ being two arbitrary real
constants of
integration. The general solution of Eq. (\ref{eq_3.026}) with coefficients (%
\ref{eq_3.028}) can be obtained in terms of the Weierstrass' elliptic
function $\wp (z;g_{2},g_{3})$ \cite{3.37,3.38}:

\begin{equation}
\zeta (z)=\zeta _{0}+\frac{\sqrt{f(\zeta _{0})}\,\frac{d}{dz}\wp
(z;g_{2},g_{3})+\frac{1}{2}f^{\prime }(\zeta _{0})\left[ \wp (z;g_{2},g_{3})-%
\frac{1}{24}f^{\prime \prime }(\zeta _{0})\right] +\frac{1}{24}f(\zeta
_{0})f^{\prime \prime \prime }(\zeta _{0})}{2\left[ \wp (z;g_{2},g_{3})-%
\frac{1}{24}f^{\prime \prime }(\zeta _{0})\right] ^{2}-\frac{1}{48}f(\zeta
_{0})f^{(IV)}(\zeta _{0})},  \label{sol01}
\end{equation}%
where $\zeta _{0}$ is an arbitrary real constant, and the prime stands for $%
d/dz$ (recall $f(\zeta )$ is defined in Eq. (\ref{eq_3.026})). Invariants $%
g_{2}$ and $g_{3}$ of function $\wp (z;g_{2},g_{3})$ are related to
coefficients of $f(\zeta )$ as \cite{3.37}%
\begin{equation}
g_{2}=\widetilde{\alpha }\widetilde{\epsilon }-4\widetilde{\beta }\widetilde{%
\delta }+3\widetilde{\gamma }^{2},\quad g_{3}=\widetilde{\alpha }\widetilde{%
\gamma }\widetilde{\epsilon }+2\widetilde{\beta }\widetilde{\gamma }%
\widetilde{\delta }-\widetilde{\alpha }\widetilde{\delta }^{2}-\widetilde{%
\gamma }^{3}-\widetilde{\epsilon }\widetilde{\beta }^{2}.  \label{eq_3.029}
\end{equation}%
The discriminant $\Delta $ of the Weierstrass' elliptic function $\wp
(z;g_{2},g_{3})$,
\begin{equation}
\Delta =g_{2}^{3}-27g_{3}^{2},  \label{eq_3.030}
\end{equation}%
is suitable to classify the behavior of the solution $\zeta (\xi )$ and to
discriminate between periodic and solitary-wave solutions \cite{3.38}. If $%
\Delta =0$, $g_{2}\geq 0$, and $g_{3}\leq 0$, $\zeta (\xi )$ is a solitary
wave given by%
\begin{equation}
\zeta (z)=\zeta _{0}+\frac{f^{\prime }(\zeta _{0})}{4\left[ e_{1}-\frac{1}{24%
}f^{\prime \prime }(\zeta _{0})+3e_{1}\text{\textrm{cosech}}^{2}\left( \sqrt{%
3e_{1}}\,z\right) \right] },  \label{eq_3.031}
\end{equation}%
where $e_{1}=\left( -g_{3}\right) ^{1/3}$. Below, for simplicity, constants
of integration are set as $\widetilde{\delta }=\widetilde{\epsilon }=0$. In
this case, $g_{2}=3\widetilde{\gamma }^{2},\quad g_{3}=-\widetilde{\gamma }%
^{3}$, and $\Delta =0$. Then Eq.~(\ref{eq_3.031}) defines solitary-wave
solutions if and only if $\widetilde{\gamma }>0$ and $2\widetilde{\beta }%
^{2}-3\widetilde{\alpha }\widetilde{\gamma }\geq 0$. The physical solution (%
\ref{eq_3.031}) must be nonnegative and bounded. Considering properties of $%
f(\zeta )$ \cite{3.39}, one obtains conditions, expressed in terms of
coefficients of the basic equation, that determine the existence of the
physical solutions, see Fig. \ref{fig08} borrowed from Ref.~\cite{3.39},
which shows phase diagrams associated to the physical solutions for $%
\widetilde{\delta }=\widetilde{\epsilon }=0$.

According to Ref. \cite{3.39}, solitary-wave solutions generated by Eq. (\ref%
{eq_3.031}) can be cast in the form of
\begin{equation}
\zeta _{\pm }(z)=\frac{\widetilde{\gamma }\left[ 2\widetilde{\beta }\pm
\sqrt{4\widetilde{\beta }^{2}-6\widetilde{\alpha }\widetilde{\gamma }}\,%
\right] \biggl[3\widetilde{\gamma }^{2}+\left( \widetilde{\gamma }%
^{2}-1\right) \cosh ^{2}\left( \sqrt{\frac{3}{2}\widetilde{\gamma }^{3}}%
\,z\right) \biggr]}{\left[ -3\widetilde{\alpha }\widetilde{\gamma }%
^{3}+\left( 4\widetilde{\beta }^{2}\pm 2\sqrt{4\widetilde{\beta }^{2}-6%
\widetilde{\alpha }\widetilde{\gamma }}-\widetilde{\alpha }\widetilde{\gamma
}(5+\widetilde{\gamma })^{2}\right) \cosh ^{2}\left( \sqrt{\frac{3}{2}%
\widetilde{\gamma }^{3}}\,z\right) \right] }.  \label{eq_3.032}
\end{equation}%
These solutions correspond to two simple roots of the polynomial $f(\zeta )$
(provided that $4\widetilde{\beta }^{2}-6\widetilde{\alpha }\widetilde{%
\gamma }>0$), and are represented by phase diagrams (b), (c), (d), (f), (g),
and (h) in Fig.~\ref{fig08} (see further details in Ref.~\cite{3.39}).

It is necessary to address the condition of the non-negativeness of
solutions (\ref{eq_3.032}). Here, two cases should be distinguished, namely,
$\widetilde{\gamma }=1$, which corresponds to a bright solitary-wave
solution, and $0<\widetilde{\gamma }\neq 1$, corresponding to both dark and
bright solitary waves.

\begin{figure}[tbp]
\centerline{\includegraphics[scale=.99]{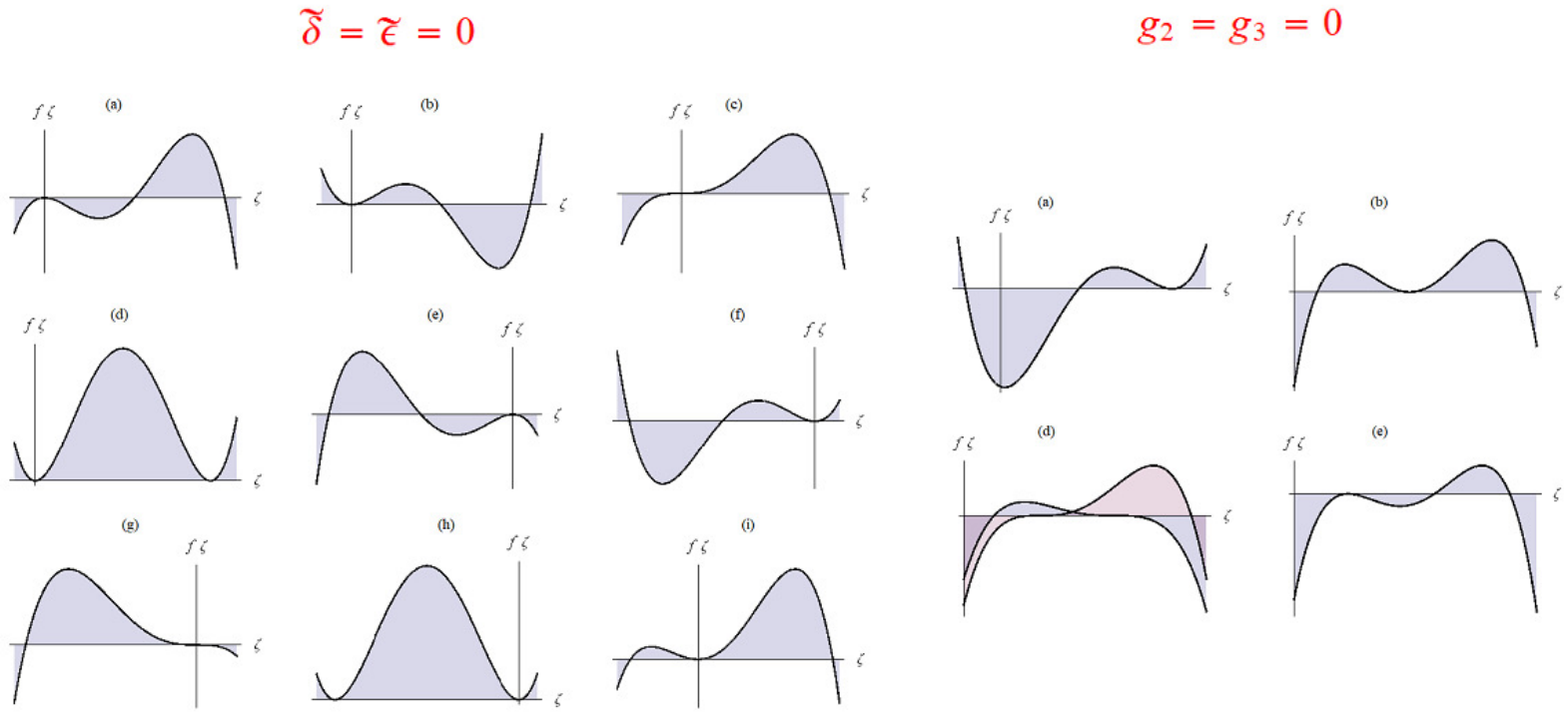}}
\caption{(Color online) Phase diagrams associated with real bounded
solutions, given by Eqs. (\protect\ref{sol01}) and (\protect\ref{eq_3.026}),
with either $\widetilde{\protect\delta }=\widetilde{\protect\epsilon }=0$
(see also Eq. (\protect\ref{eq_3.032}) for this case) or $\widetilde{\protect%
\delta }\neq 0,$ $\widetilde{\protect\epsilon }<0$, and $g_{2}=g_{3}=0$.
Further details can be found in Refs. \protect\cite{3.40} and \protect\cite%
{3.39} for the right and left plots, respectively.}
\label{fig08}
\end{figure}

\begin{figure}[tbp]
\centerline{\includegraphics[scale=.95]{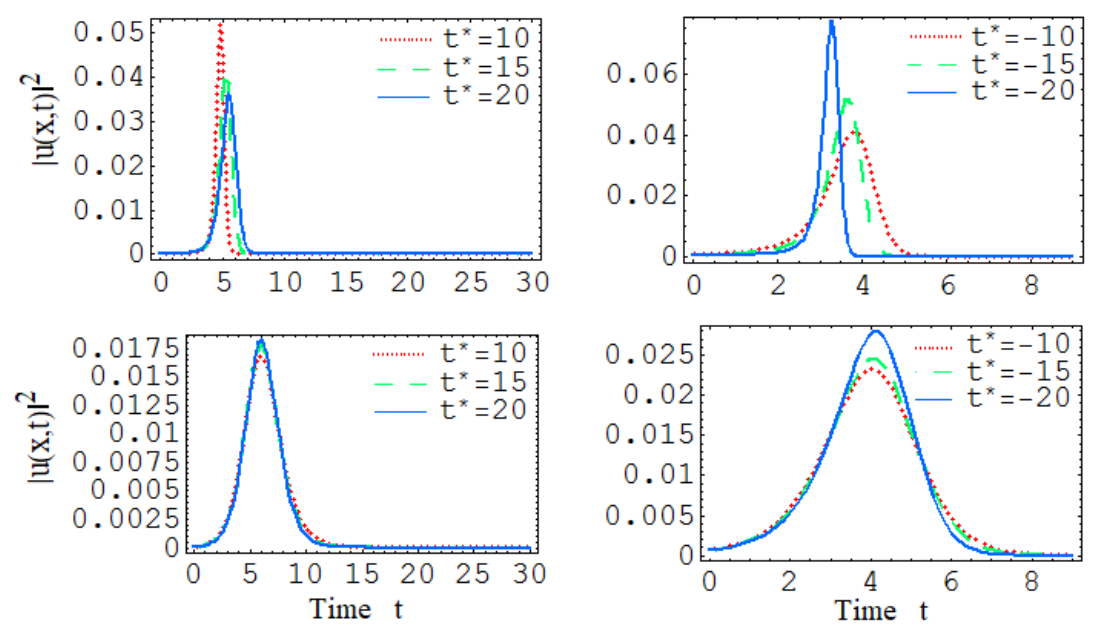}}
\caption{(Color online) Density plots $\left\vert u(x,t)\right\vert ^{2}$ of
solitary-wave solutions to Eq. (\protect\ref{eq_3.01}) at $x=3$ for three
values of parameter $t^{\ast }$ in potential (\protect\ref{mi09}), given by
expression (\protect\ref{eq_3.032}) for $\protect\zeta _{+}$. The top and
bottom panels correspond to the confining potential (with $A=-2$) and
expulsive one (with $A=1/17$), respectively, while the left and right panels
correspond to broadening shrinking traps, respectively. Values of other
parameters are given in the text. The results are reproduced from Ref.
\protect\cite{3.33b}.}
\label{fig09}
\end{figure}

\begin{enumerate}
\item[(A):] If $\widetilde{\gamma }=1$, solutions (\ref{eq_3.032}) is
nonnegative if and only if
\begin{equation}
\widetilde{\alpha }\left( 2\widetilde{\beta }^{2}\pm \sqrt{4\widetilde{\beta
}^{2}-6\widetilde{\alpha }}-18\widetilde{\alpha }\right) ^{-1}<2/3
\end{equation}%
and \newline
\begin{equation}
\left( 2\widetilde{\beta }\pm \sqrt{4\widetilde{\beta }^{2}-6\widetilde{%
\alpha }}\,\right) \left( 2\widetilde{\beta }^{2}\pm \sqrt{4\widetilde{\beta
}^{2}-6\widetilde{\alpha }}-18\widetilde{\alpha }\right) >0.
\end{equation}

An example of the bright solitary-wave solution is obtained with parameters $%
\lambda _{0}=1$, $\phi _{0}=1$, $\upsilon =0.5$, $k_{0}=1$, and $\lambda
=\allowbreak 2.0625$. With this set of parameters, $\zeta _{+}(z)$\
satisfies all the needed conditions (reality, boundedness and
non-negativeness). Figures \ref{fig09}, \ref{fig010}, and \ref{fig011},
respectively, show effects of \thinspace $t^{\ast }$, $A$ and $\beta $ on
the density profile of the solitary-wave solution, $\left\vert
u(x,t)\right\vert ^{2}$, at $x=3$; here, $\ell (0)=1$ is set. In these three
figures, left and right panels correspond, respectively, to the broadening
and shrinking traps ($t^{\ast }>0$ and $t^{\ast }<0$, respectively), while
top and bottom panels correspond to the confining and expulsive potentials $%
(A<0\ $and $A>0$, respectively). At $x=3$, Fig.~\ref{fig09} shows the time
evolution of density $\left\vert u(x,t)\right\vert ^{2}$ for three different
values of $t^{\ast }.$ As seen in the figure, in the case of the confining
potential, the amplitude of the density profile decreases as $t^{\ast }$
increases for the broadening trap, and increases with the growth of $t^{\ast
}$ for the shrinking trap. In the case of the expulsive potential, the
profile's amplitude increases with $t^{\ast }$ for both broadening and
shrinking traps. Figure \ref{fig010} depicts density $\left\vert
u(x,t)\right\vert ^{2}$ at $x=3$ for three different values of $A$. This
figure shows that, for both the confining and expulsive potentials (the top
and bottom plots, respectively), the profile's amplitude decreases as $A$
increases, which happens for both the broadening and shrinking BEC\ traps
(left and right plots, respectively). It is seen from Fig. \ref{fig011},
where density $\left\vert u(x,t)\right\vert ^{2}$ at $x=3$ is depicted for
different values of $\beta $, that, irrespective of the sign of $A$ (the
confining or expulsive potential) and the sign of $t^{\ast }$ (the
broadening or shrinking trap), the density-profile's amplitude decreases
with the increase of the imprint parameter $\beta $.

\begin{figure}[tbp]
\centerline{\includegraphics[scale=.95]{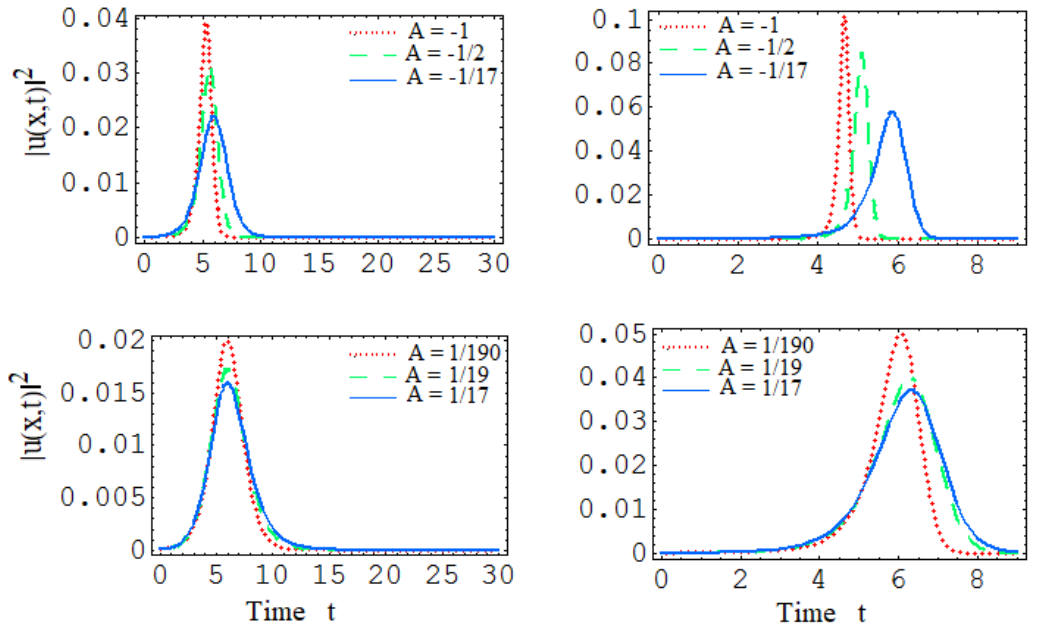}}
\caption{(Color online) The same as in Fig. \protect\ref{fig09}, but for
three different values of parameter $A$ in potential (\protect\ref{mi09}).
The top and bottom panels correspond to the confining and expulsive
potential, respectively, while the left and right panels are associated,
severally, with the broadening trap (for $t^{\ast }=10$) and the shrinking
one (for $t^{\ast }=-10$). Typical values of other parameters are given in
the text. The results are reproduced from Ref. \protect\cite{3.33b}.}
\label{fig010}
\end{figure}

\begin{figure}[tbp]
\centerline{\includegraphics[scale=.95]{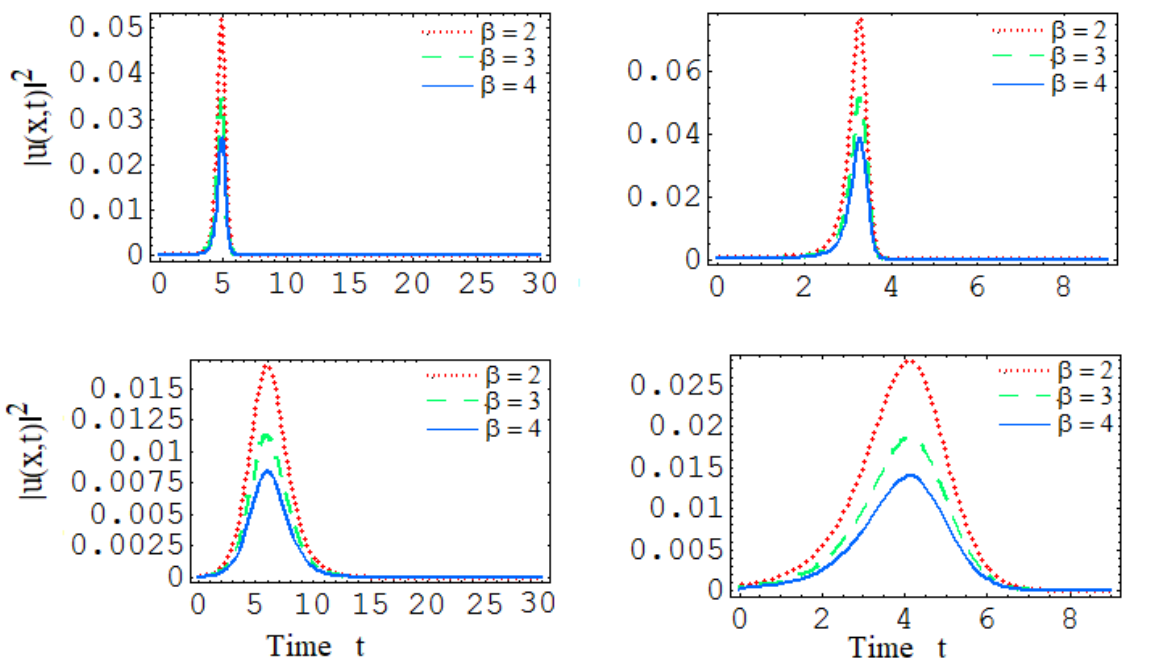}}
\caption{(Color online) The same as in Figs. \protect\ref{fig09} and \protect
\ref{fig010}, but for three different values of the imprint parameter $%
\protect\beta $ in Eq. (\protect\ref{eq_3.04b}). The top and bottom panels
correspond to the confining potential (with $A=-2$) and expulsive one (with $%
A=1/17$), respectively, while the left and right panels are associated,
severally, with the broadening trap (for $t^{\ast }=10$) and the shrinking
one (for $t^{\ast }=-10$). Typical values of other parameters are given in
the text. The results are reproduced from Ref. \protect\cite{3.33b}.}
\label{fig011}
\end{figure}

\begin{figure}[tbp]
\centerline{\includegraphics[scale=.95]{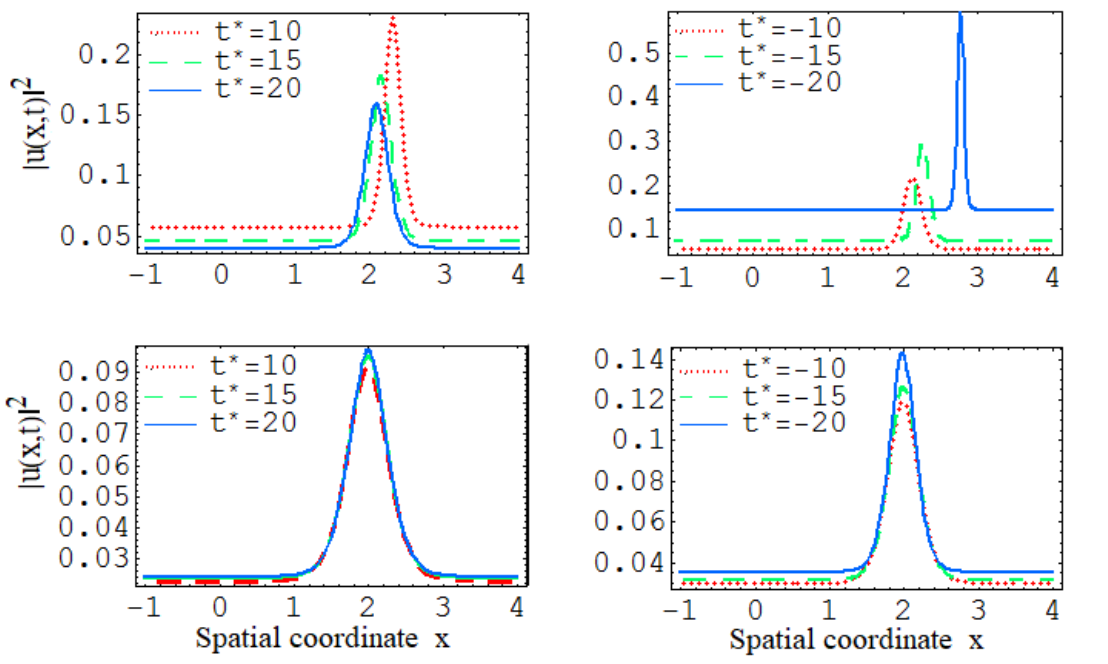}}
\caption{ (Color online) Density $\left\vert u(x,t)\right\vert ^{2}$ at $t=4$%
, as given by solution (\protect\ref{eq_3.032}), for three different values
of parameter $t^{\ast }$ appearing in potential (\protect\ref{mi09}). The
top and bottom panels correspond, severally, to the confining potential
(with $A=-2$) and the expulsive one (with $A=1/17$), while the left and
right panels are associated with the broadening and shrinking trap,
respectively. Typical values of other parameters are given in the text. The
results are reproduced from Ref. \cite{3.33b}.}
\label{fig012}
\end{figure}

\item[(B):] If $0<\widetilde{\gamma }\neq 1$, solutions (\ref{eq_3.032}) are
nonnegative if and only if the following three conditions are simultaneously
satisfied: (i) $3\widetilde{\gamma }^{2}\left( 1-\widetilde{\gamma }%
^{2}\right) ^{-1}\leq 1$, (ii) $3\widetilde{\alpha }\widetilde{\gamma }%
^{3}\left( 4\widetilde{\beta }^{2}\pm 2\sqrt{4\widetilde{\beta }^{2}-6%
\widetilde{\alpha }\widetilde{\gamma }}-\widetilde{\alpha }\widetilde{\gamma
}(5+\widetilde{\gamma })^{2}\right) ^{-1}<1$, and (iii)
\begin{equation}
\widetilde{\gamma }\left( \widetilde{\gamma }^{2}-1\right) \left( 2%
\widetilde{\beta }\pm \sqrt{4\widetilde{\beta }^{2}-6\widetilde{\alpha }%
\widetilde{\gamma }}\,\right) \left( 4\widetilde{\beta }^{2}-\widetilde{%
\alpha }\widetilde{\gamma }(5+\widetilde{\gamma })^{2}\pm 2\sqrt{4\widetilde{%
\beta }^{2}-6\widetilde{\alpha }\widetilde{\gamma }}\,\right) >0.
\end{equation}%
\newline
With parameters $\lambda _{0}=1$, $\phi _{0}=1$, $\upsilon =0.5$, $\lambda
=\allowbreak 3.5625$, and $k_{0}=1$, conditions (i)--(iii) are
simultaneously satisfied for $\zeta _{+}(z)$, providing an example of a dark
solitary-wave solution to Eq.~(\ref{eq_3.026}). For this set of parameters,
density $\left\vert u\left( x,t\right) \right\vert ^{2}$, associated with
the solitary-wave solution (\ref{eq_3.032}), is depicted at $x=3$ in Figs. %
\ref{fig09}, \ref{fig010}, and \ref{fig011}, and at $t=4$ in Figs. \ref%
{fig012}, \ref{fig013}, and \ref{fig014}, to show the effect of \thinspace $%
t^{\ast }$, $A$, and $\beta $ on the solitary-wave shape. Here, $\ell (0)=1$
is used. In these figures, left and right panels correspond to the
broadening and shrinking traps ($t^{\ast }>0$ and $t^{\ast }<0$,
respectively), while top and bottom panels correspond to the confining and
expulsive potential $(A<0\ $and $A>0$, respectively). In particular, Fig. %
\ref{fig012} shows, at time $t=4$, the spatial evolution of density $%
\left\vert u(x,t)\right\vert ^{2}$ for three different values of $t^{\ast }$%
. As seen in the figure, the peak density in the case of confining potential
decreases as $t^{\ast }$ increases for the broadening trap (the top panels),
and increases with $t^{\ast }$ for the shrinking trap. In the case of the
expulsive potential (the bottom panels), the peak density increases with $%
t^{\ast }$ for both the broadening and shrinking traps. Figure \ref{fig013}
depicts density $\left\vert u(x,t)\right\vert ^{2}$ at $t=4$ for three
different values of $A$. The figure demonstrates that, for both the
confining and expulsive potentials, the peak density decreases as $A$
increases, which happens for both the broadening and shrinking traps. Figure~%
\ref{fig014}, where density $\left\vert u(x,t)\right\vert ^{2}$ is displayed
at $t=4$ for different values of $\beta $, shows that, irrespective of the
sign of $A$ (the confining or expulsive potential) and the sign of $t^{\ast
} $ (the broadening or shrinking trap), the peak density decreases with the
increase of imprint parameter $\beta $.
\end{enumerate}

\begin{figure}[tbp]
\centerline{\includegraphics[scale=.95]{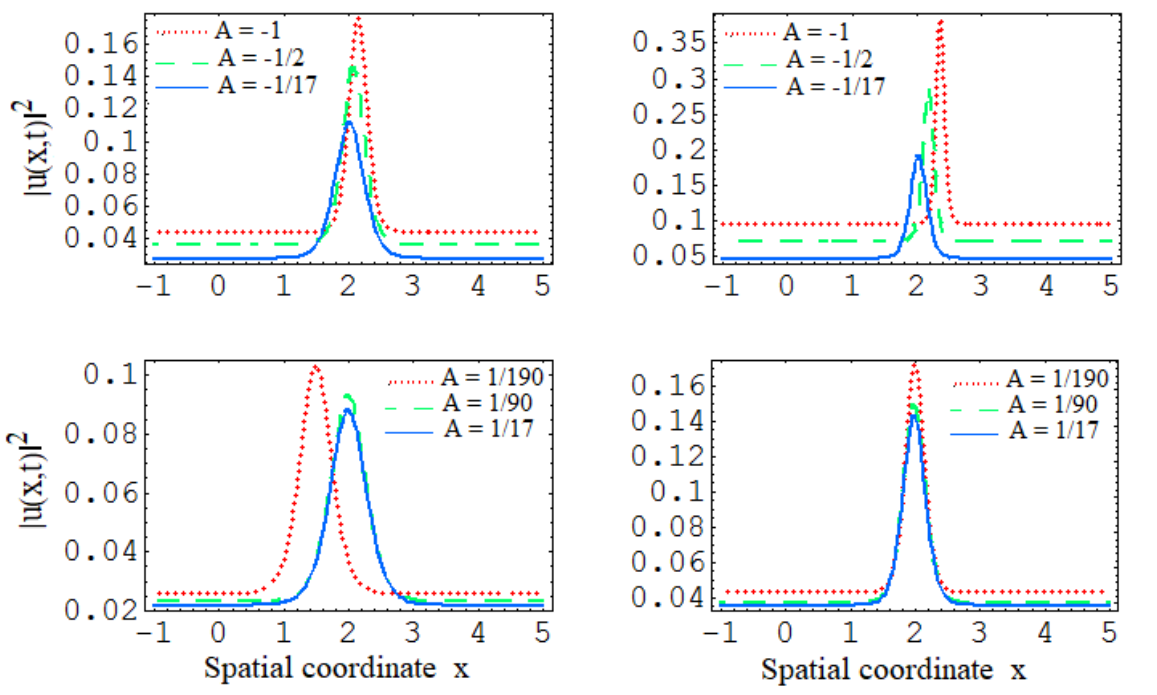}}
\caption{(Color online) The same as in Fig. \protect\ref{fig012}, but for
three different values of parameter $A$ in potential (\protect\ref{mi09}).
The top and bottom panels correspond to the confining and expulsive
potentials, respectively, while the left and right panels are associated,
severally, with the broadening trap (for $t^{\ast }=10$) and the shrinking
one (for $t^{\ast }=-10$), respectively. Values of other parameters are
given in the text. The results are reproduced from Ref. \cite{3.33b}.}
\label{fig013}
\end{figure}

\begin{figure}[tbp]
\centerline{\includegraphics[scale=.95]{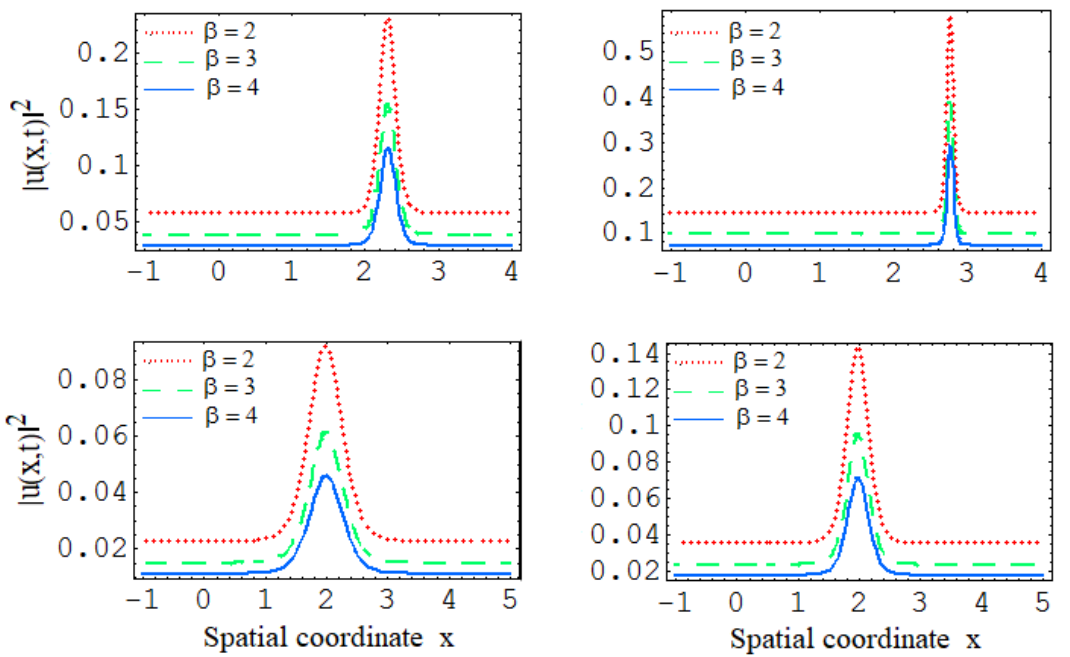}}
\caption{ (Color online) The same as in Figs. \protect\ref{fig012} and
\protect\ref{fig013}, but for three different values of the imprint
parameter $\protect\beta $ in transformation (\protect\ref{eq_3.04b}). The
upper and lower panels correspond to the confining potential (with $A=-2$)
and the expulsive one (with $A=1/17$), respectively, while the left and
right panels are associated, severally, with yjr broadening trap (for $%
t^{\ast }=10$) and the shrinking one (for $t^{\ast }=-10$),. Typical values
of other parameters are given in the text.}
\label{fig014}
\end{figure}

\subsection{Conclusion of the section}

In this section, MI of CW states is surveyed in the context of the
inhomogeneous cubic NLS equations with the external potential. The
motivation for this study was its link to BEC with the spatially modulated
local nonlinearity. To make the investigation of MI possible for both
attractive and repulsive nonlinearities, the inhomogeneous cubic NLS
equation is first transformed into the inhomogeneous cubic derivative NLS
equation, by means of the suitably designed PIT, applied to the wave
function of the original cubic NLS equation. A modified LT is then used to
cast the problem in the form in which the cubic derivative NLS equation has
constant coefficients. For the strength of the magnetic trap modulated in
time $\sim $ $(t+t^{\ast })^{-2}$ and the local time-dependent nonlinearity
coefficient being a linear function of $x$, the resulting MI gain is either
constant or time varying. The impact of both the PIT\ imprint parameter and
trap parameter $t^{\ast }$ on the MI gain is considered. In the case of the
constant MI gain, analytical matter-wave solitons of the inhomogeneous NLS
equation under the consideration are presented, and the effect of the
above-mentioned parameters on their shape is considered.

\section{ Collapse management for BEC with time-modulated nonlinearity}

In this section we address the dynamics of 2D and 3D condensates with the
nonlinearity coefficient, i.e., the scattering length of inter-atomic
collision, subject to the time modulation imposed by the FR, which makes the
coefficient a sum of constant and periodically oscillating terms. The
respective results were obtained by means of VA and systematic direct
simulations of the GP equation \cite{3.1}-\cite{Itin}. An averaging method
can be used too, in the case of the rapid time modulation \cite{Baizakov}.
In the 2D case, all these methods reveal the existence of stable
self-confined states in the free space (without an external trap), in
agreement with similar results originally reported for (2+1)D spatial
solitons in nonlinear optics \cite{4.7}. In the 3D free space, the VA also
predicts the existence of self-confined state without a trap \cite{Adhikari}%
. In this case, direct simulations demonstrate that the stability is limited
in time, eventually switching into collapse. Thus, a spatially uniform ac
magnetic field, resonantly tuned to drive the periodic temporal modulation
of the scattering length by means of FR, may play the role of an effective
trap confining the condensate, and sometimes causing its collapse.

Results collected in this section are chiefly based on original works \cite%
{4.7}, \cite{3.1}, and \cite{Baizakov}. Although these works were published
quite some time ago, the findings reported in them are highly relevant to
the topic of the present review article.

\subsection{The model and VA (variational approximation)}

The starting point is the mean-field GP equation for the single-particle
wave function in its usual form, cf. Eq. (\ref{eq_2.1}):

\begin{equation}
i\hbar \frac{\partial \psi (\mathbf{r,}t\mathbf{)}}{\partial t}=\left[ -%
\frac{\hbar ^{2}}{2m}\nabla ^{2}+g\left\vert \psi (\mathbf{r,}t\mathbf{)}%
\right\vert ^{2}\right] \psi (\mathbf{r,}t\mathbf{)}  \label{eq_4.1}
\end{equation}%
with $g=4\pi \hbar ^{2}a_{s}/m$, where $a_{s}$ and $m$ are the atomic
scattering length and mass. Throughout this section, its is assumed the
scattering length to be modulated in time, so that the nonlinearity
coefficient in Eq. (\ref{eq_4.1}) takes the form of $g=g_{0}+g_{1}\sin
\left( \chi t\right) $, where $g_{0}$ and $g_{1}$ are the amplitudes of the
dc and ac parts, and $\chi $ is the ac-modulation frequency.

To stabilize the condensate, an external trapping potential is usually
included. Nevertheless, it is omitted in Eq. (\ref{eq_4.1}) because it does
not play an essential role in the present context. This is also the case in
many other situations -- for example, the formation of stable Skyrmions in
two-component condensates is possible in the free space \cite{4.1}. Indeed,
it is demonstrated in some detail below that the temporal modulation of the
nonlinearity coefficient, combining the dc and ac parts as in Eq. (\ref%
{eq_4.2}), may, in a certain sense, replace the trapping potential.

Equation (\ref{eq_4.1}) is cast in a normalized form by introducing a
typical frequency, $\Omega =2gn_{0}/\hbar $, where $n_{0}$ is the largest
value of the condensate density, and rescaling the time and space variables
as $t^{\prime }=\Omega t$, $\mathbf{r}^{\prime }=\mathbf{r}\sqrt{2m\Omega
/\hbar }$. This leads to the scaled equation for isotropic states (in which
only the radial coordinate is kept, and the primes are omitted):

\begin{equation}
i\frac{\partial \psi (\mathbf{r,}t\mathbf{)}}{\partial t}=-\left( \frac{%
\partial ^{2}}{\partial r^{2}}+\frac{D-1}{r}\frac{\partial }{\partial r}%
\right) \psi (r,t\mathbf{)}-\left[ \lambda _{0}+\lambda _{1}\sin \left(
\omega t\right) \right] \left\vert \psi (r,t\mathbf{)}\right\vert ^{2}\psi
(r,t\mathbf{)}.  \label{eq_4.2}
\end{equation}%
Here $D=2$ or $3$ is the spatial dimension, and $\lambda _{0,1}\equiv
-g_{0,1}/\left( \Omega \hbar \right) $, $\omega \equiv \chi /\Omega $. Note
that $\lambda _{0}>0$ and $\lambda _{0}<0$ in Eq. (\ref{eq_4.2}) correspond
to the self-focusing and self-defocusing nonlinearity, respectively.
Additionally rescaling field $\psi $, $\left\vert \lambda _{0}\right\vert =1$
is set, so that $\lambda _{0}$ remains a sign-defining parameter.

As the next step, one applies VA to Eq. (\ref{eq_4.2}). This approximation
was originally proposed by Anderson \textit{et al}. \cite{4.2,Anderson,4.4}
for 1D temporal solitons, then for matter-wave solitons in BEC\ \cite{4.3},
and later developed for multidimensional models \cite{Desaix,4.7,3.1}. To
apply the VA in the present case, one should use the Lagrangian density
generating Eq. (\ref{eq_4.2}),%
\begin{equation}
\mathcal{L}\left( \psi \right) =\frac{i}{2}\left( \frac{\partial \psi }{%
\partial t}\psi ^{\ast }-\frac{\partial \psi ^{\ast }}{\partial t}\psi
\right) -\left\vert \frac{\partial \psi }{\partial r}\right\vert ^{2}+\frac{1%
}{2}\lambda (t)\left\vert \psi \right\vert ^{4},  \label{eq_4.3}
\end{equation}%
where $\lambda (t)=\lambda _{0}+\lambda _{1}\sin \left( \omega t\right) $,
and the asterisk stands for the complex conjugation. The variational ansatz
for the wave function is chosen as the Gaussian \cite{4.2}:%
\begin{equation}
\psi _{G}(r,t)=A(t)\exp \left[ -\frac{r^{2}}{2a^{2}(t)}+\frac{1}{2}%
ib(t)r^{2}+i\delta (t),\right]  \label{eq_4.4}
\end{equation}%
where $A$, $a$, $b$, and $\delta $ are, respectively, the amplitude, width,
chirp, and overall phase, which are assumed to be real functions of $t$. In
fact, the Gaussian is the single type of the trial wave function which makes
it possible to develop VA in the fully analytical form. A caveat is that, if
the frequency of the ac drive resonates with a transition between the ground
state of the condensate and a possible excited one (it definitely exists if
the model includes a trapping potential), the single GP equation should be
replaced by a system of coupled ones for the resonantly interacting states.

Following Ref. \cite{4.4}, one inserts the ansatz in the Lagrangian density (%
\ref{eq_4.3}) and calculates the respective effective Lagrangian,%
\begin{equation}
L_{\text{eff}}=C_{D}\int_{0}^{\infty }\mathcal{L}\left( \psi _{g}\right)
r^{D-1}dr,  \label{eq_4.5}
\end{equation}%
where $C_{D}=2\pi $ or $4\pi $ in the 2D or 3D cases, respectively. Finally,
the evolution equations for the time-dependent parameters of ansatz (\ref%
{eq_4.4}) are derived from $L_{\text{eff}}$ using the corresponding
Euler-Lagrange equations. Subsequent analysis, as well as the results of
direct numerical simulations, is presented separately for the 2D and 3D
cases.

\subsection{The two-dimensional case}

We start the consideration with the 2D case, presenting, consecutively,
results produced by VA, application of the averaging method to the GP
equation, and results produced by numerical simulations. In particular, in
the case of the high-frequency modulation, it is possible to apply the
averaging method to the 2D equation (\ref{eq_4.2}), without using VA \cite%
{3.1,Baizakov,Abdullaev}. The averaging method may also be applied to the 2D
NLS equation with a potential rapidly varying in space, rather than in time,
the main result being renormalization of parameters of the equation and a
shift of the collapse threshold \cite{4.5}. As shown below, rapid temporal
modulation of the nonlinearity coefficient in the GP equation leads to
nontrivial effects, such as generation of additional nonlinear-dispersive
and higher-order nonlinear terms in the corresponding effective NLS
equation, see, Eq. (\ref{eq_4.22}) below. These terms may essentially affect
the dynamics of the collapsing condensate.

\subsubsection{The variational approximation}

The calculation of the effective Lagrangian (\ref{eq_4.5}) for the 2D GP
equation yields%
\begin{equation}
L_{\text{eff}}^{(2D)}=\pi \left( -\frac{1}{2}a^{4}A^{2}\frac{db}{dt}%
-a^{2}A^{2}\frac{d\delta }{dt}-A^{2}-a^{4}A^{2}b^{2}+\frac{1}{4}\lambda
(t)a^{2}A^{4}\right) .  \label{eq_4.6}
\end{equation}%
The Euler-Lagrange equations following from this Lagrangian yield the
conservation of the total number of atoms $N$ in the condensate (represented
by norm $N$ of the mean-field wave function),
\begin{equation}
\pi a^{2}A^{2}\equiv N=\text{ const},  \label{eq_4.7}
\end{equation}%
expressions for the chirp and width,%
\begin{equation}
\frac{da}{dt}=2ab\text{, }\frac{db}{dt}=\frac{2}{a^{4}}-2b^{2}-\frac{\lambda
(t)N}{2\pi a^{4}},
\end{equation}%
and a closed-form evolution equation for the width:%
\begin{equation}
\frac{d^{2}a}{dt^{2}}=\frac{2\left( 2-\lambda (t)N/2\pi \right) }{a^{3}},
\label{eq_4.8}
\end{equation}%
which may be rewritten as%
\begin{equation}
\frac{d^{2}a}{dt^{2}}=\frac{-\Lambda +\epsilon \sin \left( \omega t\right) }{%
a^{3}},  \label{eq_4.9}
\end{equation}%
where
\begin{equation}
\Lambda \equiv 2\left[ \lambda _{0}N/(2\pi )-2\right] ,\text{ \ }\epsilon
=-\lambda _{1}N/\pi .  \label{eq_4.10}
\end{equation}

In the absence of the ac component, i.e., $\epsilon =0$, Eq. (\ref{eq_4.9})
conserves the energy, $E_{\mathrm{2D}}=(1/2)\left[ \left( da/dt\right)
^{2}-\Lambda a^{-2}\right] $. Obviously, $E_{\mathrm{2D}}\rightarrow -\infty
$ as $a\rightarrow 0$, if $\Lambda >0$, and $E_{\mathrm{2D}}\rightarrow
+\infty $ as $a\rightarrow 0$, if $\Lambda <0$. This means that, in the
absence of the ac component, the 2D pulse is expected to collapse at $%
\Lambda >0$, and spread out at $\Lambda <0$. The case of $\Lambda =0$
corresponds to the critical norm, realized by the above-mentioned TS \cite%
{Townes}. Note that a numerically exact value of the critical norm is (in
the present notation) $N=1.862$ \cite{Fibich}, while the variational
equation (\ref{eq_4.10}) yields $N=2$ (if $\lambda _{0}=+1$) \cite{Anderson}.

If the ac component of the nonlinearity coefficient oscillates at a high
frequency, one can set $a(t)=\overline{a}+\delta a$, with $\left\vert \delta
a\right\vert <<\left\vert \overline{a}\right\vert $, where, $\overline{a}$
varies on a slow time scale and $\delta a$ is a rapidly varying function
with zero mean value. Then, Eq. (\ref{eq_4.9}) can be treated analytically
by means of the Kapitsa averaging method \cite{3.1,Baizakov,Abdullaev}.
After straightforward manipulations, an ODE system is derived for the slow
and rapid variables:
\begin{subequations}
\begin{eqnarray}
\frac{d^{2}\overline{a}}{dt^{2}} &=&-\Lambda \left( \overline{a}^{-3}+6%
\overline{a}^{-5}\langle \delta a^{2}\rangle \right) -3\epsilon \langle
\delta a\sin \left( \omega t\right) \rangle \overline{a}^{-4},
\label{eq_4.11} \\
\frac{d^{2}}{dt^{2}}\delta a &=&3\delta a\Lambda \overline{a}^{-4}+\epsilon
\sin \left( \omega t\right) \overline{a}^{-3},  \label{eq_4.12}
\end{eqnarray}%
where $\langle \cdots \rangle $ stands for averaging over period $2\pi
/\omega $. Equation (\ref{eq_4.12}) admits an obvious solution,
\end{subequations}
\begin{equation}
\delta a(t)=-\frac{\epsilon \sin \left( \omega t\right) }{\overline{a}%
^{3}\left( \omega ^{2}+3\overline{a}^{-4}\Lambda \right) }.  \label{eq_4.13}
\end{equation}%
Substituting Eq. (\ref{eq_4.13}) into Eq. (\ref{eq_4.11}) yields the
following final evolution equation for the slow variable:%
\begin{equation}
\frac{d^{2}\overline{a}}{dt^{2}}=\overline{a}^{-3}\left[ -\Lambda -\frac{%
3\Lambda \epsilon ^{2}}{\left( \omega ^{2}\overline{a}^{4}+3\Lambda \right)
^{2}}+\frac{3}{2}\frac{\epsilon ^{2}}{\omega ^{2}\overline{a}^{4}+3\Lambda }%
\right] .  \label{eq_4.14}
\end{equation}

To examine whether the collapse is enforced or inhibited by the ac component
of the nonlinearity, one may look at Eq. (\ref{eq_4.14}) in the limit of $%
\overline{a}\rightarrow 0$, reducing the equation to
\begin{equation}
\frac{d^{2}\overline{a}}{dt^{2}}=\left( -\Lambda +\frac{\epsilon ^{2}}{%
6\Lambda }\right) \overline{a}^{-3}.  \label{eq_4.15}
\end{equation}%
It follows from Eq. (\ref{eq_4.15}) that, if the amplitude of the
high-frequency ac component is large enough, \textit{viz}., $\epsilon
>6\Lambda ^{2}$, the behavior of the condensate (in the limit of small $%
\overline{a}$) is exactly opposite to that which would be expected in the
presence of the dc component only: in the case $\Lambda >0$, rebound occurs
rather than the collapse, and vice versa in the case $\Lambda <0$.

On the other hand, in the limit of large $\overline{a}$ , Eq. (\ref{eq_4.14}%
) takes the asymptotic form of $d^{2}\overline{a}/dt^{2}=-\Lambda \overline{a%
}^{-3}$, which shows that the condensate remains self-confined in the case
of $\Lambda >0$, i.e., if the norm exceeds the critical value. This
consideration is relevant if $\overline{a}$, although being large, remains
smaller than the limit imposed by an external trapping potential, should it
be added to the model. Thus, these asymptotic results guarantee that Eq. (%
\ref{eq_4.14}) give rise to a stable behavior of the condensate, the
collapse and decay (spreading out) being ruled out if
\begin{equation}
\epsilon >\sqrt{6}\Lambda >0.  \label{eq_4.16}
\end{equation}

To illustrate the above results in terms of the experimentally relevant
setting -- for example, for the condensate of $^{7}$\textrm{Li} with the
critical number $\sim 1500$ atoms -- one concludes that, for $1800$ atoms
(i.e., $N/2\pi =2.2$) the stabilization requires to add the periodic
modulation with amplitude $\epsilon =0.98$ (see Eq. (\ref{eq_4.10}) to the
constant coefficient $\lambda _{0}=1$. In fact, conditions (\ref{eq_4.16})
ensure that the right-hand side of Eq. (\ref{eq_4.14}) is positive for small
$\overline{a}$ and negative for large $\overline{a}$, hence Eq. (\ref%
{eq_4.14}) must give rise to a stable fixed point. Indeed, when conditions (%
\ref{eq_4.16}) hold, the right-hand side of Eq. (\ref{eq_4.14}) vanishes at
exactly one fixed point,%
\begin{equation}
\omega ^{2}\overline{a}^{4}=\frac{3\epsilon ^{2}}{4\Lambda }+\sqrt{3\left(
\frac{3\epsilon ^{4}}{16\Lambda ^{2}}-1\right) }-3\Lambda ,  \label{eq_4.17}
\end{equation}%
which can be easily checked to be stable through the calculation of an
eigenfrequency of small oscillations around it.

Direct numerical simulations of Eq. (\ref{eq_4.9}) produce results that are
in exact agreement with those provided by the averaging method, i.e., a
stable state with $\alpha (t)$ performing small oscillations around point (%
\ref{eq_4.17}) \cite{3.1,Abdullaev}. However, the 3D situation shows a
drastic difference, see below.

For the sake of comparison with the results obtained in the 3D case, one
also needs an approximate form of Eq. (\ref{eq_4.14}) valid in the limit of
small $\Lambda $ (i.e., when norm is close to the critical value) and very
large $\omega $:%
\begin{equation}
\frac{d^{2}\overline{a}}{dt^{2}}=-\frac{\Lambda }{\overline{a}^{3}}+\frac{3}{%
2}\frac{\epsilon ^{2}}{\omega ^{2}\overline{a}^{7}}.  \label{eq_4.18}
\end{equation}

To estimate the value of the amplitude of the high-frequency ac component
necessary to stop the collapse, note that a characteristic trap frequency is
$\Omega \sim 100$ \textrm{Hz}, in physical units. Then, for a typical high
modulation frequency $\sim 3$ \textrm{kHz},\ the scaled one is $\omega
\simeq 30$. If the norm is, for instance, $N/2\pi =2.2$, so that, according
to Eq. (\ref{eq_4.10}), $\Lambda =0.4$ (this corresponds to the condensate
of $^{7}$\textrm{Li} with $\simeq 1800$ atoms, the critical number being $%
\simeq 1500$), and the modulation parameters are $\lambda _{0}=1$, $\lambda
_{1}=2.3$, $\epsilon =10$, then the stationary value of the condensate's
width, as obtained from Eq. (\ref{eq_4.17}), is $a_{\mathrm{st}}=0.8l$,
where $l=\sqrt{m\Omega /\hbar }$ is the healing length.

Thus the analytical approach, based on VA and the assumption that the norm
slightly exceeds the critical value, leads to an important prediction: in
the case of the 2D GP equation, the ac component of the nonlinearity, acting
jointly with the dc one corresponding to the attraction, may replace the
collapse by a stable soliton-like oscillatory state that confines itself
without the trapping potential. It is relevant to mention, once again, that
a qualitatively similar result, \textit{viz}., the existence of stable
periodically oscillating spatial cylindrical solitons in a bulk
nonlinear-optical medium consisting of alternating layers with opposite
signs of the Kerr coefficient, was reported in Ref. \cite{4.7}, where this
result was obtained in a completely analytical form on the basis of the
variational approximation, and was confirmed by direct simulations.

\subsubsection{Averaging of the 2D GP (Gross-Pitaevskii) equation and
Hamiltonian}

Here, it is assumed that the ac frequency $\omega $ is large, and the 2D GP
equation (\ref{eq_4.2}) is rewritten in the following form,%
\begin{equation}
i\frac{\partial \psi }{\partial t}+\nabla ^{2}\psi +\lambda (\omega
t)\left\vert \psi \right\vert ^{2}\psi =0.  \label{eq_4.19}
\end{equation}%
To derive an equation governing the slow variations of the field, one can
use the multiscale approach, writing the solution as an expansion in powers
of $1/\omega $ and introducing slow temporal variables, $T_{k}\equiv \omega
^{-k}t$, $k=0,$ $1,$ $2,$ $...$, while the fast time is $\zeta =\omega t$.
Thus, the solution is sought for as%
\begin{equation}
\psi (r,t)=A(r,T_{k})+\omega ^{-1}u_{1}(\zeta ,A)+\omega ^{-2}u_{2}(\zeta
,A)+\cdots ,  \label{eq_4.20}
\end{equation}%
with $\langle u_{k}\rangle =0$, where $\langle \cdots \rangle $ stands for
the average over the period of the rapid modulation, and $\lambda _{0}=1$ is
assumes (i.e., the dc part of the nonlinearity coefficient corresponds to
attraction between the atoms).

Following the procedure developed in Ref. \cite{4.8}, one first finds the
first and second corrections,
\begin{gather}
u_{1}=-i\left( \mu _{1}-\langle \mu _{1}\rangle \right) \left\vert
A\right\vert ^{2}A,\text{ \ }\mu _{1}\equiv \int_{0}^{\zeta }\left[ \lambda
(\tau )-\langle \lambda _{1}\rangle \right] d\tau ,  \notag \\
u_{2}=\left( \mu _{2}-\langle \mu _{2}\rangle \right) \left[ 2i\left\vert
A\right\vert ^{2}A_{t}+iA^{2}A_{t}^{\ast }+\nabla ^{2}(\left\vert
A\right\vert ^{2}A)\right]  \label{eq_4.21} \\
-\left\vert A\right\vert ^{4}A\left\{ \frac{1}{2}\left[ \left( \mu
_{1}-\langle \mu _{1}\rangle \right) ^{2}-2M\right] +\langle \lambda \rangle
\left( \mu _{2}-\langle \mu _{2}\rangle \right) \right\} .  \notag
\end{gather}%
In Eq. (\ref{eq_4.21}), $\mu _{2}\equiv \int_{0}^{\zeta }\left( \mu
_{1}-\langle \mu _{1}\rangle \right) ds$, and $M=\left( 1/2\right) \left(
\langle \mu _{1}^{2}\rangle -\langle \mu _{1}\rangle ^{2}\right) =\left(
1/2\right) \left( \langle \lambda ^{2}\rangle -1\right) $ (recall $\lambda
_{0}=1$ was set). Using these results, one obtains the following evolution
equation for the slowly varying field $A(x,T_{0})$, derived at order $\omega
^{-2}:$%
\begin{equation}
i\frac{\partial A}{\partial t}+\nabla ^{2}A+\left\vert A\right\vert
^{2}A+2M\left( \frac{\epsilon }{\omega }\right) ^{2}\left( \left\vert
A\right\vert ^{6}A-3\left\vert A\right\vert ^{4}\nabla ^{2}A\right)
+2\left\vert A\right\vert ^{2}\nabla ^{2}\left( \left\vert A\right\vert
^{2}A\right) +A^{2}\nabla ^{2}\left( \left\vert A\right\vert ^{2}A^{\ast
}\right) =0,  \label{eq_4.22}
\end{equation}%
where $\epsilon $ is the same amplitude of the ac component as in Eq. (\ref%
{eq_4.10}). Note that Eq. (\ref{eq_4.22}) is valid in both 2D and 3D cases.
In either case, it can be represented in the quasi-Hamiltonian form%
\begin{equation}
\left[ 1+6M\left( \frac{\epsilon }{\omega }\right) ^{2}\left\vert
A\right\vert ^{4}\right] \frac{\partial A}{\partial t}=-i\frac{\delta H_{q}}{%
\delta A^{\ast }},  \label{eq_4.23}
\end{equation}%
\begin{equation}
H_{q}=\int dV\left[ \left\vert \nabla A\right\vert ^{2}-2M\left( \frac{%
\epsilon }{\omega }\right) ^{2}\left\vert A\right\vert ^{8}-\frac{1}{2}%
\left\vert A\right\vert ^{4}+4M\left( \frac{\epsilon }{\omega }\right)
^{2}\left\vert \nabla \left( \left\vert A\right\vert ^{2}A\right)
\right\vert \right] ,  \label{eq_4.24}
\end{equation}%
where $dV$ is the infinitesimal volume in the 2D or 3D space. It immediately
follows from Eq. (\ref{eq_4.23}) and the reality of the (quasi-)Hamiltonian (%
\ref{eq_4.24}) that it is a dynamical invariant, i.e., $dH_{q}/dt=0$.

For further analysis of the 2D case, it is relevant apply the modulation
theory developed in Ref. \cite{4.9}, according to which the solution is
searched for in the form of a modulated TS. The above-mentioned TS is a
solution of the 2D NLS equation in the form of $\psi (r,t)=\exp \left(
it\right) R_{T}(r)$, where function $R_{T}(r)$ is the solution of the
boundary-value problem,%
\begin{equation}
\frac{d^{2}R_{T}}{dr^{2}}+\frac{1}{r}\frac{dR_{T}}{dr}-R_{T}+R_{T}^{3}=0,%
\text{ }\left. \frac{dR_{T}}{dr}\right\vert _{r=0}=0,\text{ }\left.
R_{T}(r)\right\vert _{r=\infty }=0.  \label{eq_4.25}
\end{equation}%
For this solution, norm $N$ and the Hamiltonian $H$ take the well-known
values,%
\begin{eqnarray}
N_{T} &\equiv &\int_{0}^{\infty }R_{T}^{2}(r|)rdr=N_{c}\approx 1.862,  \notag
\\
H_{T} &=&\int_{0}^{\infty }\left[ \left( \frac{dR_{T}}{dr}\right) ^{2}-\frac{%
1}{2}R_{T}^{4}(r)\right] rdr=0.  \label{eq_4.26}
\end{eqnarray}%
The averaged variational equation (\ref{eq_4.22}) indicates an increase of
the critical norm for the collapse, as opposed to the classical value in Eq.
(\ref{eq_4.26}). Using relation (\ref{eq_4.20}), we find
\begin{equation}
N_{\text{crit}}=\int_{0}^{\infty }\left\vert \psi \right\vert ^{2}rdr=N_{T}+2%
\mathrm{I}M\left( \frac{\epsilon }{\omega }\right) ^{2},
\end{equation}%
where $\mathrm{I}=11.178$. This increase of the critical norm is similar to
the well-known energy enhancement of dispersion-managed solitons in optical
fibers with periodically modulated dispersion \cite{4.6}.

Another nontrivial perturbative effect is the appearance of a nonzero value
of the phase \textit{chirp} in the soliton. The mean value of the chirp is
defined as%
\begin{equation}
b=\frac{\int_{0}^{\infty }\text{\textrm{Im}}\left[ \left( \partial \psi
/\partial r\right) \psi ^{\ast }\right] rdr}{\int_{0}^{\infty
}r^{2}dr\left\vert \psi \right\vert ^{2}}.
\end{equation}%
Making use of expression (\ref{eq_4.21}) for the first correction leads to%
\begin{eqnarray}
b &=&-\left( \epsilon /\omega \right) BM\left( \mu _{1}-\langle \mu
_{1}\rangle \right) , \\
B &\equiv &3\frac{\int_{0}^{\infty }rdrR^{2}(R^{\prime
})^{2}-(1/4)\int_{0}^{\infty }drR^{4}}{\int_{0}^{\infty }r^{2}drR^{2}}%
\approx 0.596.
\end{eqnarray}

To develop a general analysis, it is assumed that the solution with the norm
close to the critical value may be approximated as a modulated TS, i.e.,
\begin{equation}
A(r,t)\approx \left[ a(t)\right] ^{-1}R_{T}\left[ r/a(t)\right] \exp \left(
iS\right) ,\text{ \ }S=\sigma (t)+\frac{r^{2}}{4a}\frac{da}{dt},\text{ \ }%
\frac{d\sigma }{dt}=a^{-2}  \label{eq_4.27}
\end{equation}%
with some function $a(t)$. If the initial norm is close to the critical
value, i.e., when $\left\vert N-N_{c}\right\vert <<N_{c}$, the method worked
out in Ref. \cite{4.9} makes it possible to derive an evolution equation for
$a(t)$, starting from approximation (\ref{eq_4.27}). The evolution equation
for the width is%
\begin{equation}
a^{3}a_{tt}=-\beta _{0}+\frac{\epsilon ^{2}}{4M_{0}\omega ^{2}}f_{1}(t),
\label{eq_4.28}
\end{equation}%
where
\begin{equation}
\beta _{0}=\beta (0)-\frac{\epsilon ^{2}}{4M_{0}\omega ^{2}}f_{1}(0),\text{
\ }\beta (0)=\frac{N-N_{c}}{M_{0}},\text{ \ }M_{0}\equiv \frac{1}{4}%
\int_{0}^{\infty }r^{3}drR_{T}^{2}\approx 0.55,
\end{equation}%
with an auxiliary function%
\begin{equation}
f_{1}(t)=2a(t)\text{\textrm{Re}}\left[ \frac{1}{2\pi }\int \int
dxdyF(A_{T})\exp \left[ -iS\right] \left\{ R_{T}+\rho \nabla R_{R}(\rho
)\right\} \right] .  \label{eq_4.29}
\end{equation}%
For the harmonic modulation, the equation in the lowest-order approximation
takes the form of
\begin{equation}
\frac{d^{2}a}{dt^{2}}=-\frac{\Lambda _{1}}{a^{3}}+\frac{C\epsilon ^{2}}{%
\omega ^{2}a^{7}},  \label{eq_4.30}
\end{equation}%
where $\Lambda _{1}=\left( N-N_{c}\right) /M_{0}-C\epsilon ^{2}/(\omega
^{2}a_{0}^{4})$ and $C$ is defined as \cite{3.1}
\begin{equation}
C\equiv \frac{3}{M_{0}}\int_{0}^{\infty }d\rho \left[ 2\rho R_{T}^{4}\left(
R_{T}^{\prime }\right) ^{2}-\rho ^{2}R_{T}^{3}\left( R_{T}^{\prime }\right)
^{3}-\frac{1}{8}\rho R_{T}^{8}\right] \approx 39.  \label{eq_4.31}
\end{equation}%
Thus the averaged equation predicts the arrest of the collapse by the rapid
modulations of the nonlinear term in the 2D GP equation. The comparison of
Eq. (\ref{eq_4.30}) with its counterpart (\ref{eq_4.18}), which was derived
by means of averaging the VA-generated equation (\ref{eq_4.9}), shows that
both approaches lead to the same behavior near the collapse threshold,
numerical coefficients in the second terms being different due to the
different profiles of the Gaussian and TS.

It is relevant to estimate the fixed point as per numerical simulations
performed in Ref. \cite{4.10}. In that work, the stable propagation of
solitons has been observed for two-step modulation of the nonlinearity
coefficient in the 2D NLS equation: $\lambda =1+\epsilon $ at $0<t<T$, and $%
\lambda =1-\epsilon $ at $T<t<2T$. The parameters in the numerical
simulations were taken as $T=\epsilon =0.1,$ $N/(2\pi )=11.726/(2\pi )$,
with the critical number $N_{c}=11.68/(2\pi )$. For these values, one has $%
a_{c}=0.49$, which agrees with the value $a_{c}\approx 0.56$ produced by
numerical simulations.

Instead of averaging Eq. (\ref{eq_4.2}), one can apply the averaging
procedure, also based on representation (\ref{eq_4.20}) for the wave
function, directly to the Hamiltonian of Eq. (\ref{eq_4.2}). As a result,
the averaged Hamiltonian is found in the form of
\begin{equation}
\overline{H}=\int \int dxdy\left[ \left\vert \nabla A\right\vert
^{2}+2M\left( \frac{\epsilon }{\omega }\right) ^{2}\left\vert \nabla \left(
\left\vert A\right\vert ^{2}A\right) \right\vert ^{2}-\frac{1}{2}\left\vert
A\right\vert ^{4}-6M\left( \frac{\epsilon }{\omega }\right) ^{2}\left\vert
A\right\vert ^{8}\right] .  \label{eq_4.32}
\end{equation}%
A possibility to arrest the collapse, in the presence of the rapid periodic
modulation of the nonlinearity strength, can be explained with the help of
this Hamiltonian. To this end, following the pattern of the virial estimates
\cite{4.11}, one notes that, if a given field configuration has compressed
itself to a spot with size $\rho $, where the amplitude of the $A$ field is $%
\sim \aleph $, the conservation of norm $N$ (which may be applied to field $%
A $ through relation (\ref{eq_4.20})) yields relation%
\begin{equation}
\aleph ^{2}\rho ^{D}\sim N,  \label{eq_4.33}
\end{equation}%
$D$ being the space dimension. On the other hand, the same estimate for the
strongest collapse-driving and collapse-arresting terms (the fourth and
second terms, respectively, in expression (\ref{eq_4.32})), $H_{-}$ and $%
H_{+}$, in the Hamiltonian yields%
\begin{equation}
H_{-}\sim -\left( \frac{\epsilon }{\omega }\right) ^{2}\aleph ^{8}\rho ^{D},%
\text{ \ \ }H_{+}\sim \left( \frac{\epsilon }{\omega }\right) ^{2}\aleph
^{6}\rho ^{D-2}.  \label{eq_4.34}
\end{equation}%
Eliminating the amplitude from Eqs. (\ref{eq_4.34}) by means of relation (%
\ref{eq_4.33}), one concludes that, in the case of the catastrophic
self-compression of the field in the 2D space, $\rho \rightarrow 0$, both
terms $H_{\pm }$ asymptotically scale as $\rho ^{-5}$, hence the collapse
may be arrested, depending on details of the configuration. However, in the
3D case the collapse-driving term diverges as $\rho ^{-9}$, while the
collapse-arresting one scales $\sim \rho ^{-8}$ at $\rho \rightarrow 0$,
hence in this case the collapse cannot be prevented.

Lastly, it is relevant to mention that, although the quasi-Hamiltonian (\ref%
{eq_4.24}) is not identical to the averaged Hamiltonian (\ref{eq_4.32}), the
virial estimate applied to $H_{q}$ yields exactly the same result: the
collapse can be arrested in the 2D case, but not in the 3D one.

\subsubsection{Numerical simulations}

\begin{figure}[tbp]
\centerline{\includegraphics[scale=.85]{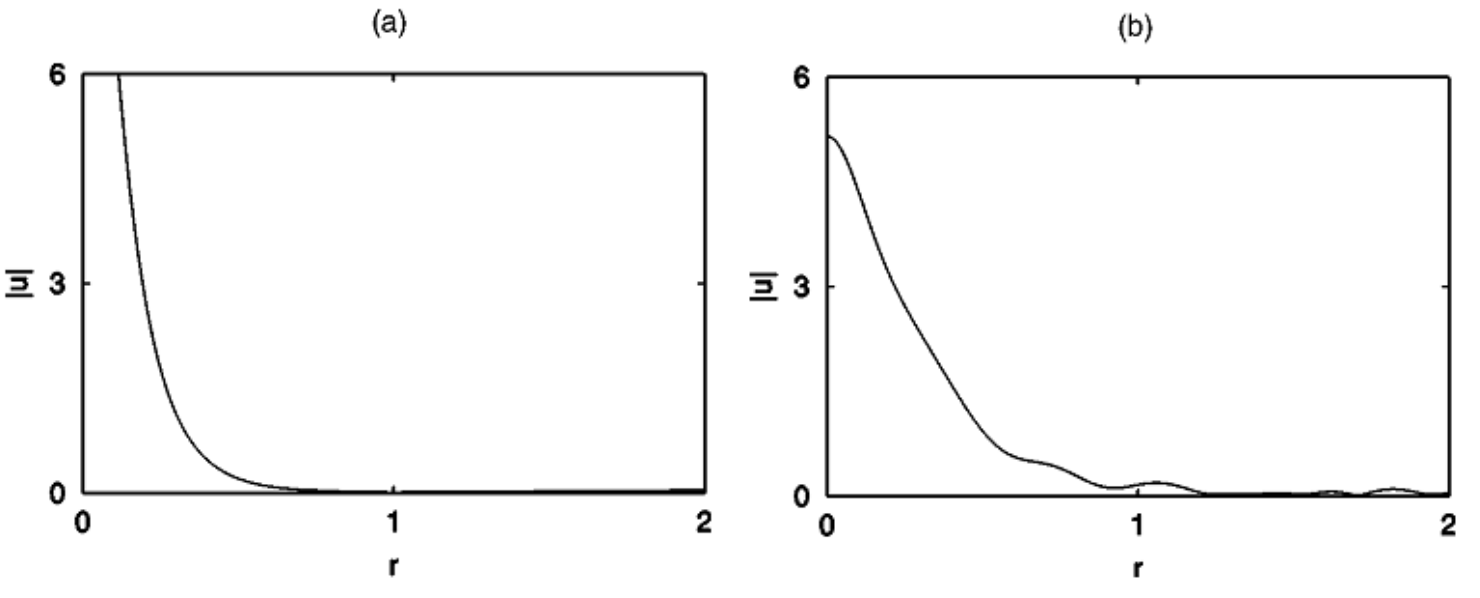}} \caption{A
typical example of the formation of a self-confined condensate in
simulations of the 2D equation (\protect\ref{eq_4.2}). Panel (a)
shows the collapsing state in the absence of the ac modulation at
$t\approx 0.3$. Panel (b) shows the radial profile of the stable
state formed by the same input at $t\approx 0.6$, in the presence
of the ac term in the nonlinearity
coefficient. The parameters are $\protect\lambda _{0}=2.4$, $\protect\lambda %
_{1}=0.85$, $\protect\omega =100\protect\pi $, and $N=5$. The results are
reproduced from Ref. \cite{3.1}.}
\label{fig015}
\end{figure}

The existence of stable self-confined soliton-like oscillating states,
predicted by means of the analytical approximations in region (\ref{eq_4.16}%
), when the dc part of the nonlinearity corresponds to self-attraction, and
the amplitude of the ac component is not too small, was checked by
simulations of the 2D equation (\ref{eq_4.2}). It was quite easy to confirm
this prediction (in the case $\lambda _{0}=-1$, i.e., when the dc component
of the nonlinearity corresponds to repulsion, the direct simulations always
show a decay (spreading out) of the condensate, which also agrees with the
above predictions). A typical example of the formation of a self-confined
state, supported by the combination of the self-focusing dc and sufficiently
strong ac components of the nonlinearity in the absence of an external trap,
is displayed in Fig. \ref{fig015}. The left panel shows the radial profile
of a collapsing state at $t\approx 0.3$ in the absence of the ac term. In
the presence of this term, the right panel shows that the pulse is
stabilized for about $40$ ac-modulation periods, after which it decays
(possible eventual decay on a very long time scale was numerically
investigated, in a detailed form, in Ref. \cite{Itin}). Note the generation
of radiation, attached to the main body as a tail, which is a result of
self-adjustment of the input to the modulation.

\subsection{The 3D GP model}

To address the situation in the case of the 3D equation (\ref{eq_4.2}), it
is relevant, similar to the 2D case, to separately consider results produced
by the analytical approximations, i.e., VA and averaging, and by direct
simulations.

\subsubsection{VA and averaging}

Calculating effective Lagrangian (\ref{eq_4.5}) in the case of the 3D GP
equation yields%
\begin{equation}
L_{\text{eff}}^{(3D)}=\frac{1}{2}\pi ^{\frac{3}{2}}A^{2}a^{3}\left[ -\frac{3%
}{2}a^{2}\frac{db}{dt}-2\frac{d\delta }{dt}+\frac{1}{2\sqrt{2}}\lambda
(t)A^{2}-\frac{3}{a^{2}}-3b^{2}a^{2}\right] .  \label{eq_4.35}
\end{equation}%
The Euler-Lagrange equations produced by this Lagrangian yield the norm
conservation,
\begin{equation}
\pi ^{\frac{3}{2}}A^{2}a^{3}\equiv N=\text{ const,}
\end{equation}%
relations including the \textit{chirp},
\begin{equation}
\frac{da}{dt}=2ab\text{, }\frac{db}{dt}=\frac{2}{a^{4}}-2b^{2}-\frac{\lambda
(t)N}{2\sqrt{2}\pi ^{\frac{3}{2}}a^{5}},
\end{equation}%
and the evolution equation for the width of the condensate,%
\begin{equation}
\frac{d^{2}a}{dt^{2}}=\frac{4}{a^{3}}-\frac{\lambda (t)}{2\sqrt{2}\pi ^{%
\frac{3}{2}}}\frac{N}{a^{4}},  \label{eq_4.36}
\end{equation}%
which is different from its counterpart (\ref{eq_4.8}) corresponding to the
2D case.

Proceeding as in the 2D case, one renormalizes the amplitudes of
the dc and
ac components of the nonlinearity as $\Lambda \equiv \lambda _{0}N/\sqrt{%
2\pi ^{3}}$ and $\epsilon \equiv -\lambda _{1}N/\sqrt{2\pi ^{3}}$, and cast
Eq. (\ref{eq_4.36}) in the scaled form%
\begin{equation}
\frac{d^{2}a}{dt^{2}}=\frac{4}{a^{3}}+\frac{-\Lambda +\epsilon \sin \left(
\omega t\right) }{a^{4}}.  \label{eq_4.37}
\end{equation}%
In the absence of the ac term, i.e., $\epsilon =0$, Eq. (\ref{eq_4.37})
conserves the energy%
\begin{equation}
E_{\text{\textrm{3D}}}=\frac{1}{2}\left( \frac{da}{dt}\right) ^{2}+2a^{-2}-%
\frac{1}{3}\Lambda a^{-3}.
\end{equation}%
This expression shows that $E_{\text{\textrm{3D}}}\rightarrow -\infty $ as $%
a\rightarrow 0$, if $\Lambda >0$, and $E_{\text{\textrm{3D}}}\rightarrow
+\infty $ as $a\rightarrow 0$ if $\Lambda <0$, which corresponds to the
collapse or decay of the pulse, respectively.

\begin{figure}[tbp]
\centerline{\includegraphics[scale=.85]{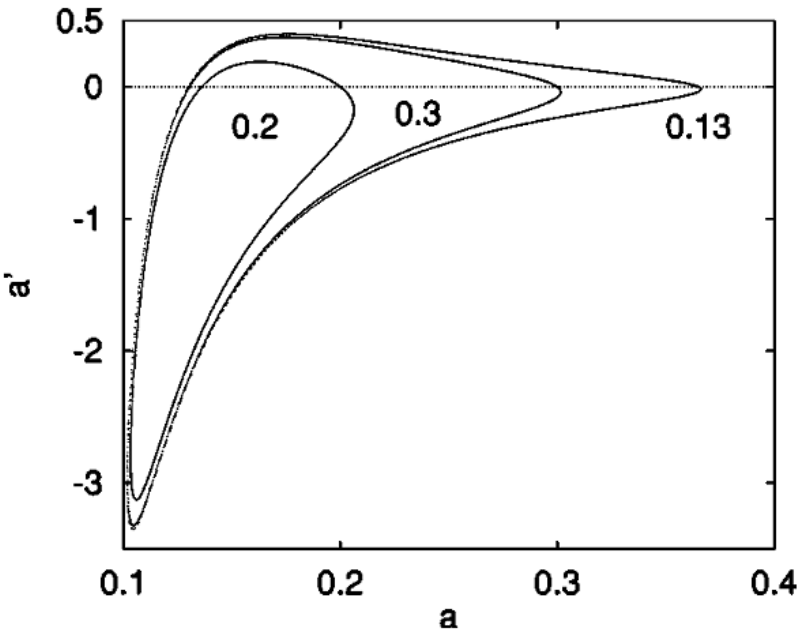}}
\caption{The Poincar\'{e} section in the plane of ($a,da/dt$ ) for $\Lambda
=-1$, $\protect\epsilon =100$, $\protect\omega =10^{4}\protect\pi $,
generated by the numerical solution of the variational equation (\protect\ref%
{eq_4.37}) with different initial conditions (see the text). The results are
reproduced from Ref. \protect\cite{3.1}.}
\label{fig016}
\end{figure}

Prior to applying the averaging procedure, Eq. (\ref{eq_4.37}) can be solved
numerically, without averaging, to show that there is a region in the
parameter space where the condensate, which would decay under the action of
the repulsive dc nonlinearity ($\Lambda <0$), may be stabilized by the ac
component, provided that its amplitude is sufficiently large. The behavior
of the solutions is displayed in Fig. \ref{fig016}, by means of the Poincar%
\'{e} section in the phase plane of ($a,da/dt$), for $\Lambda =-1$, $%
\epsilon =100$, $\omega =10^{4}\pi $, and initial conditions $a(t=0)=0.3$, $%
0.2$, or $0.13$ and $da/dt(t=0)=0$. As seen in Fig. \ref{fig016}, in all the
cases the solution remains bounded with quasiperiodic oscillations, avoiding
the collapse or decay.

\begin{figure}[tbp]
\centerline{\includegraphics[scale=.85]{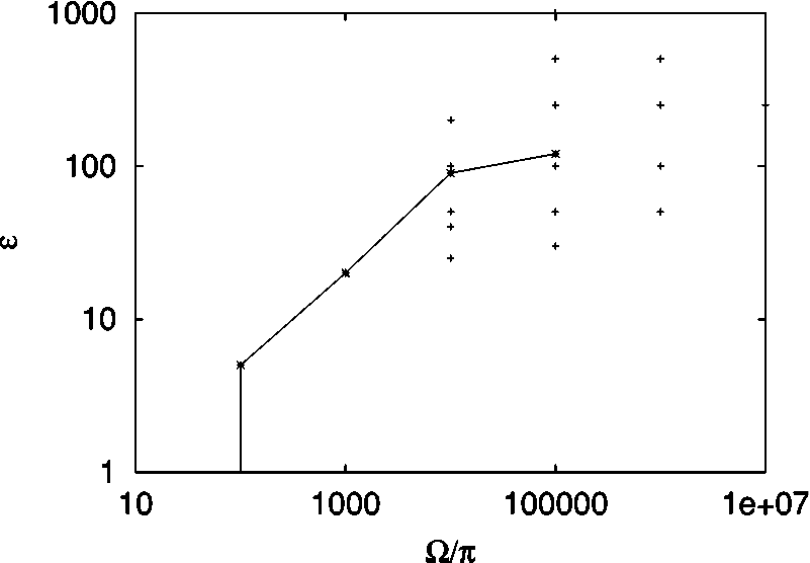}}
\caption{The region in the ($\protect\epsilon ,\protect\omega /\protect\pi $%
) parameter plane where the numerical solution of Eq. (\protect\ref{eq_4.37}%
) with $\Lambda =-1$ predicts stable quasiperiodic solutions in the 3D case.
Crosses mark points where stable solutions were actually obtained. Stars
correspond to minimum values of the ac-component's amplitude $\protect%
\epsilon $ eventually leading to the collapse of the solution of the full
partial differential equation (\protect\ref{eq_4.2}) with $\Lambda =-1$. The
results are reproduced from Ref. \protect\cite{3.1}.}
\label{fig017}
\end{figure}

In fact, the corresponding stability region in the parameter plane ($\omega
/\pi ,\epsilon $) is small, see Fig. \ref{fig017}. It is also seen that the
frequency and amplitude of the ac component need to be large enough to
maintain the stability. Note that, for frequencies larger than $10^{6}\pi $,
the width of the condensate $a(t)$ assumes very small values in the course
of the evolution (as predicted by VA), suggesting that the collapse may
occur in the solution of the full 3D equation (\ref{eq_4.2}). The stability
is predicted by VA only for $\Lambda <9$, i.e., for the repulsive dc
component of the nonlinearity coefficient. In the opposite case, VA predicts
solely the collapse.

As $\omega $ is large enough in the stability region shown in Fig. \ref%
{fig017}, it seems natural to apply the averaging method to this case too.
Similar to how it is outlined above in the 2D case, the rapidly oscillating
correction $\delta a(t)$ to the solution can be found, cf. Eq. (\ref{eq_4.13}%
):%
\begin{equation}
\delta a=-\frac{\epsilon \sin \left[ \omega t\right] \overline{a}}{\omega
^{2}\overline{a}^{5}-12\overline{a}+4\Lambda },  \label{eq_4.38}
\end{equation}%
and the resulting evolution equation for the slow variable $\overline{a}(t)$
is%
\begin{equation}
\frac{d^{2}\overline{a}}{dt^{2}}=\overline{a}^{-4}\left[ 4\overline{a}%
^{4}-\Lambda +\frac{2\epsilon ^{2}}{\omega ^{2}\overline{a}^{5}-12\overline{a%
}+4\Lambda }+\epsilon ^{2}\frac{6\overline{a}-5\Lambda }{\left( \omega ^{2}%
\overline{a}^{5}-12\overline{a}+4\Lambda \right) ^{2}}\right] .
\label{eq_4.39}
\end{equation}%
In the limit $\overline{a}\rightarrow 0$, Eq. (\ref{eq_4.39}) takes the form
of%
\begin{equation}
\frac{d^{2}\overline{a}}{dt^{2}}=\left[ -\Lambda +\frac{3\epsilon ^{2}}{%
16\Lambda }\right] \overline{a}^{-4}.  \label{eq_4.40}
\end{equation}%
Equation (\ref{eq_4.40}) predicts one property of the 3D model correctly,
\textit{viz}., in the case $\Lambda <0$ and with a sufficiently large
amplitude of the ac component ($\epsilon >(4/\sqrt{3})\left\vert \Lambda
\right\vert $), it follows from Eq. (\ref{eq_4.40}) that collapse takes
place instead of spreading out. However, other results following from the
averaged equation (\ref{eq_4.39}) are wrong, as compared to those produced
by simulations of the full variational equation (\ref{eq_4.37}), which are
displayed above (see Figs. \ref{fig016} and \ref{fig017}). In particular,
detailed analysis of the right-hand side of Eq. (\ref{eq_4.39}) shows that
it does not predict a stable fixed point for $\Lambda <0$, and does predict
it for $\Lambda >0$, exactly opposite to what is revealed by the simulations
of Eq. (\ref{eq_4.37}). The failure of the averaging approach in this case
may be explained by the existence of singular points in Eqs. (\ref{eq_4.38})
and (\ref{eq_4.39}) (for both $\Lambda >0$ and $\Lambda <0$), at which the
denominator $\omega ^{2}\overline{a}^{5}-12\overline{a}+4\Lambda $ vanishes.
Note that, in the 2D case with $\Lambda >0$, for which the stable state was
found (see Eq. (\ref{eq_4.16})), the corresponding Eq. (\ref{eq_4.14}) did
not have singularities.

\subsubsection{Direct simulations of the GP equation in the 3D case}

Comparison of the results produced by VA with direct simulations of the 3D
version of the radial equation (\ref{eq_4.2}) is necessary.

\begin{figure}[tbp]
\centerline{\includegraphics[scale=.95]{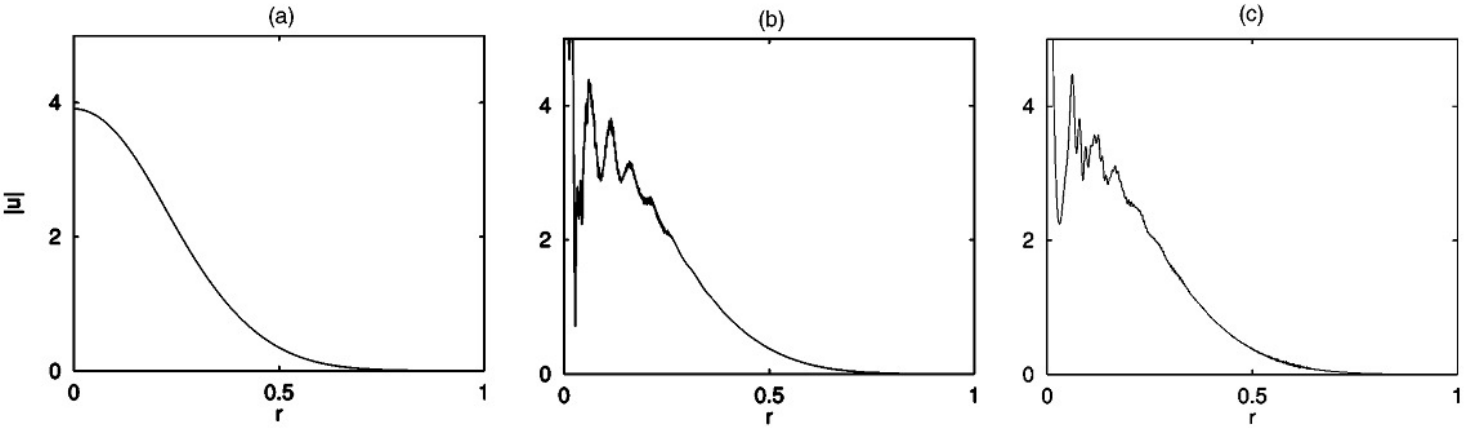}}
\caption{The evolution of the density radial profile, $\left\vert
u(r)\right\vert ^{2}$, in the presence of the strong and fast ac modulation (%
$\protect\omega =10^{4}\protect\pi $, $\protect\epsilon =90$). The profiles
of $\left\vert u(r)\right\vert ^{2}$ are shown at different times: $t=0.007$
(a), $0.01$ (b), and $0.015$ (c). The results are reproduced from Ref.
\protect\cite{3.1}.}
\label{fig018}
\end{figure}

In the absence of the ac component, i.e., $\epsilon =0$ and under the
condition $\Lambda <0$, the simulations show straightforward decay. If an ac
component with a sufficiently large amplitude is added, transient
stabilization of the condensate takes place, roughly similar to how it is
predicted by the solution of the variational equation (\ref{eq_4.37}).
However, the stabilization is not permanent: the wave function begins to
develop small-amplitude short-scale variations around the center, and after
about $50$ periods of the ac modulation, the collapse takes place.

An example of this behavior is displayed in Fig. \ref{fig018}, for $N=1$, $%
\Lambda =-1$, and $\omega =10^{4}\pi $. Figure \ref{fig018} shows radial
density profiles $\left\vert u(r)\right\vert ^{2}$ at different instants of
time. Results presented in Fig. \ref{fig018} is typical for the 3D case with
$\Lambda <0$. The eventual collapse that takes place in this case is a
nontrivial finding, as it occurs despite the fact that the dc part of the
nonlinearity drives the condensate towards spreading out. Therefore, a basic
characteristic of the system is a dependence of the minimum ac amplitude $%
\epsilon $, which gives rise to the collapse at fixed $\Lambda =-1$, versus
ac frequency $\omega $. Several points marked by stars show this dependence
in Fig. \ref{fig017}. It is natural that the minimum value of $\epsilon $
necessary for the collapse grows with $\omega $. On the other hand, for $%
\omega $ not too large, this minimum value is small, as even small $\epsilon
$ is sufficient to push the state into the collapse during the relatively
long half-period when the sign of the net nonlinearity coefficient $\lambda
(t)$ is positive, see Eq. (\ref{eq_4.19}).

In the case of $\Lambda >0$ the collapse cannot be predicted, in agreement
with the analysis developed above on the basis of the Hamiltonian of the
averaged version of the GP equation, which showed that the collapse could
not be arrested in the 3D case, provided that the amplitude of the ac
component was large enough. This result also accords with the findings of
direct simulations of the propagation of localized 3D spatiotemporal pulses
in the above-mentioned model of the nonlinear-optical medium consisting of
alternating layers with opposite signs of the Kerr coefficient: on the
contrary to the stable 2D spatial solitons, the 3D \textquotedblleft light
bullets\textquotedblright\ can never be stable in that model.

\subsection{Conclusion of the section}

The dynamics of both 2D and 3D BEC is considered in the case when the
nonlinearity coefficient in the GP equation contains constant (dc) and
time-variable (ac) parts. This may be achieved in the experiment by means of
a resonantly tuned ac magnetic field, through FR. Using VA, directly
simulating the GP equation, and applying the averaging procedure to it
without the use of VA, one concludes that, in the 2D case, the ac component
of the nonlinearity makes it possible to maintain the condensate in a stable
self-confined state without an external trap, which agrees with known
results reported for spatial solitons in nonlinear optics. In the 3D case,
VA also predicts a stable self-confined state of the condensate without a
trap, provided that the constant part of the nonlinearity corresponds to
repulsion between atoms. However, in this case direct simulations reveal
that the stability of the self-confined condensate is limited in time.
Eventually, collapse takes place, despite the fact that the dc component of
the nonlinearity is repulsive. Thus, the conclusion is that the spatially
uniform ac magnetic field, resonantly tuned to affect the scattering length
via FR, may readily play the role of an effective trap that confines the
condensate, and sometimes enforces its collapse.

\section{Soliton stability in two-component BEC under temporal modulation}

In the present section, we address the dynamics of a binary condensate in an
expulsive time-varying HO potential with time-varying intrinsic attractive
interactions, governed by an integrable system of two 1D GP equations. In
this model, the strength of the attractive nonlinearity exponentially decays
with time, hence solitons are also subject to decay. Nevertheless, it is
shown that the robustness of bright solitons can be enhanced, making their
lifetime longer, by matching the time dependence of the interaction strength
to the time modulation of the strength of the HO potential. The nonlinearity
coefficient can be made time-dependent through FR, which helps to choose the
time-modulation pattern corresponding to the integrable system. A conclusion
is that the expulsive time-modulated HO potential, combined with the
modulated nonlinearity, may sustain stable BEC, while it quickly decays in
the time-independent potential. The analytical results predicting this
behavior are confirmed by numerical simulations. Furthermore, it is
demonstrated that the addition of noise does not impact the BEC\ stability
in the expulsive time-dependent potential.

Results presented in this section are based on Ref. \cite{add8}.

\subsection{The physical model and the Lax pair}

\begin{figure}[tbp]
\centerline{\includegraphics[scale=.95]{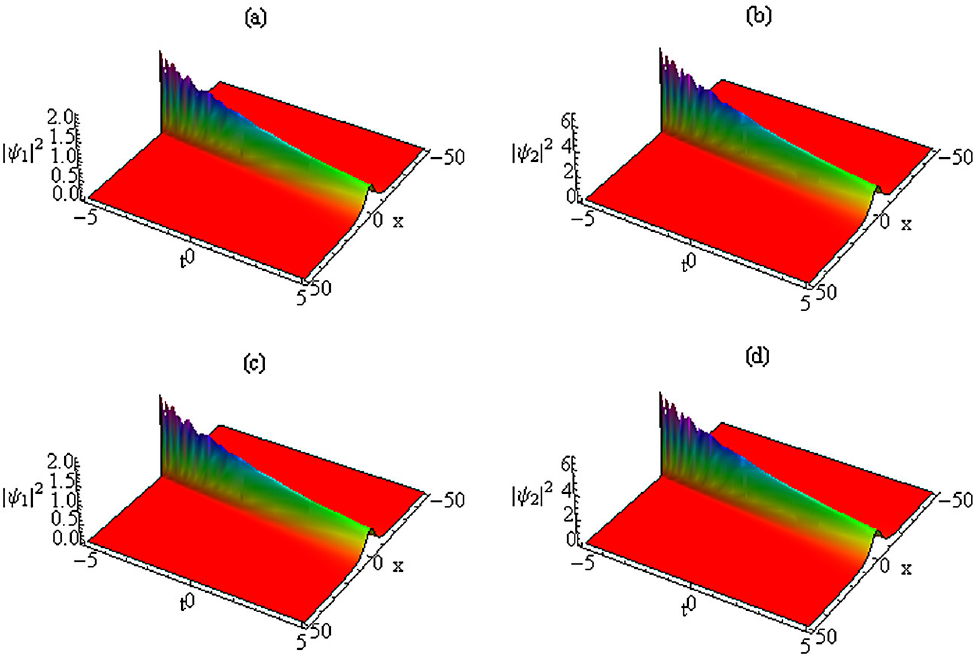}}
\caption{(Color online) (a,b): The analytical soliton solution of Eqs. (%
\protect\ref{eq_5.2}) under the integrability condition (\protect\ref{eq_5.8}%
), with $g(t)$ taken as per Eq. (\protect\ref{g(t)}) (and $\protect\lambda %
^{2}=1/16$). (c,d): The numerically generated counterpart of the
same solution. The results are reproduced from Ref.
\protect\cite{add8}.} \label{fig019}
\end{figure}

The model to be studied in this section pertains to two-component BEC, with
equal atomic masses and attractive interactions in both components (such as
the condensate composed of two hyperfine states of $^{7}$\textrm{Li} \cite%
{4.12} or $^{85}$\textrm{Rb} \cite{4.13} atoms), trapped in the HO
potential. The mean-field evolution of the setting is governed by coupled 1D
GP equations, written in the scaled form as%
\begin{equation}
i\frac{\partial \psi _{j}}{\partial t}=\left( -\frac{1}{2}\frac{\partial ^{2}%
}{\partial x^{2}}+b_{jj}\left\vert \psi _{j}\right\vert
^{2}+b_{j3-j}\left\vert \psi _{3-j}\right\vert ^{2}+\frac{1}{2}\Omega
_{j}^{2}(t)x^{2}\right) \psi _{j}=0,\text{\ }j=1,2,  \label{eq_5.1}
\end{equation}%
where $\psi _{j}$ is the mean-field wave function of the $j$-\textrm{th}
component subject to the normalization condition $\int_{-\infty }^{+\infty
}\left\vert \psi _{1}\right\vert ^{2}dx=1$ and $\int_{-\infty }^{+\infty
}\left\vert \psi _{2}\right\vert ^{2}dx=N_{2}/N_{1}$. The intrinsic
interactions in Eq. (\ref{eq_5.1}) are controlled by the SPM and XPM
coefficients, $b_{jj}=4a_{jj}N_{i}/r_{\perp }$ and $b_{jk}=4a_{jk}N_{i}/r_{%
\perp }$, where $a_{jj}$ and $a_{jk}$ are the respective scattering lengths,
and $r_{\perp }$ is the transverse-component radius. In the section, the
symmetric system is considered, with $b_{11}=b_{22}=b_{21}=b_{12}\equiv -g$,
and $\Omega _{j}^{2}(t)=\omega _{j}^{2}(t)/\omega _{\perp }^{2}$, where $%
\omega _{j}$ and $\omega _{\perp }$ represent, respectively, frequencies of
the trapping potential in the longitudinal and transverse directions. Time $t
$ and coordinate $x$ are measured in units of $2/\omega _{\perp }$ and $%
r_{\perp }$ $=\sqrt{\hbar /\left( m\omega _{\perp }\right) }$, respectively.

Further, assuming $\Omega _{1}^{2}(t)=\Omega _{2}^{2}(t)\equiv -\lambda
^{2}(t)$ (i.e., the confining and expulsive signs of the potential
correspond to $\lambda ^{2}<0$ and $\lambda ^{2}>0$, respectively), and
allowing the interaction coefficient $g(t)$ and potential strength $\lambda
^{2}(t)$ to vary in time, Eq. (\ref{eq_5.1}) takes the following form (with
additional rescaling $t\rightarrow 2t$):%
\begin{equation}
i\frac{\partial \psi _{j}}{\partial t}+\frac{\partial ^{2}\psi _{j}}{%
\partial x^{2}}+2g(t)\left( \left\vert \psi _{j}\right\vert ^{2}+\left\vert
\psi _{3-j}\right\vert ^{2}\right) \psi _{j}+\lambda ^{2}(t)x^{2}\psi _{j}=0,%
\text{ }j=1,2.  \label{eq_5.2}
\end{equation}

\begin{figure}[tbp]
\centerline{\includegraphics[scale=1.05]{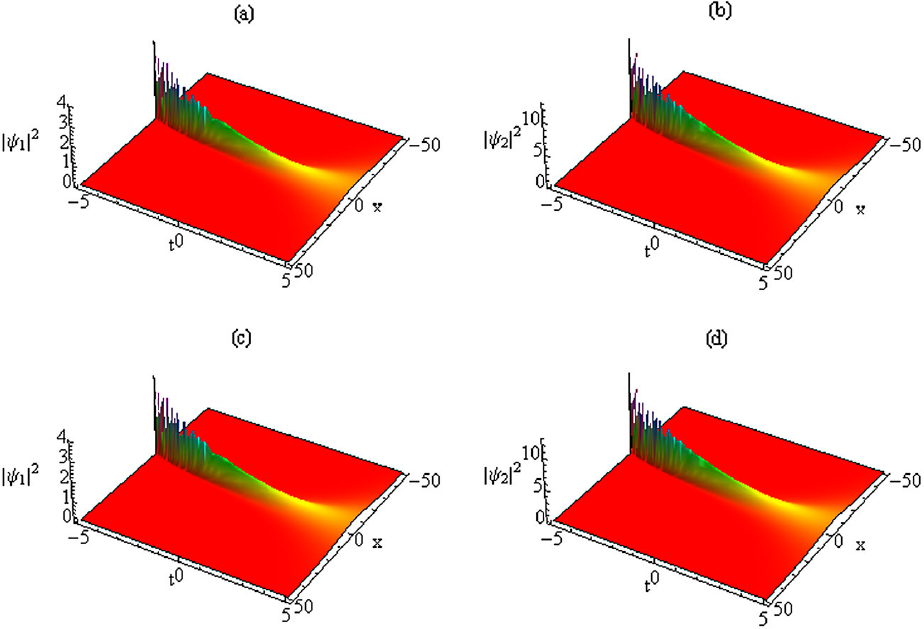}}
\caption{(Color online) The same as in Fig. \protect\ref{fig019},
but $g(t)$ given by Eq. (\protect\ref{gg(t)}), with
$\protect\lambda ^{2}=1/4$. The results are reproduced from Ref.
\protect\cite{add8}.} \label{fig020}
\end{figure}
In this system the nonlinearity takes the Manakov's form \cite{4.14}, which
is a well-known necessary condition for the integrability of the system. In
what follows, we consider the case of $g(t)>0$, which corresponds to the
attractive sign of the SPM and XPM interactions. Under the special
integrability condition imposed on $g(t)$ and $\lambda (t)$, system (\ref%
{eq_5.2}) admit a representation in the form of the Lax pair,%
\begin{equation}
\left.
\begin{array}{c}
\Phi _{x}+U\Phi =0, \\
\Phi _{t}+V\Phi =0,%
\end{array}%
\right.  \label{eq_5.3}
\end{equation}%
where $\Phi =(\phi _{1},\phi _{2},\phi _{3})^{T}$ is the three-component
Jost function, and the operators are
\begin{subequations}
\begin{eqnarray}
U &=&\left(
\begin{array}{ccc}
i\zeta (t) & Q_{1} & Q_{2} \\
-Q_{1}^{\ast } & -i\zeta (t) & 0 \\
-Q_{2}^{\ast } & 0 & -i\zeta (t)%
\end{array}%
\right) ,  \label{eq_5.4} \\
V &=&\left(
\begin{array}{ccc}
v_{11} & v_{12} & v_{13} \\
v_{21} & v_{22} & v_{23} \\
v_{31} & v_{32} & v_{33})%
\end{array}%
\right) ,  \label{eq_5.5}
\end{eqnarray}%
with
\end{subequations}
\begin{eqnarray}
v_{11} &=&-i\zeta ^{2}(t)+i\Gamma (t)x\zeta (t)+\frac{i}{2}Q_{1}Q_{1}^{\ast
}+\frac{i}{2}Q_{2}Q_{2}^{\ast }, \\
v_{12} &=&\Gamma (t)xQ_{1}-\zeta (t)Q_{1}+\frac{i}{2}\frac{\partial Q_{1}}{%
\partial x}, \\
v_{13} &=&\Gamma (t)xQ_{2}-\zeta (t)Q_{2}+\frac{i}{2}\frac{\partial Q_{2}}{%
\partial x},
\end{eqnarray}%
\begin{eqnarray}
v_{21} &=&\Gamma (t)xQ_{1}^{\ast }+\zeta (t)Q_{1}^{\ast }+\frac{i}{2}\frac{%
\partial Q_{1}^{\ast }}{\partial x}, \\
v_{22} &=&i\zeta ^{2}(t)-i\Gamma (t)x\zeta (t)-\frac{i}{2}Q_{1}Q_{1}^{\ast },
\\
v_{23} &=&-\frac{i}{2}Q_{2}Q_{1}^{\ast },
\end{eqnarray}%
\begin{eqnarray}
v_{31} &=&-\Gamma (t)xQ_{2}^{\ast }+\zeta (t)Q_{2}^{\ast }+\frac{i}{2}\frac{%
\partial Q_{2}^{\ast }}{\partial x}, \\
v_{32} &=&-\frac{i}{2}Q_{1}Q_{2}^{\ast }, \\
v_{33} &=&i\zeta ^{2}(t)-i\Gamma (t)x\zeta (t)-\frac{i}{2}Q_{2}Q_{2}^{\ast },
\end{eqnarray}%
\begin{eqnarray}
Q_{1} &=&\frac{1}{\sqrt{g(t)}}\psi _{1}(x,t)\exp \left( \frac{i}{2}\Gamma
(t)x^{2}\right) , \\
Q_{2} &=&\frac{1}{\sqrt{g(t)}}\psi _{2}(x,t)\exp \left( \frac{i}{2}\Gamma
(t)x^{2}\right) .
\end{eqnarray}%
The compatibility condition, $(\Phi _{x})_{t}=(\Phi _{t})_{x}$, leads to the
zero-curvature equation, $U_{t}-V_{x}+[U,V]$ $=0$, which is tantamount to
the integrable system (\ref{eq_5.2}), provided that the spectral parameter $%
\zeta (t)$ obeys the following nonisospectral condition:%
\begin{equation}
\zeta (t)=\mu \exp \left( -\int \Gamma (t)dt\right) ,  \label{eq_5.6}
\end{equation}%
where $\mu $ is a hidden complex constant and $\Gamma (t)$ is an arbitrary
function of time, which is related to the trap's strength:%
\begin{equation}
\lambda ^{2}(t)=\Gamma ^{2}(t)-\frac{d\Gamma }{dt},  \label{eq_5.7}
\end{equation}%
and $\lambda (t)$ is related to the variable interaction strength $g(t)$ by
the integrability condition:%
\begin{equation}
\lambda ^{2}(t)g^{2}(t)=2\left( \frac{dg}{dt}\right) ^{2}-g(t)\frac{d^{2}g}{%
dt^{2}}.  \label{eq_5.8}
\end{equation}%
Thus, the system of GP equations (\ref{eq_5.2}) is integrable if $\lambda
(t) $ and $g(t)$ satisfy the integrability condition (\ref{eq_5.8}). For the
time-independent trap, $\lambda (t)=$const $\equiv $ $c_{1}$, Eq. (\ref%
{eq_5.8}) yields $g(t)=\exp \left( c_{1}t\right) $. In what follows, we
focus on the consideration of the integrable system satisfying this
condition. If the rigorous integrability condition is slightly broken, the
result depends on the accumulation of the deviation from the integrability
over a characteristic time scale, $T$, of the dynamical regime. Namely, if
the deviation from the integrability is characterized by difference $\Delta
\lambda (t)$ from the value imposed by Eq. (\ref{eq_5.8}), the condition for
the system to remain close to the integrability is%
\begin{equation}
\left\vert \int_{0}^{T}\Delta \lambda (t)dt\right\vert \ll 1.  \label{eq_5.9}
\end{equation}

\begin{figure}[tbp]
\centerline{\includegraphics[scale=.95]{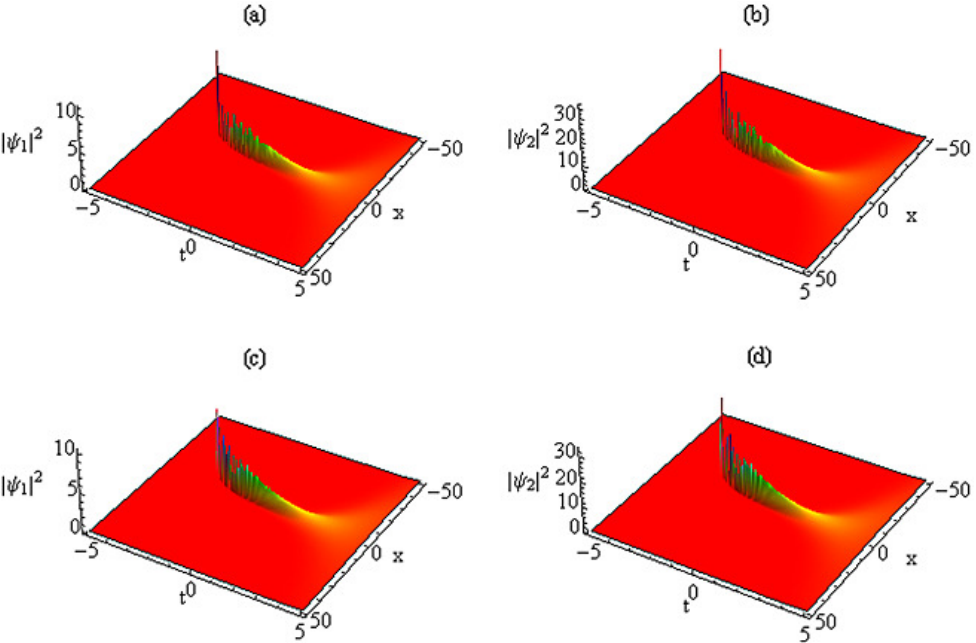}}
\caption{(Color online) The same as in Fig. \protect\ref{fig019}, but for $%
g(t)$ given by Eq. (\protect\ref{ggg(t)}), with $\protect\lambda ^{2}=0.81$.
The figure is borrowed from Ref. \protect\cite{add8}.}
\label{fig021}
\end{figure}

It is important to mention that one can convert the integrable system of the
coupled GP equations, defined by Eqs. (\ref{eq_5.2}) and (\ref{eq_5.8}),
into the usual Manakov's model with constant coefficients (\ref{eq_5.2}) by
a suitable transformation. Nevertheless, the direct formulation of the
integrability formalism in the form of Eqs. (\ref{eq_5.4}), (\ref{eq_5.5}),
and (\ref{eq_5.8}) is quite useful, as it makes it possible to apply the
gauge-transformation method for generating multisoliton solutions, and it
may also be used for the search of more general integrable systems.

\subsection{Analytical and numerical results for two-component bright
solitons in the integrable system}

With the help of the gauge-transformation approach, bright solitons of the
system of GP equations (\ref{eq_5.2}), subject to the integrability
condition (\ref{eq_5.8}), can be looked for as
\begin{subequations}
\begin{eqnarray}
\psi _{1}^{(1)} &=&\frac{2}{\sqrt{g(t)}}\varepsilon _{1}^{(1)}\frac{\beta
_{1}(t)}{\cosh \theta _{1}}\exp \left[ i\left( -\xi _{1}+\Gamma
(t)x^{2}/2\right) \right] ,  \label{eq_5.10} \\
\psi _{2}^{(1)} &=&\frac{2}{\sqrt{g(t)}}\varepsilon _{2}^{(1)}\frac{\beta
_{2}(t)}{\cosh \theta _{1}}\exp \left[ i\left( -\xi _{1}+\Gamma
(t)x^{2}/2\right) \right] ,  \label{eq_5.11}
\end{eqnarray}%
where
\end{subequations}
\begin{eqnarray}
\theta _{1} &=&2\beta _{1}x+4\int \alpha _{1}\beta _{1}dt-2\delta _{1}, \\
\xi _{1} &=&2\alpha _{1}x+2\int (\alpha _{1}^{2}-\beta _{1}^{2})dt-2\chi
_{1},
\end{eqnarray}%
with $\alpha _{1}=\alpha _{10}\exp \left[ \int \Gamma (t)dt\right] $, $\beta
_{1}=\beta _{10}\exp \left[ \int \Gamma (t)dt\right] $, while $\delta _{1}$
and $\chi _{1}$ are arbitrary parameters, and $\varepsilon _{1,2}^{(1)}$ are
coupling coefficients, which are subject to constraint $\left\vert
\varepsilon _{1}^{(1)}\right\vert ^{2}+\left\vert \varepsilon
_{2}^{(1)}\right\vert ^{2}=1$.

\begin{figure}[tbp]
\centerline{\includegraphics[scale=1.07]{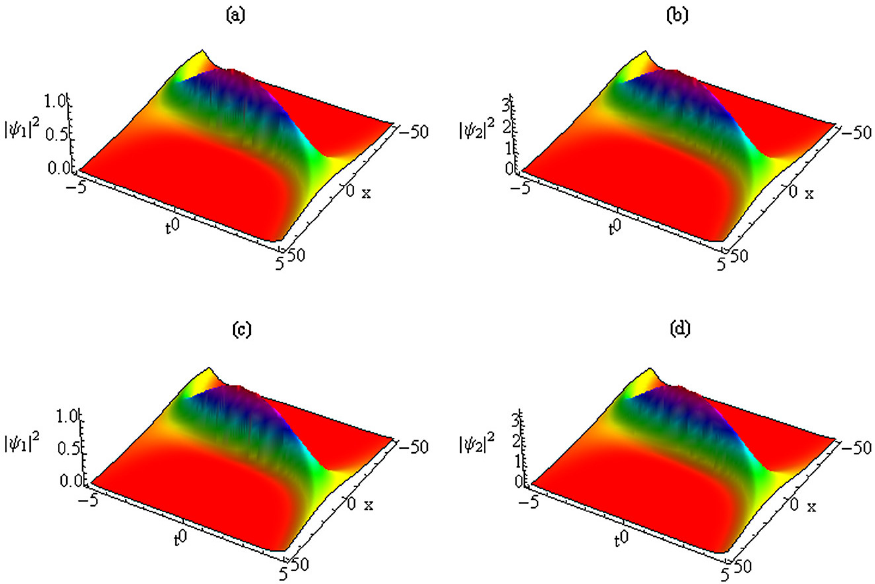}}
\caption{(Color online) (a,b): The analytical soliton solution of Eqs. (%
\protect\ref{eq_5.2}), given by Eqs. (\protect\ref{eq_5.10}) and (\protect
\ref{eq_5.11}), with $g(t)$ and $\protect\lambda ^{2}(t)$ taken as per Eqs. (%
\protect\ref{g(tt)}) and (\protect\ref{lambda^2}). (c,d): The numerically
generated counterpart of the same solution. The results are reproduced from
Ref. \protect\cite{add8}.}
\label{fig022}
\end{figure}

To start the analysis of particular solutions relevant for the physical
realization of the system, one can switch off the time dependence of the HO
trap and adopt the dependence of the nonlinearity coefficient in the form of
\begin{equation}
g(t)=0.5\exp \left( -0.25t\right) ,  \label{g(t)}
\end{equation}%
for which Eq. (\ref{eq_5.8}) renders the HO potential time-independent and
expulsive, with $\lambda ^{2}=1/16>0$. The corresponding density profile of
the analytical solution, given by Eqs. (\ref{eq_5.7}) and (\ref{eq_5.10}), (%
\ref{eq_5.11}), is displayed in Figs. \ref{fig019}(a,b). Its counterpart
produced by numerical simulations of Eq. (\ref{eq_5.2}) is shown in Figs. %
\ref{fig019}(c,d). In these figures one observes perfect agreement between
the analytical and numerical solutions, which also implies stability of the
analytical solution. If the time-modulated interaction strength (\ref{g(t)})
is replaced by%
\begin{equation}
g(t)=0.5\exp \left( -0.5t\right) ,  \label{gg(t)}
\end{equation}%
for which Eq. (\ref{eq_5.8}) yields a stronger expulsive potential, with $%
\lambda ^{2}=1/4$, the condensate quickly spreads out, as shown by the
analytical solution in Figs. \ref{fig020}(a,b), and its numerical
counterpart in Figs. \ref{fig020}(c,d). As seen in Fig. \ref{fig021}, this
trend (including the stability of the analytical solution) continues if the
time dependence (\ref{gg(t)} is replaced by
\begin{equation}
g(t)=0.5\exp \left( -0.9t\right) ,  \label{ggg(t)}
\end{equation}%
for which Eq. (\ref{eq_5.8}) yields $\lambda ^{2}=0.81$.

\begin{figure}[tbp]
\centerline{\includegraphics[scale=1.07]{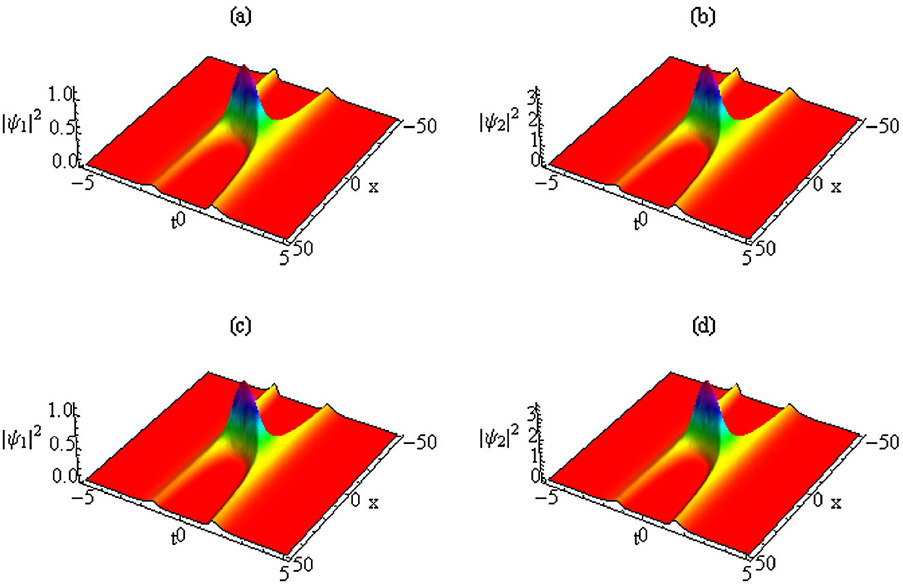}}
\caption{The same as in Fig. \protect\ref{fig022}, but for $g(t)$ and $%
\protect\lambda ^{2}(t)$ taken as per Eqs. (\protect\ref{gg(tt)}) and (%
\protect\ref{lambda^^2}). Note the essential difference of the
spatiotemporal shape of the solution in comparison with that displayed in
Fig. \protect\ref{fig022}. The results are reproduced from Ref. \protect\cite%
{add8}.}
\label{fig023}
\end{figure}

To enhance the stability of the condensates, one can switch on the time
dependence of the HO strength, taking
\begin{equation}
g(t)=0.5\exp \left( -0.125t^{2}\right) ,  \label{g(tt)}
\end{equation}%
for which Eq. (\ref{eq_5.8}) yields the expulsive potential with strength%
\begin{equation}
\lambda ^{2}=1/4+t^{2}/16.  \label{lambda^2}
\end{equation}%
The corresponding analytical solution, generated by Eqs. (\ref{eq_5.10}) and
(\ref{eq_5.11}) is displayed in Figs. \ref{fig022}(a,b). The correctness and
stability of the analytical solution is corroborated by its numerical
counterpart, as shown in Figs. \ref{fig022}(c,d). Further, for the much
steeper modulation of the interaction coefficient, with
\begin{equation}
g(t)=0.5\exp \left( -2.5t^{2}\right) ,  \label{gg(tt)}
\end{equation}%
the integrability condition (\ref{eq_5.8}) yields%
\begin{equation}
\lambda ^{2}=5+25t^{2}.  \label{lambda^^2}
\end{equation}%
The corresponding analytical solution, given by Eqs. (\ref{eq_5.10}) and (%
\ref{eq_5.11}), and its numerical counterpart are displayed in Fig. \ref%
{fig023}. Note that this solution demonstrates non-monotonous evolution in
time: the soliton's wave fields shrink and then expand.

\begin{figure}[tbp]
\centerline{\includegraphics[scale=1.02]{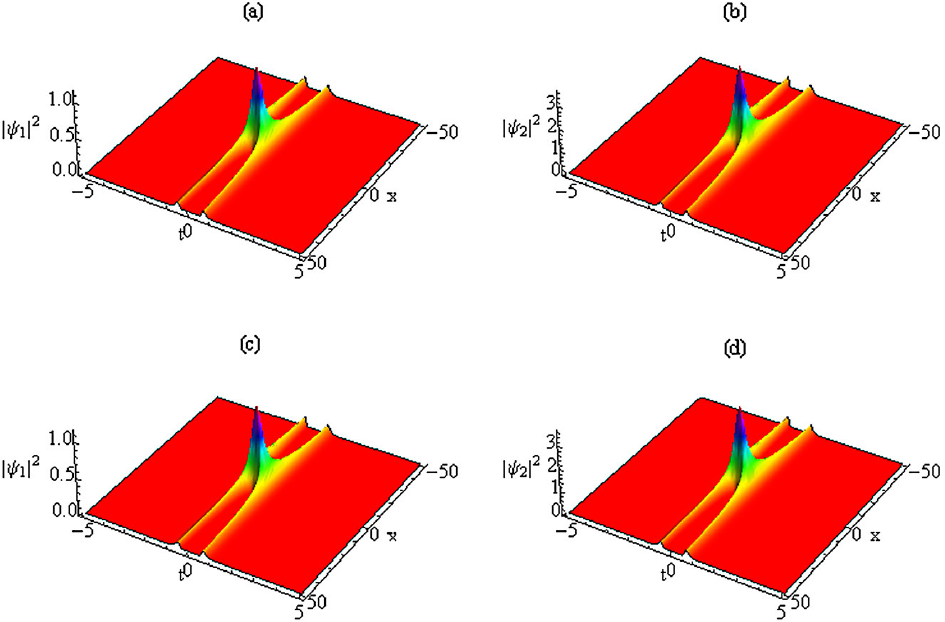}}
\caption{The same as in Fig. \protect\ref{fig023}, but for $g(t)$ and $%
\protect\lambda ^{2}(t)$ given by Eqs. (\protect\ref{ggg(tt)}) and (\protect
\ref{lambda2}). The results are reproduced from Ref. \protect\cite{add8}.}
\label{fig024}
\end{figure}

\begin{figure}[tbp]
\centerline{\includegraphics[scale=1.05]{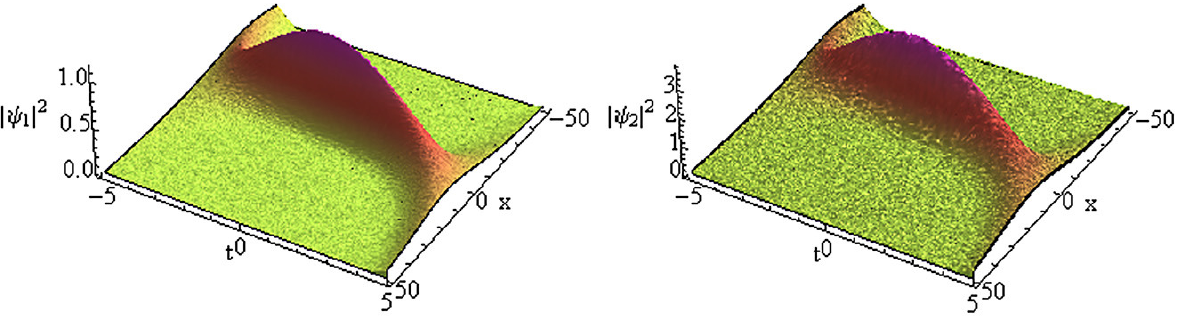}}
\caption{(Color online) The same as in Fig. \protect\ref{fig022}(c,d), but
in the case when strong white random noise, with a standard spectral width,
is added to the simulations. The results are reproduced from Ref.
\protect\cite{add8}. }
\label{fig025}
\end{figure}
A still steeper modulation of the interaction coefficient, with
\begin{equation}
g(t)=0.5\exp \left( -12.5t^{2}\right) ,  \label{ggg(tt)}
\end{equation}%
and%
\begin{equation}
\lambda ^{2}=5+25t^{2},  \label{lambda2}
\end{equation}%
given by integrability relation (\ref{eq_2.8}), leads to the exact solution
displayed in Fig. \ref{fig024}, whose shape remains qualitatively similar to
that in Fig. \ref{fig023}.

\begin{figure}[tbp]
\centerline{\includegraphics[scale=1.05]{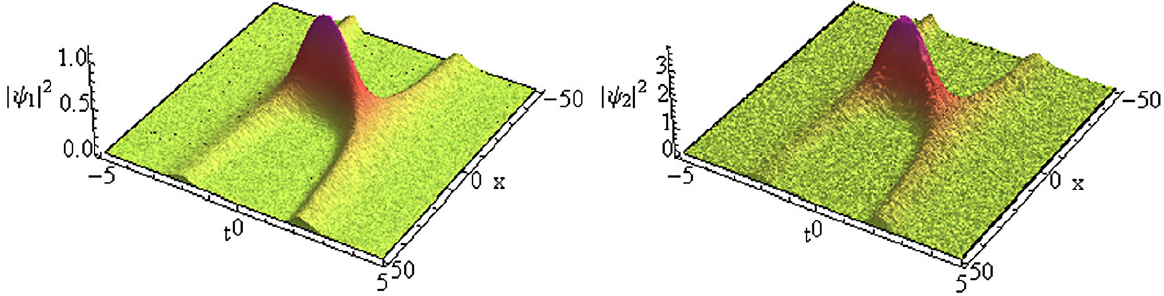}}
\caption{(Color online) The same as in Fig.
\protect\ref{fig023}(c,d), but in the case when strong random
noise is added to the simulations. The results are reproduced from
Fig. \cite{add8}.} \label{fig026}
\end{figure}

Thus, the two-component BEC with attractive interactions in the
time-dependent expulsive trap, adjusted to the integrable system based on
Eqs. (\ref{eq_5.2}) and (\ref{eq_5.8}), is more long-lived, as its lifetime
can be increased by means of the FR management, in comparison with the
condensates in the time-independent expulsive trap, which quickly decays in
the course of the evolution.

\begin{figure}[tbp]
\centerline{\includegraphics[scale=1.05]{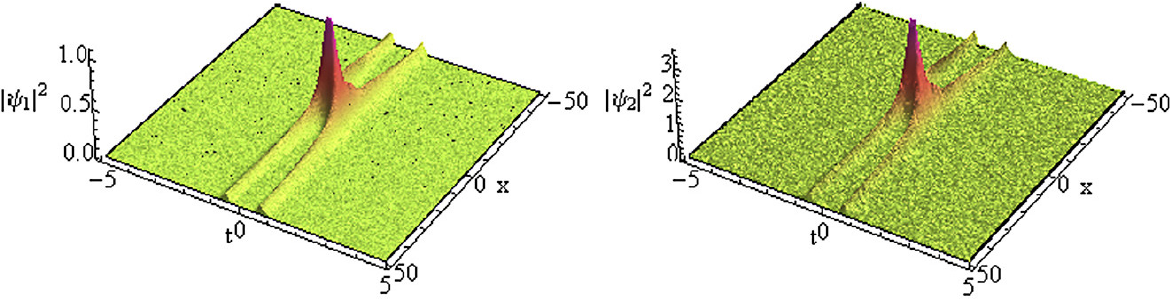}}
\caption{(Color online) The same as in Fig. \protect\ref{fig024}(c,d), but
in the case when strong random noise is added to the simulations. The
results are reproduced from Ref. \protect\cite{add8}.}
\label{fig027}
\end{figure}

To further confirm the stability of the exact soliton solution produced by
Eqs. (\ref{eq_5.2}) and (\ref{eq_5.8}), one can add random noise to the
simulations (while rigorous stability analysis for these solutions in an
experimentally relevant form is not straightforward). The respective
numerically simulated density profiles demonstrate, in Figs. \ref{fig025}-%
\ref{fig027}, that the white noise does not break the stability of the
evolving condensates. Additional numerical results demonstrate that the
stability does not depend on the particular correlation structure of the
noise (white or colored), nor on its spectral width. This conclusion further
corroborates that the two-component condensate in the time-dependent
expulsive HO, governed by the specially devised integrable system, is more
robust in comparison with its counterpart in the time-independent trap. The
above results indicate the possibility of increasing the life span of the
two-component BEC with the time-modulated attractive interactions in the
time-dependent HO potential. These predictions may be realized
experimentally in condensates composed of $^{39}$\textrm{K} \cite{4.15}, $%
^{85}$\textrm{Rb} \cite{4.13} and $^{7}$\textrm{Li} \cite{3.30} atoms.

\subsection{Conclusion of the section}

It is shown that the temporal modulation of the scattering length by means
of FR can be efficiently used to design conditions under which the
mean-field dynamics of two-component condensates in the time-dependent
expulsive HO potential is governed by the integrable model. The solitons
generated by this integrable system stay stable for a reasonably large
interval of time, compared to the condensate in the time-independent
expulsive HO potential. The analytical results and their stability are
corroborated by comparison with numerical simulations, including the case
when strong random noise is added to the simulations.

\section{Soliton motion in a binary Bose-Einstein condensate}

This section deals with the analytical study of dynamics of matter-wave
solitons governed by an integrable system of coupled GP equations of the
Manakov's type \cite{4.14}, modeling a two-component BEC. By means of the
Darboux transform (DT), analytical solutions of these equations can be
obtained for a soliton set on top of a plane-wave background. These
solutions with and without the XPM interaction between the two components
are considered. In the presence of XPM, the solutions exhibit properties
different from those in the single-component GP, such as restriction of MI
and soliton splitting.

The results presented in this section are essentially based on Ref. \cite%
{5.1}.

\subsection{The model and analysis}

\begin{figure}[tbp]
\centerline{\includegraphics[scale=1.15]{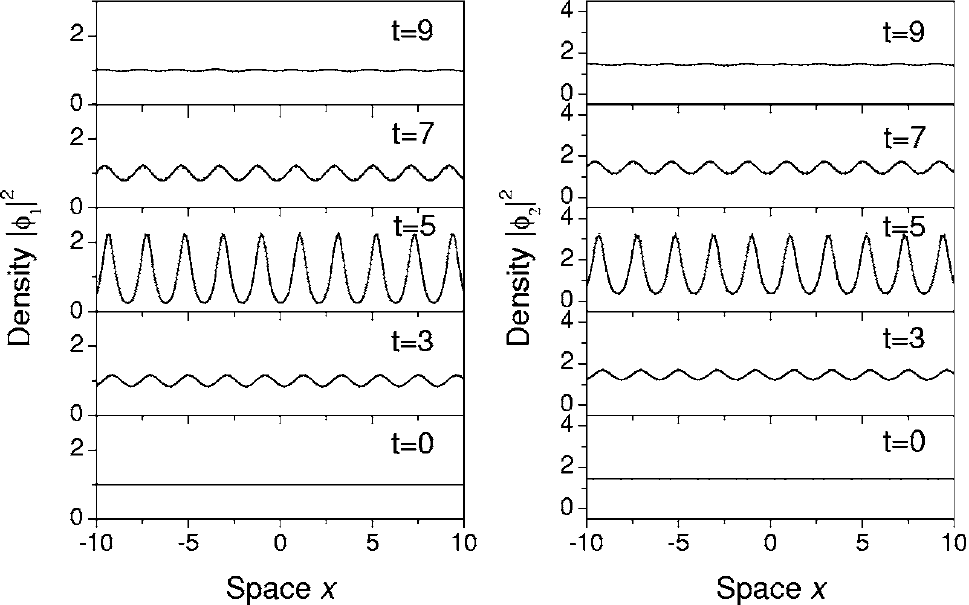}}
\caption{(Color online) A set of snapshots showing exact solution (\protect
\ref{eq_6.8}) by dotted curves, and numerical results, obtained under nearly
integrable condition (\protect\ref{eq_6.15}), by solid curves (in fact, the
dotted and solid plots completely overlap). Parameters are $\protect\eta %
=0.8 $, $k=-0.1$,$g=1$, $A_{1}=1$, $A_{2}=1.2$, $\protect\theta _{10}=6$,
the latter one representing amplitude (\protect\ref{eq_6.13}) of the initial
perturbation which triggers the onset of MI, with $\protect\epsilon %
=2.4788\times 10^{-3}$ and $\protect\varphi _{10}=0$. The results are
reproduced from Ref. \protect\cite{5.1}.}
\label{fig028}
\end{figure}

The subject on this section is the system of coupled 1D GP equations for
two-component self-attractive BEC with negative scattering lengths. In the
scaled form, these equations are%
\begin{equation}
i\frac{\partial \phi _{j}}{\partial t}=-\frac{1}{2}\frac{\partial ^{2}\phi
_{j}}{\partial x^{2}}-g\left( \frac{a_{ij}}{a_{12}}\left\vert \phi
_{j}\right\vert ^{2}+\left\vert \phi _{3-j}\right\vert ^{2}\right) \phi _{j},%
\text{\ }j=1,2,  \label{eq_6.1}
\end{equation}%
where $a_{jj}$ and $a_{12}$ are, respectively, the negative scattering
lengths of intra- and inter-component atomic collisions, coordinate $x$ is
measured in units of $x_{0}\sim 1~\mathrm{\mu }$m, $t$ in units of $%
mx_{0}^{2}/\hbar $, with atomic mass $m$, $\phi _{j}$ in units of $\sqrt{%
n_{0}}$ ($n_{0}$ is the maximum density of the initial distribution of the
condensate), and the interaction constant is defined as $g=4\pi
n_{0}x_{0}^{2}\left\vert a_{12}\right\vert $.

As is well known, coupled equations (\ref{eq_6.1}) are integrable in the
Manakov's case, with $a_{11}=a_{22}=a_{12}$ \cite{4.14}. In this case, by
means of DT, one can generate an exact two-component solution which accounts
for a soliton interacting with a plane wave:%
\begin{equation}
\phi _{j}=A_{j}\exp \left( i\varphi \right) \left( 1+\frac{4i\eta G_{1}}{F}%
\right) +\frac{2\eta \overline{C}_{j}G_{2}}{\sqrt{g}}\exp \left( \frac{i}{2}%
\varphi \right) ,  \label{eq_6.2}
\end{equation}%
where $j=1,2$, and
\begin{subequations}
\begin{eqnarray}
F &=&D\exp \theta _{1}+A^{2}D\exp \left( -\theta _{1}\right) -2iA^{2}\left(
L-\overline{L}\right) \sin \varphi _{1}+C\exp \left( -\theta _{2}\right) ,
\label{eq_6.3} \\
G_{1} &=&L\exp \theta _{1}-A^{2}\overline{L}\exp \left( -\theta _{1}\right)
-\left\vert L\right\vert ^{2}\exp \left( -i\varphi _{1}\right) +A^{2}\exp
\left( i\varphi _{1}\right) ,  \label{eq_6.4} \\
G_{2} &=&L\exp \left[ -\frac{i}{2}\left( \varphi _{1}+\varphi _{2}\right) +%
\frac{1}{2}\left( \theta _{1}-\theta _{2}\right) \right] +A^{2}\exp \left[
\frac{i}{2}\left( \varphi _{1}-\varphi _{2}\right) -\frac{1}{2}\left( \theta
_{1}+\theta _{2}\right) \right] ;  \label{eq_6.5}
\end{eqnarray}%
\begin{eqnarray}
\theta _{1} &\equiv &M_{I}x+\frac{1}{2}\left[ \left( \xi -k\right)
M_{I}+\eta M_{R}\right] t-\theta _{10},  \label{eq_6.6} \\
\theta _{2} &\equiv &\eta x+\eta \xi t+\theta _{10},  \label{eq_6.7}
\end{eqnarray}%
\end{subequations}
\begin{eqnarray}
\varphi _{1} &\equiv &M_{R}x+\frac{1}{2}\left[ \left( \xi -k\right)
M_{R}-\eta M_{I}\right] t+\varphi _{10}, \\
\varphi _{2} &\equiv &\xi x+\frac{1}{2}\left( \xi ^{2}-\eta ^{2}\right)
t-\varphi _{10}, \\
\varphi &\equiv &kx+\left[ g\left( A_{1}^{2}+A_{2}^{2}\right) -\frac{1}{2}%
k^{2}\right] t;
\end{eqnarray}%
\begin{eqnarray}
L &\equiv &M_{R}+\xi +k+i\left( M_{I}+\eta \right) ,M\equiv \sqrt{\left(
k+\xi +i\eta \right) ^{2}+A^{2}}=\text{\ }M_{R}+iM_{I}, \\
D &\equiv &\left\vert L\right\vert ^{2}+A^{2},A\equiv \sqrt{4g\left(
A_{1}^{2}+A_{2}^{2}\right) },C=\left\vert C_{1}\right\vert ^{2}+\left\vert
C_{2}\right\vert ^{2},
\end{eqnarray}%
$k$, $\theta _{10}$, $\varphi _{10}$, $A_{1,2}$ are arbitrary real
constants, and $C_{1,2}$ are arbitrary complex constants subject to
constraint
\begin{equation}
A_{1}C_{1}+A_{2}C_{2}=0.  \label{ACAC}
\end{equation}%
In the case of zero background amplitude, $A_{1}=A_{2}=0$, solution (\ref%
{eq_6.2}) amounts to $\phi _{j}=\left[ \eta \epsilon _{j}/\left( \sqrt{g}%
\cosh \theta \right) \right] \exp \left( -i\varphi _{2}\right) $, where $%
\theta =\eta \left( x+\xi \right) -\theta _{0}$, $\theta _{0}$ is an
arbitrary real constant, and $\epsilon _{1,2}$ are arbitrary complex
constants obeying relation $\left\vert \epsilon _{1}\right\vert
^{2}+\left\vert \epsilon _{2}\right\vert ^{2}=1$. The latter solution is a
stable two-component soliton with velocity $V_{\text{sol}}=-\xi $, width $%
\eta ^{-1}$, and amplitudes $A_{s1,2}=\left\vert \eta \epsilon
_{1,2}\right\vert /\sqrt{g}$ which satisfy relation $A_{s1}^{2}+A_{s2}^{2}=%
\eta ^{2}/g$. The total norm of the two-component soliton (proportional to
the number of atoms in the binary condensate) is $Q\equiv
Q_{1}+Q_{2}=\int_{-\infty }^{+\infty }\left( \left\vert \phi _{1}\right\vert
^{2}+\left\vert \phi _{2}\right\vert ^{2}\right) dx=2\left\vert \eta
\right\vert /g,$ with $Q_{j}=2\left\vert \eta \right\vert \left\vert
\epsilon _{j}\right\vert ^{2}/g$. Further, its momentum and Hamiltonian of
the system are
\begin{equation}
M\equiv M_{1}+M_{2}=-\frac{i}{2}\int_{-\infty }^{+\infty }\left[ \left(
\overline{\phi }_{1}\phi _{1,x}-\phi _{1}\overline{\phi }_{1,x}\right)
+\left( \overline{\phi }_{2}\phi _{2,x}-\phi _{1}\overline{\phi }%
_{2,x}\right) \right] dx=V_{\text{sol}}Q
\end{equation}%
and
\begin{equation}
H\equiv \frac{1}{2}\int_{-\infty }^{+\infty }\left[ \left( \left\vert \phi
_{1,x}\right\vert ^{2}+\left\vert \phi _{2,x}\right\vert ^{2}\right)
-g\left( \left\vert \phi _{1}\right\vert ^{2}+\left\vert \phi
_{2}\right\vert ^{2}\right) ^{2}\right] dx=\frac{M^{2}}{2Q}-\frac{g^{2}}{24}%
Q^{3}.
\end{equation}

\begin{figure}[tbp]
\centerline{\includegraphics[scale=1.17]{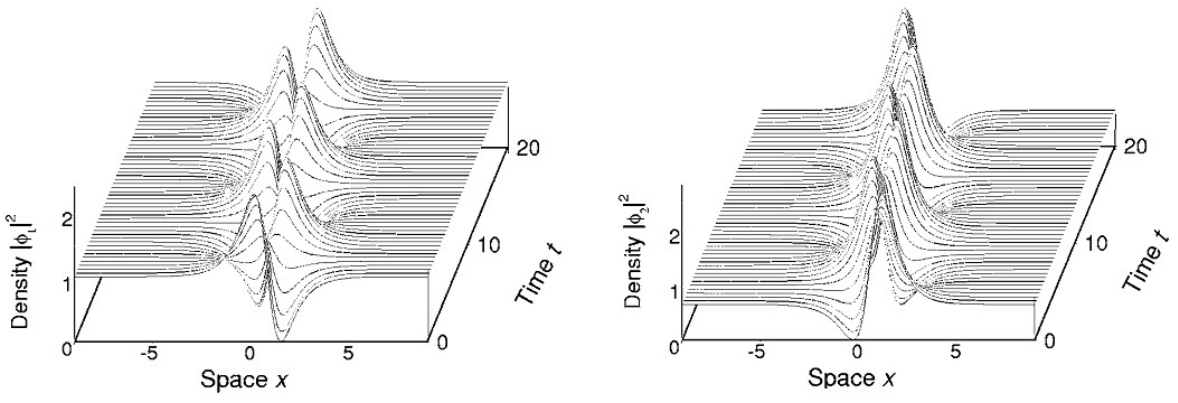}}
\caption{Spatiotemporal distribution of densities of the two components in
solution (\protect\ref{eq_6.16}), with phase difference $\protect\pi $
between the components. Parameters are $k=-0.1$, $g=1$, $A_{1}=1$, $%
A_{2}=0.8 $, $\protect\theta _{10}=-2$, and $\protect\varphi _{10}=0$. The
results are reproduced from Ref. }
\label{fig029}
\end{figure}

On the other hand, when the soliton's amplitudes vanish, $A_{1s,2s}=0$
(e.g., $\eta =0$), solution (\ref{eq_6.2}) reduces to the plane wave (alias
CW), $\phi _{1,2}=A_{1}\exp \left( i\varphi \right) $, with amplitudes $%
A_{1,2}$, wave number $k$, and frequency $\Omega =g\left(
A_{1}^{2}+A_{2}^{2}\right) -k^{2}/2$. Thus, in the general case exact
solution (\ref{eq_6.2}) represents a two-component soliton embedded in the
CW background. Condition (\ref{ACAC}) determines the XPM interaction between
the two components of the plane-wave background. If $C_{1}=C_{2}=0$, the
second term in solution (\ref{eq_6.2}) vanishes, each soliton component
being embedded in its own background, see a more detailed consideration of
this special case below. If $C_{1}C_{2}\neq 0$, the second term in solution (%
\ref{eq_6.2}) is different from zero, hence condition (\ref{ACAC}) implies
that coefficients $C_{1}$ and $C_{2}$ depend on the CW amplitudes $A_{2}$
and $A_{1}$.

\begin{figure}[tbp]
\centerline{\includegraphics[scale=.8]{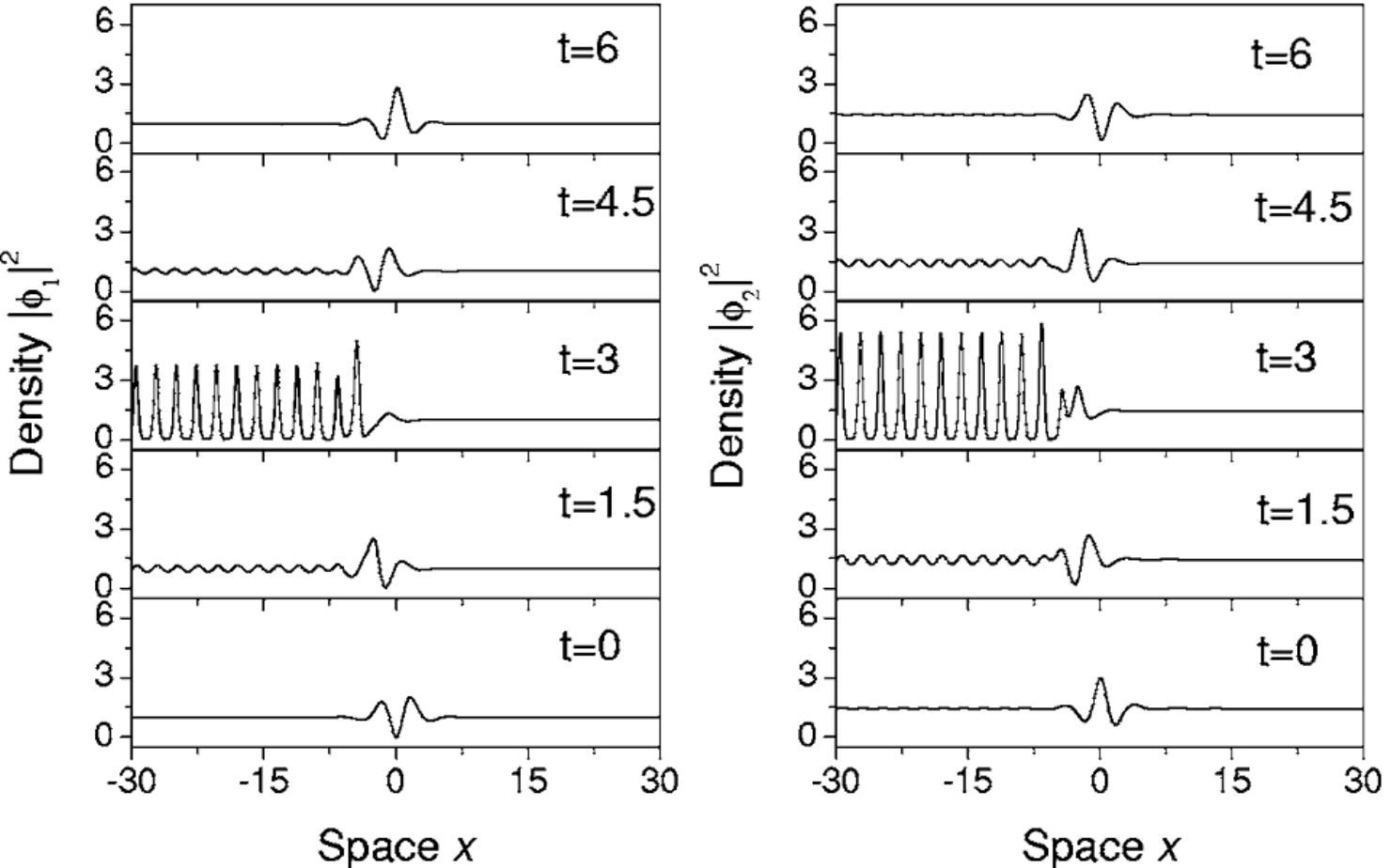}}
\caption{A set of snapshots representing exact solution (\protect\ref%
{eq_6.18}) for $\protect\eta =-1.5$, $k=-0.1$, $g=1$, $A_{1}=1$, $A_{2}=1.2$%
, $\protect\theta _{10}=-6$, $\protect\varphi _{10}=0$, and $\protect\theta %
_{20}=-5$, $\protect\varphi _{20}=0$. The results are reproduced from Ref.
\protect\cite{5.1}.}
\label{fig030}
\end{figure}

To understand the behavior of solution (\ref{eq_6.2}), it is relevant to
start with the above-mentioned case of $C_{1}=C_{2}=0$, when solution (\ref%
{eq_6.2}) simplifies to
\begin{subequations}
\begin{eqnarray}
\phi _{j} &=&A_{j}\exp \left( i\varphi \right) \left( 1-2\eta W\right) ,
\label{eq_6.8} \\
W &\equiv &\frac{1}{A}\left( \frac{a\cosh \theta _{1}+\sin \varphi _{1}}{%
\cosh \theta _{1}+a\sin \varphi _{1}}-i\frac{b\sinh \theta _{1}+c\cos
\varphi _{1}}{\cosh \theta _{1}+a\sin \varphi _{1}}\right) ,  \label{eq_6.9}
\end{eqnarray}%
with $a=-iA\left( L-\overline{L}\right) /D,$ $b=A\left( L+\overline{L}%
\right) /D$, and $c=\left( A^{2}-\left\vert L\right\vert ^{2}\right) /D$.
Note that expressions (\ref{eq_6.8}) and (\ref{eq_6.9}) do not include $%
\theta _{2}$ and $\varphi _{2}$, and the two components share a common
shape. This solution is similar to that for the single-component BEC. As $%
M_{I}\neq 0$, from Eq. (\ref{eq_6.8}) one can extract the total norm of the
soliton component of the solution, which is defined as the norm above the CW
level:
\end{subequations}
\begin{subequations}
\begin{equation}
\int_{-\infty }^{+\infty }\left[ \left( \left\vert \phi _{1}\right\vert
^{2}-A_{1}\right) ^{2}+\left( \left\vert \phi _{2}\right\vert
^{2}-A_{2}^{2}\right) \right] dx=\frac{\eta \left( b^{2}+c^{2}\right) }{%
g\left\vert M_{I}\right\vert }I,  \label{eq_6.10}
\end{equation}%
\begin{equation}
I\equiv \int_{-\infty }^{+\infty }\frac{\eta +A\left( \cosh x\right) \sin
\left( Bx+\Delta \right) }{\cosh x+a\sin \left( Bx+\Delta \right) }dx,
\label{eq_6.11}
\end{equation}%
where $\Delta =-(1/2)\eta M_{I}\left( 1+B^{2}\right) t+B\theta _{10}+\varphi
_{10}$ and $B=M_{R}/M_{I}$. In deriving the above expression, relations $%
\left\vert \phi _{1}(\pm \infty ,t)\right\vert ^{2}=A_{1}^{2}$ and $%
\left\vert \phi _{2}(\pm \infty ,t)\right\vert ^{2}=A_{2}^{2}$ for
asymptotic values of the fields are used. It can be verified that integral (%
\ref{eq_6.11}) does not depend on $\Delta $, hence the soliton's total norm
is a conserved quantity.

\begin{figure}[tbp]
\centerline{\includegraphics[scale=1.08]{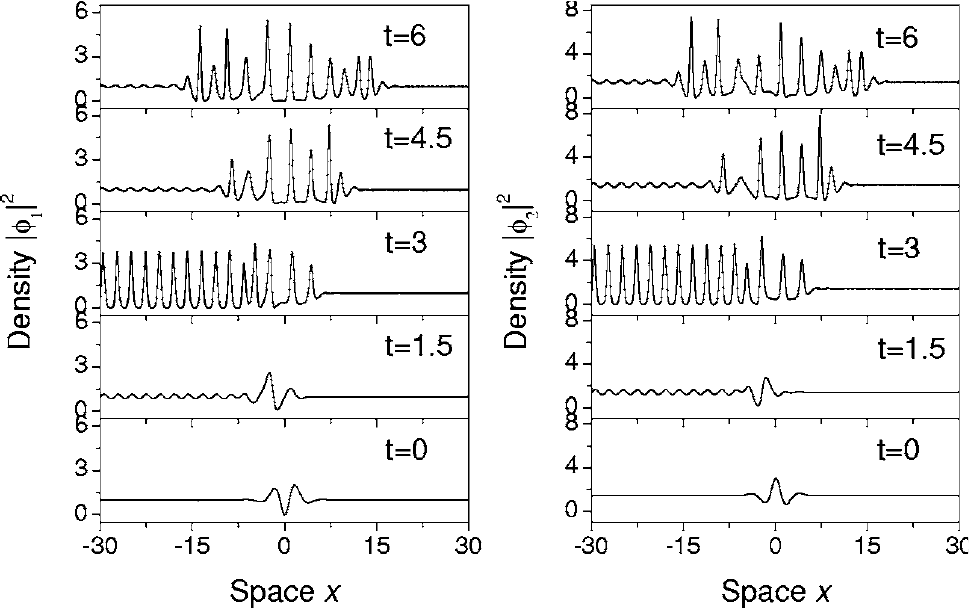}}
\caption{The same as in Fig. \protect\ref{fig030}, but obtained from a
numerical solution of Eqs. (\protect\ref{eq_6.1}) with the nonlinearity
constants taken as per Eq. (\protect\ref{eq_6.15}). Other parameters are
identical to those in Fig. \protect\ref{fig030}. The results are reproduced
from Ref. \protect\cite{5.1}.}
\label{fig031}
\end{figure}

When $M_{I}=0$, the soliton's velocity vanishes, i.e., the CW background
completely traps the soliton \cite{5.1}. Setting, in particular, $\xi =-k$
and $A^{2}>\eta ^{2}$ yields $M_{I}=0$, and $W$ in solution (\ref{eq_6.8})
becomes
\end{subequations}
\begin{equation}
W=\frac{1}{A}\frac{\eta \cosh \theta _{1}+A\sin \varphi _{1}-iM_{R}\sinh
\theta _{1}}{A\cosh \theta _{1}+\eta \sin \varphi _{1}},  \label{eq_6.12}
\end{equation}%
with $\theta _{1}\equiv (1/2)\eta M_{R}t-\theta _{10}$ and $\varphi
_{1}\equiv M_{R}\left( x-kt\right) +\varphi _{10}$, where $M_{R}=\sqrt{%
A^{2}-\eta ^{2}},$ and $\theta _{10}$ and $\varphi _{10}$ are arbitrary real
constants. Expression (\ref{eq_6.12}) is periodic in $x$, with period $%
\Lambda =2\pi /M_{R}$, and aperiodic in $t$. To better understand the MI
development provided by this solution, one may assume that%
\begin{equation}
\epsilon =\exp \left( -\theta _{10}\right)  \label{eq_6.13}
\end{equation}%
is small (it then plays the role of a small amplitude of the initial
perturbation that triggers the onset of the MI), and linearize the initial
form of solution (\ref{eq_6.8}) at $t=0$ with respect to $\epsilon $, taking
Eq. (\ref{eq_6.12}) into account. This yields%
\begin{equation}
\phi _{j}(x,0)\approx A_{j}\exp \left( i\varphi \right) \left[ \rho
-\epsilon \chi \sin \left( M_{R}x+\varphi _{10}\right) \right] ,
\label{eq_6.14}
\end{equation}%
where $\rho \equiv \left( A^{2}-2\eta ^{2}-2i\eta M_{R}\right) /A^{2}$ with $%
\left\vert \rho \right\vert ^{2}=1$, $\chi \equiv 4\eta M_{R}\left(
M_{R}-i\eta \right) /A^{3}$. Direct numerical simulations of the underlying
equations (\ref{eq_6.1}) demonstrate that the evolution of initial
configuration (\ref{eq_6.14}), which is a plane wave with a small
modulational perturbation added to it, indeed closely follows the exact
solution provided by Eqs. (\ref{eq_6.8}) and (\ref{eq_6.12}) \cite{5.1}.

\subsection{Results}

The above results were obtained under the Manakov integrability conditions $%
a_{11}=a_{22}=a_{12}$, which may not be exactly satisfied in reality.
However, the actual difference of the scattering lengths in a BEC mixture of
two different hyperfine states of the same atomic species is very small,
therefore the Manakov's system may be used as a good approximation. For
instance, taking%
\begin{equation}
a_{11}=-1.03,\text{ }a_{12}=-1,\text{ }a_{22}=-0.97,  \label{eq_6.15}
\end{equation}%
numerical solution of Eqs. (\ref{eq_6.1}) yields the picture of the MI
development shown in Fig. \ref{fig028}. Comparing it with the exact solution
obtained for $a_{11}=a_{22}=a_{12}$ demonstrates that the two solutions are
virtually indistinguishable.

\begin{figure}[tbp]
\centerline{\includegraphics[scale=0.8]{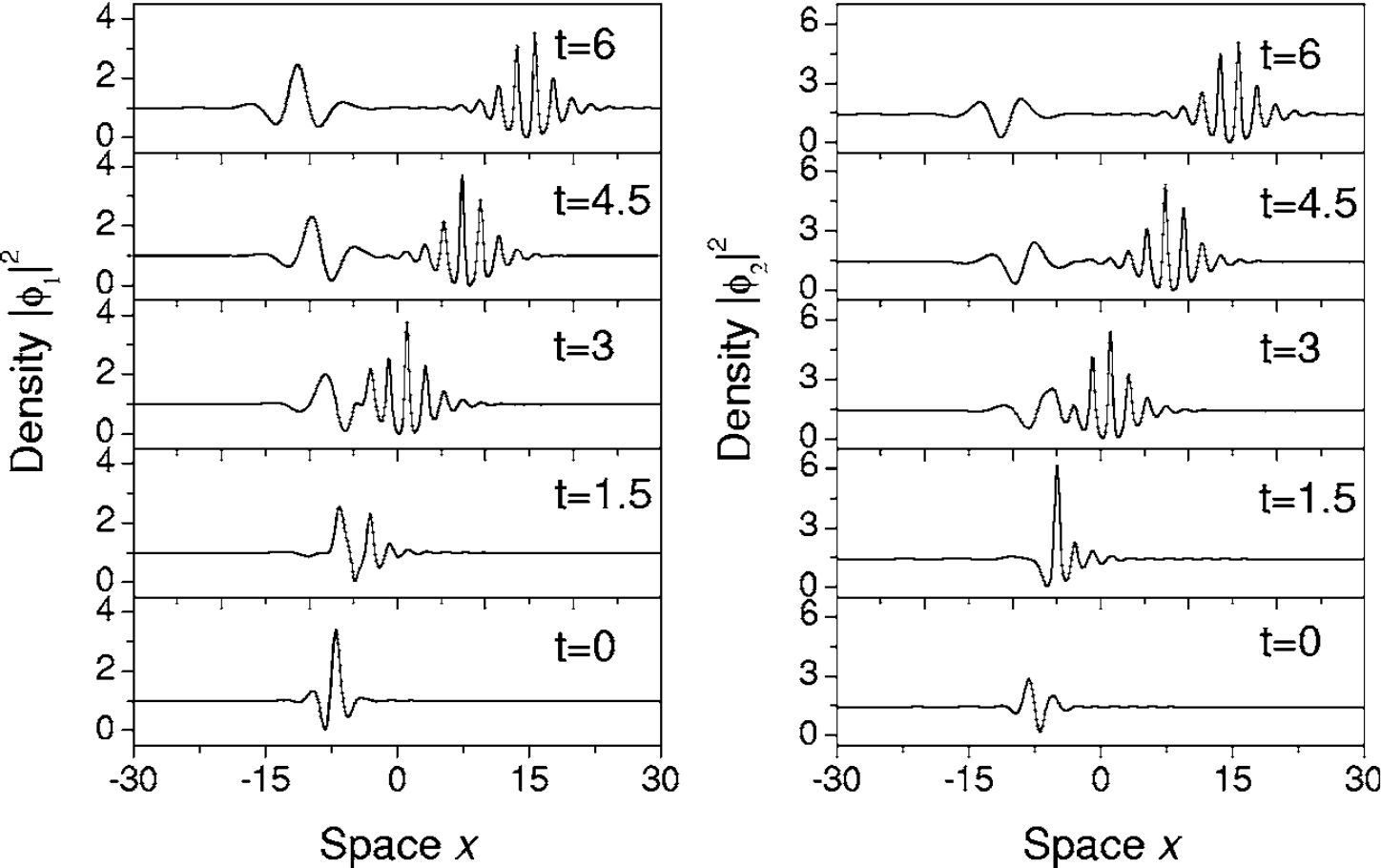}}
\caption{A set of snapshots of the general-form analytical solution (\protect
\ref{eq_6.2}). Parameters are $\protect\eta =1.5$, $\protect\xi =-0.4$, $%
k=-0.5$, $g=1$, $A_{1}=1$, $A_{2}=1.2$, $10=5$, $10=0$, $C_{1}=1$, and $%
C_{2} $ is determined by constraint (\protect\ref{ACAC}). The results are
reproduced from Ref. \protect\cite{5.1}.}
\label{fig032}
\end{figure}

The consideration of solution (\ref{eq_6.2}) in a more general situation,
with nonzero coefficients $C_{1,2}$ and $A_{1,2}$, subject to condition (\ref%
{ACAC}), make it possible to explicitly consider effects of the XPM
interaction between the plane waves in the two-component BEC. In this case,
to satisfy condition (\ref{ACAC}), one sets $C_{1}=4\sqrt{g}A_{2}C$ and $%
C_{2}=-4\sqrt{g}A_{1}C$, where $C$ is an arbitrary complex constant.
Further, one fixes $\xi =-k$ in solution (\ref{eq_6.2}) to analyze two
following representative situations in detail.

(i) Setting $A^{2}=\eta ^{2}$, i.e., $4\left( A_{1}^{2}+A_{2}^{2}\right)
=A_{s1}^{2}+A_{s2}^{2}$, solution (\ref{eq_6.2}) can be written as
\begin{equation}
\phi _{j}=-\exp \left( i\varphi \right) \left[ A_{j}\tanh \left( \theta
_{2}/2\right) +(-1)^{j}\sqrt{2}A_{3-j}\cosh ^{-1}\left( \theta _{2}/2\right)
\exp \left( i\varphi _{3}\right) \right] ,  \label{eq_6.16}
\end{equation}%
where $\theta _{2}\equiv \eta \left( x-kt\right) +\theta _{10},$ $\varphi
\equiv kx+\left( \eta ^{2}/4-k^{2}/2\right) t,$ $\varphi _{3}\equiv \eta
^{2}t/8+\varphi _{10}$, and $\theta _{10}$ and $\varphi _{10}$ are arbitrary
real constants. The density distribution corresponding to this solution is $%
\left\vert \phi _{1}\right\vert ^{2}+\left\vert \phi _{2}\right\vert ^{2}=$ $%
\left( A_{2}^{2}+A_{1}^{2}\right) \left[ 1+\mathrm{sech}^{2}\left( \theta
_{2}/2\right) \right] $, see Fig. \ref{fig029}. Solution (\ref{eq_6.16}) may
be regarded as a superposition of bright and dark solitons, produced by the
action of XPM. In particular, the solution with $A_{1}=0$ or $A_{2}=0$ is a
complex consisting of bright and dark solitons in the first and second
species, or vice versa. From the imposed condition $A^{2}=$ $\eta ^{2}$, it
follows that $\left\vert \eta \right\vert =\sqrt{4g\left(
A_{1}^{2}+A_{2}^{2}\right) }$, showing that the width of the soliton is
controlled by the amplitude of the CW background. In Fig. \ref{fig029} one
observes a shift of the soliton's peak due to the action of XPM.

(ii) In the case when $A^{2}>\eta ^{2}$, i.e.,
\begin{equation}
4\left( A_{1}^{2}+A_{2}^{2}\right) >A_{s1}^{2}+A_{s2}^{2},  \label{eq_6.17}
\end{equation}%
solution (\ref{eq_6.2})\ can be written as
\begin{equation}
\phi _{j}=A_{j}\exp \left( i\varphi \right) \left( 1-2\eta W_{1}\right)
-\left( -1\right) ^{j}2\eta A_{3-j}W_{2},  \label{eq_6.18}
\end{equation}%
\begin{subequations}
\begin{eqnarray}
W_{1} &=&\frac{1}{A}\frac{\eta \cosh \theta _{1}+A\sin \varphi
_{1}-iM_{R}\sinh \theta _{1}}{A\cosh \theta _{1}+\eta \sin \varphi
_{1}+A\exp \left( -\theta _{2}\right) },  \label{eq_6.19} \\
W_{2} &=&\frac{1}{A}\frac{\left( M_{R}+i\eta \right) \exp \left[ \left(
-i\varphi _{1}+\theta _{1}\right) /2\right] +A\exp \left[ \left( i\varphi
_{1}-\theta _{1}\right) \right] }{A\cosh \theta _{1}+\eta \sin \varphi
_{1}+A\exp \left( -\theta _{2}\right) }  \notag \\
&&\times \exp \left[ \frac{-\theta _{2}+i\left( \varphi -\varphi _{2}\right)
}{2}\right] ,  \label{eq_6.20}
\end{eqnarray}%
with $\theta _{1}\equiv \left( \eta /2\right) M_{R}t-\theta _{10}$, $\theta
_{2}\equiv $ $x-kt+\theta _{20}$, $\varphi _{1}\equiv $ $M_{R}\left(
x-kt\right) $ $+$ $\varphi _{10}$, $\varphi _{2}\equiv -kx+\frac{1}{2}\left(
k^{2}-\eta ^{2}\right) t-\varphi _{20}$, and $M_{R}=\sqrt{A^{2}-\eta ^{2}}$,
while $\theta _{10}$, $\theta _{20}$, $\varphi _{10}$, and $\varphi _{20}$
are arbitrary real constants. From expressions (\ref{eq_6.19}) and (\ref%
{eq_6.20}) one can see that $W_{2}\rightarrow 0$ at $\theta _{2}\rightarrow
\pm \infty $, and $W_{1}\rightarrow W$ at $\theta _{2}\rightarrow +\infty $,
$W_{1}\rightarrow 0$ at $\theta _{2}\rightarrow -\infty $. Thus, taking into
regard the form of solution (\ref{eq_6.8}) and expression (\ref{eq_6.12}),
one concludes that solution (\ref{eq_6.18}) describes partial MI, because
the growth of the instability is restrained in the limit of $\theta
_{2}\rightarrow -\infty $, as illustrated by Fig. \ref{fig030}. However, if
the nonlinearity constants $a_{11}$, $a_{12}$, and $a_{22}$ slightly deviate
from the integrable case $a_{11}=a_{12}=a_{22}$, a numerically found
counterpart of solution (\ref{eq_6.18}) is conspicuously different from it,
see Fig. \ref{fig031}.

Finally, one can consider solution (\ref{eq_6.2}) in the general case. From
expressions (\ref{eq_6.6}) and (\ref{eq_6.7}) one sees that the solution
contains terms with different velocities, $V_{1}=-(1/2)\left[ \left( \xi
-k\right) +\eta M_{R}/M_{I}\right] $ and $V_{2}=-\xi $, which leads to
splitting of the soliton part of the solution on top of the CW background in
two wave packets, high- and low-frequency ones, as shown in Fig. \ref{fig032}%
. Further, in Fig. \ref{fig033} this exact analytical solution, obtained in
the integrable case with $a_{11}=a_{12}=a_{22}=1$, is compared to a
numerically found one, produced by the generic nonintegrable version of Eqs.
(\ref{eq_6.1}), with the normalized scattering lengths chosen as in Eq. (\ref%
{eq_6.15}). Figure \ref{fig033} demonstrates that the soliton part of the
numerically found solution, built on top of the CW background, also splits
in two packets, stable high-frequency and unstable low-frequency ones.

\begin{figure}[tbp]
\centerline{\includegraphics[scale=0.8]{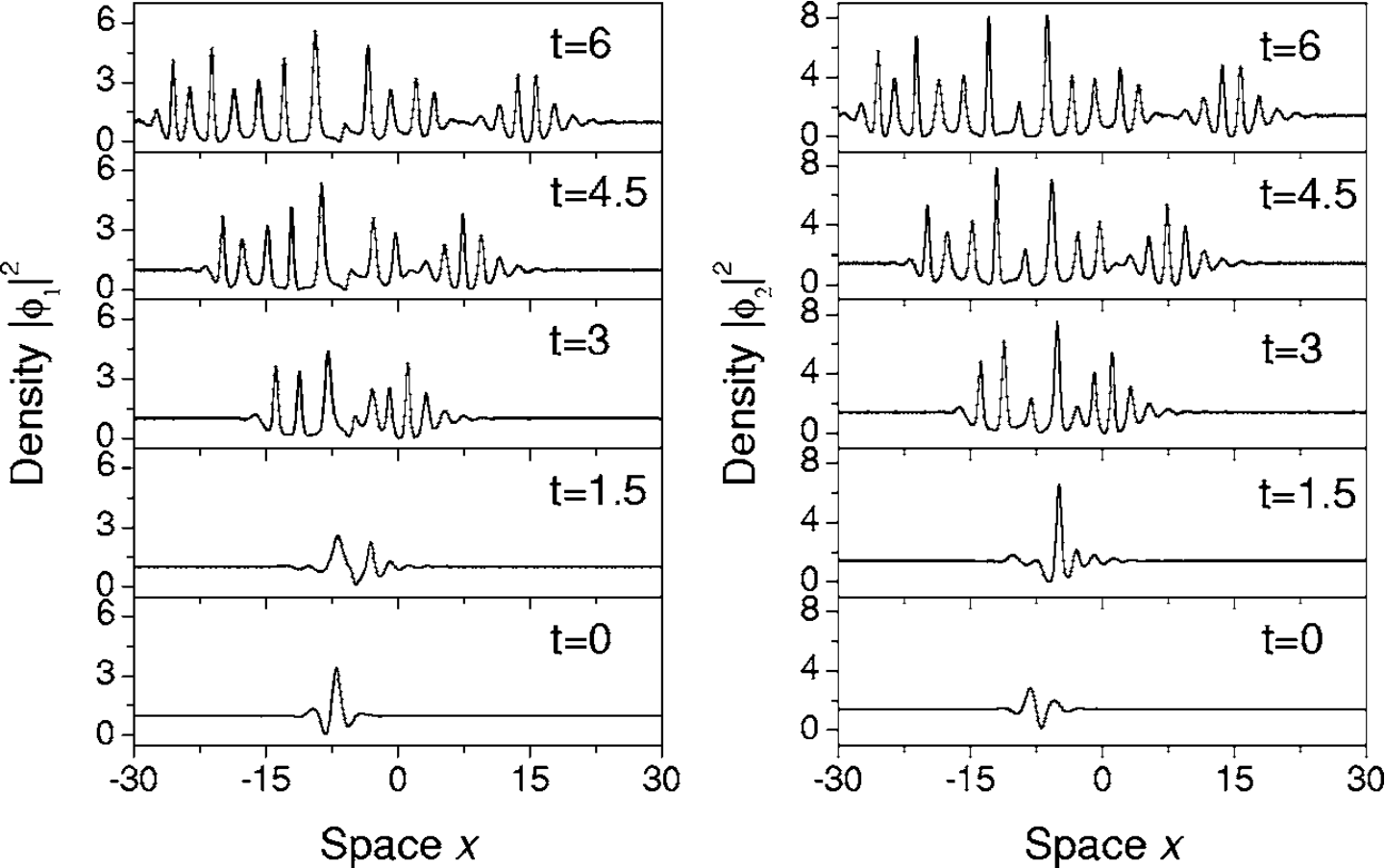}}
\caption{The same as in Fig. \protect\ref{fig032}, but for a numerically
found solution of Eqs. (\protect\ref{eq_6.1}) with $a_{11}=-1.03$, $%
a_{12}=-1 $, $a_{22}=-0.97$. Other parameters are identical to those in Fig.
\protect\ref{fig032}. The results are reproduced from Ref. \protect\cite{5.1}%
.}
\label{fig033}
\end{figure}

\subsection{Conclusion of the section}

Exact solutions are presented for coupled GP equations describing binary
BEC, in the form of a soliton placed on top of a CW (plane-wave) background.
It is shown that, when the intensity of the CW background exceeds a quarter
of the soliton's peak intensity, which is expressed by condition (\ref%
{eq_6.17}), the exact solution, produced by the Manakov system, describes
development of MI, which can be used in physical applications to generate a
soliton train. This wave complex can be generated, as a robust one, even if
parameters of the coupled equations do not exactly satisfy the Manakov's
integrability condition. The XPM interaction in the CW background of the
two-components system has been considered too. It was shown that the
interaction between two CW amplitudes gives rise to a phase difference
between the components, and, in the most general case, it causes splitting
of the soliton and formation of a complex pattern. In particular, under
condition (\ref{eq_6.17}) as mentioned above, the XPM interaction between
the two plane waves helps to effectively restrain the MI.

Note that the coupled underlying equations (\ref{eq_6.1}) are also
integrable in the case of $g<0$ and $a_{11}=a_{22}=a_{12}$ (a binary
self-repulsive condensate), and also, with either sign of $g$, for $%
a_{11}=a_{22}=-a_{12}$ \cite{5.2}, which corresponds to a mixture of two
self-repulsive condensates that attract each other, or two self-attractive
ones which repel each other. In the former case, when single-component
bright solitons are impossible due to the intraspecies repulsion, the
so-called symbiotic solitons exist and are stable, being supported by the
inter-species attraction \cite{5.3,symbio2}. However, direct attempts to
construct bound states of plane waves and solitons by means of DT lead to
singular solutions in these cases \cite{5.1}.

\section{Dynamics of matter-wave solitons in spinor BEC}

This section addresses the dynamics of matter-wave solitons in the general
nonintegrable model of a three-component spinor BEC, based on a system of
three nonlinearly coupled GP equation. First, one-, two-, and
three-component solitons of the polar and ferromagnetic (FM) types are
produced. Next, BdG equations for small perturbations are used to study the
stability of the solitons by means of direct simulations and, in a part,
analytically. Global stability of the solitons is considered by means of
comparison of energy for different states. As the main result, ground-state
and metastable soliton states of the FM and polar types are identified.
Considering the special case of the integrable version of the system, DT is
applied to find analytical solutions that display full nonlinear evolution
of MI of CW states, seeded by a small spatially periodic perturbation. Also,
it is demonstrated that solitons of both the polar and FM types, found in
the integrable system, are robust against random changes of the nonlinearity
coefficient in time (i.e., random deviations form the integrability). The
latter result demonstrates structural stability of the solitons.

The results collected in this section are essentially based on Ref. \cite%
{6.1}.

\subsection{Formulation of the model}

\begin{figure}[tbp]
\centerline{\includegraphics[scale=0.75]{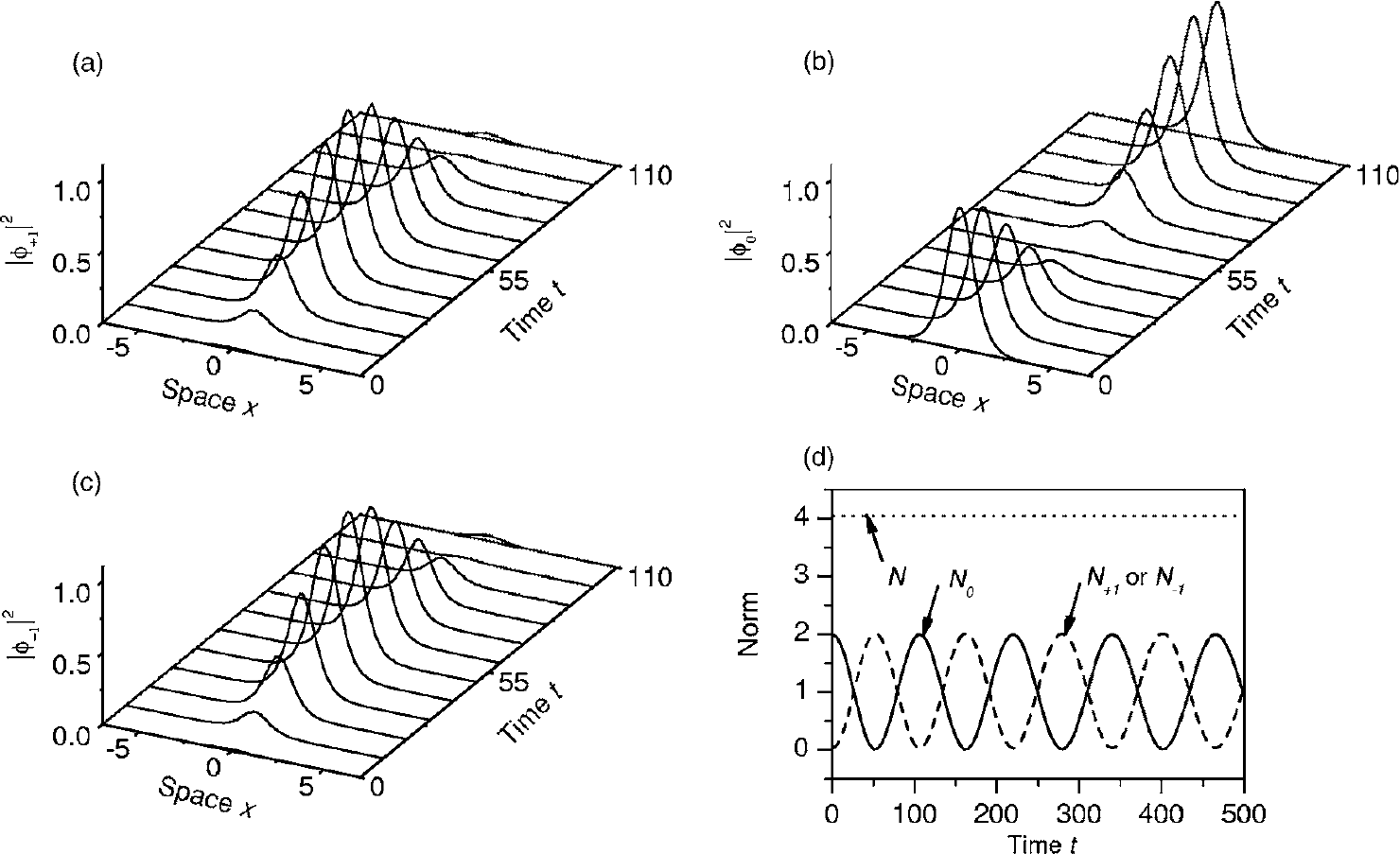}}
\caption{The evolution of the single-component polar soliton (\protect\ref%
{eq_7.11}) with a small random perturbation added, at $t=0$, to the $\protect%
\phi _{+1}$ and $\protect\phi _{-1}$ components. Parameters are $\protect\nu %
=1$, $a=-0.5$, and $\protect\mu =-1$. The results are reproduced from Ref.
\protect\cite{6.1}}
\label{fig034}
\end{figure}

The object considered here is an effectively 1D BEC loaded in a cigar-shaped
trap, which is elongated in $x$ and tightly confined in the transverse plane
$\left( y,z\right) $ \cite{3.5}. In the mean-field approximation, atoms in
the hyperfine state with atomic spin $F=1$ are described by a 1D
three-component wave function, $\mathbf{\Phi }(x,t)=\left[ \Phi
_{+1}(x,t),\Phi _{0}(x,t),\Phi _{-1}(x,t)\right] ^{\mathrm{T}}$, with the
components corresponding to the three values of the vertical spin
projection, $m_{F}=+1,$ $0,$ $-1$ \cite{Ho}. The wave functions obey the
corresponding three-component system of GP equations \cite{Ho,Ueda1,Ueda2}:
\end{subequations}
\begin{eqnarray}
i\hslash \frac{\partial \Phi _{\pm 1}}{\partial t} &=&-\frac{\hslash ^{2}}{2m%
}\frac{\partial ^{2}\Phi _{\pm 1}}{\partial x^{2}}+\left( c_{0}+c_{2}\right)
\left( \left\vert \Phi _{\pm 1}\right\vert ^{2}+\left\vert \Phi
_{0}\right\vert ^{2}\right) \Phi _{\pm 1}+\left( c_{0}-c_{2}\right)   \notag
\\
&&\times \left\vert \Phi _{\mp 1}\right\vert ^{2}\Phi _{\pm 1}+c_{2}\Phi
_{\mp 1}^{\ast }\Phi _{0}^{2},  \notag \\
i\hslash \frac{\partial \Phi _{0}}{\partial t} &=&-\frac{\hslash ^{2}}{2m}%
\frac{\partial ^{2}\Phi _{0}}{\partial x^{2}}+\left( c_{0}+c_{2}\right)
\left( \left\vert \Phi _{+1}\right\vert ^{2}+\left\vert \Phi
_{-1}\right\vert ^{2}\right) \Phi _{0}  \label{eq_7.1} \\
&&+c_{0}\left\vert \Phi _{0}\right\vert ^{2}\Phi _{0}+2c_{2}\Phi _{+1}\Phi
_{-1}\Phi _{0}^{\ast },  \notag
\end{eqnarray}%
where $c_{0}=\left( g_{0}+2g_{2}\right) /3$ and $c_{2}=\left(
g_{2}-g_{0}\right) /3$ denote, respectively, effective constants of the
spin-independent and spin-exchange interactions \cite{6.1,6.2}. In terms of
similar optics models, $c_{2}$ is the coefficient of the four-wave mixing,
while $c_{0}\pm c_{2}$ account for the SPM and XPM interactions. Here $g_{f}$
$=4\hslash ^{2}a_{f}/\left[ ma_{\bot }^{2}\left( 1-ca_{f}/a_{\bot }\right) %
\right] $ , with $f=0,2$, are effective 1D coupling constants, $a_{f}$ is
the \textit{s}-wave scattering length in the channel with total hyperfine
spin $f$, $a_{\bot }$ is the size of the transverse ground state, $m$ is the
atomic mass, and $c=-\zeta \left( 1/2\right) \approx 1.46$. Redefining the
wave function as $\mathbf{\Phi }\rightarrow \left( \phi _{+1},\sqrt{2}\phi
_{0},\phi _{-1}\right) ^{\mathrm{T}}$ and measuring time and length in units
of $\hslash /\left\vert c_{0}\right\vert $ and $\sqrt{\hslash
^{2}/2m\left\vert c_{0}\right\vert }$, respectively, one casts system (\ref%
{eq_7.1}) in the scaled form:%
\begin{eqnarray}
i\frac{\partial \phi _{\pm 1}}{\partial t} &=&-\frac{\partial ^{2}\phi _{\pm
1}}{\partial x^{2}}-\left( \nu +a\right) \left( \left\vert \phi _{\pm
1}\right\vert ^{2}+2\left\vert \phi _{0}\right\vert ^{2}\right) \phi _{\pm
1}-\left( \nu -a\right) \left\vert \phi _{\mp 1}\right\vert _{\pm 1}^{2}\phi
_{\pm 1}  \notag \\
&&-2a\phi _{\mp 1}^{\ast }\phi _{0}^{2},  \label{eq_7.2} \\
i\frac{\partial \phi _{0}}{\partial t} &=&-\frac{\partial ^{2}\phi _{0}}{%
\partial x^{2}}-2\nu \left\vert \phi _{0}\right\vert ^{2}\phi _{0}-\left(
\nu +a\right) \left( \left\vert \phi _{+1}\right\vert ^{2}+\left\vert \phi
_{-1}\right\vert ^{2}\right) \phi _{0}-2a\phi _{+1}\phi _{-1}\phi _{0}^{\ast
},  \notag
\end{eqnarray}%
where $\nu \equiv -$\textrm{sgn}$\left( c_{0}\right) $ and $a\equiv
-c_{2}/c_{0}$.

System (\ref{eq_7.2}) can be derived from the Hamiltonian,%
\begin{eqnarray}
H &=&\int_{-\infty }^{+\infty }dx\left\{ \left\vert \frac{\partial \phi _{+1}%
}{\partial x}\right\vert ^{2}+\left\vert \frac{\partial \phi _{-1}}{\partial
x}\right\vert ^{2}-\frac{1}{2}\left( \nu +a\right) \left( \left\vert \phi
_{+1}\right\vert ^{4}+\left\vert \phi _{-1}\right\vert ^{4}\right) \right.
\notag \\
&&-\left( \nu -a\right) \left\vert \phi _{+1}\right\vert ^{2}\left\vert \phi
_{-1}\right\vert ^{2}+2\left[ \left\vert \frac{\partial \phi _{0}}{\partial x%
}\right\vert ^{2}-\nu \left\vert \phi _{0}\right\vert ^{4}-\left( \nu
+a\right) \right.  \label{eq_7.3} \\
&&\left. \left. \times \left( \left\vert \phi _{+1}\right\vert
^{2}+\left\vert \phi _{-1}\right\vert ^{2}\right) \left\vert \phi
_{0}\right\vert ^{2}-a\left( \phi _{+1}^{\ast }\phi _{-1}^{\ast }\phi
_{0}^{2}+\phi _{+1}\phi _{-1}\left( \phi _{0}^{\ast }\right) ^{2}\right)
\right] \right\} ,  \notag
\end{eqnarray}%
which is a dynamical invariant of the model ($dH/dt=0$). Moreover, system (%
\ref{eq_7.2}) conserve the momentum, i.e.,%
\begin{equation}
P=i\int_{-\infty }^{+\infty }\left( \phi _{+1}^{\ast }\frac{\partial \phi
_{+1}}{\partial x}+\phi _{-1}^{\ast }\frac{\partial \phi _{-1}}{\partial x}%
+2\phi _{0}^{2}\frac{\partial \phi _{0}}{\partial x}\right) dx,
\end{equation}%
the solution's norm, proportional to total number of atoms,%
\begin{equation}
N=\int_{-\infty }^{+\infty }\left[ \left\vert \phi _{+1}(x,t)\right\vert
^{2}+\left\vert \phi _{-1}(x,t)\right\vert ^{2}+2\left\vert \phi
_{0}(x,t)\right\vert ^{2}\right] dx,  \label{eq_7.4}
\end{equation}%
and the total magnetization,%
\begin{equation}
\mathbf{M}=\int_{-\infty }^{+\infty }\left[ \left\vert \phi
_{+1}(x,t)\right\vert ^{2}-\left\vert \phi _{-1}(x,t)\right\vert ^{2}\right]
dx.  \label{M}
\end{equation}%
These conservation laws are generated by symmetries of the system of GP
equations (\ref{eq_7.2}). In particular, the conservation of the
magnetization is related to the invariance of the system with respect to
rotation of the atomic spin.

\begin{figure}[tbp]
\centerline{\includegraphics[scale=0.75]{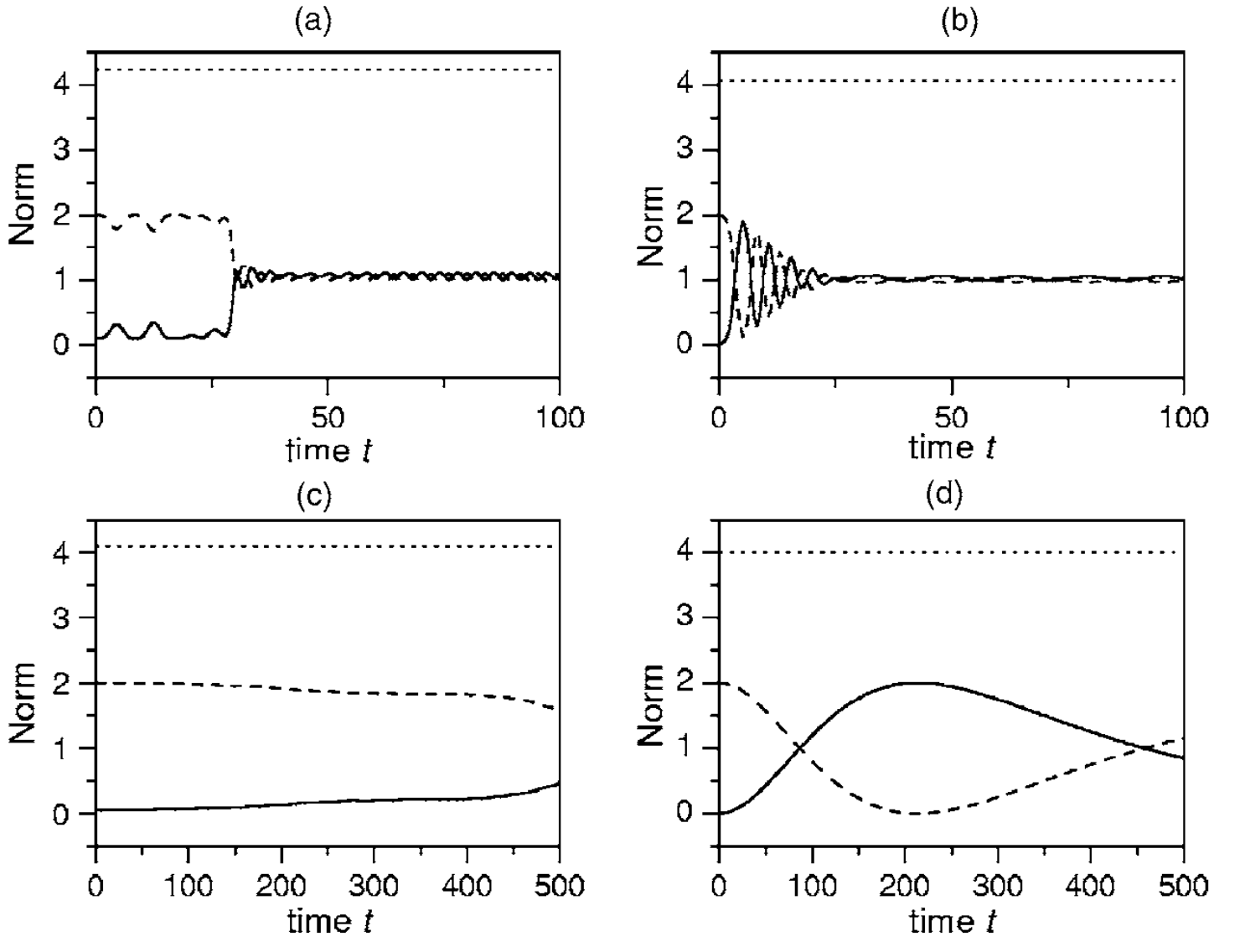}}
\caption{Solid, dashed, and dotted curves show, respectively, the evolution
of norms $N_{0}$, $N_{\pm }$ of the $\protect\phi _{0}$ and $\protect\phi %
_{\pm 1}$ components, and total norm $N$ (see Eqs. (\protect\ref{eq_7.15})
and (\protect\ref{eq_7.4})) in the two-component polar soliton (\protect\ref%
{eq_7.16}) perturbed by a small random perturbation introduced in the $%
\protect\phi _{0}$ component. Parameters are $\protect\nu =1$ and $\protect%
\mu =-1$ and (a) $a=1.5$ for $\protect\phi _{+1}\protect\phi _{-1}^{\ast }<0$%
; (b) $a=1.5$ for $\protect\phi _{+1}\protect\phi _{-1}^{\ast }>0$; (c) $%
a=-1.5$ for $\protect\phi _{+1}\protect\phi _{-1}^{\ast }<0$; (d) $a=-1.5$
for $\protect\phi _{+1}\protect\phi _{-1}^{\ast }>0$. The results are
reproduced from Ref. \protect\cite{6.1}.}
\label{fig035}
\end{figure}

An obvious reduction of system (\ref{eq_7.2}) can be obtained by setting $%
\phi _{0}=0$. In this case, the model reduces to a system of two equations
which are tantamount to those describing light transmission in bimodal
nonlinear optical fibers, with the two modes representing either different
wavelengths or two orthogonal polarizations \cite{6.1}:
\begin{equation}
i\frac{\partial \phi _{\pm 1}}{\partial t}=-\frac{\partial ^{2}\phi _{\pm 1}%
}{\partial x^{2}}-\left( \nu +a\right) \left\vert \phi _{\pm 1}\right\vert
^{2}\phi _{\pm 1}-\left( \nu -a\right) \left\vert \phi _{\mp 1}\right\vert
^{2}\phi _{\pm 1}.  \label{eq_7.5}
\end{equation}%
In particular, the case of two linear polarizations in the birefringent
fiber corresponds to $\left( \nu -a\right) /\left( \nu +a\right) $ $=2/3$,
i.e., $a=\nu /5$ and two different carrier waves of circular polarizations
correspond to $\left( \nu -a\right) /\left( \nu +a\right) =2$, i.e., $a=-\nu
/3$. The two nonlinear terms in system (\ref{eq_7.5}) account for,
respectively, the SPM and XPM interactions of the two waves. The MI of CW
(uniform) states in the system of two XPM-coupled equations (\ref{eq_7.5})
has been studied in detail \cite{6.3}. If the nonlinearity in system (\ref%
{eq_7.5}) is self-defocusing, i.e., $\nu =-1$ and $1\pm a>0$ (in other
words, $\left\vert a\right\vert <$ $1$), the single NLS equation would of
course show no MI, but the XPM-coupled system of NLS equations (\ref{eq_7.5}%
) gives rise to MI, provided that the XPM interaction is stronger than SPM,
i.e., $0<a<1$ \cite{Agrawal-XPM}.

\subsection{Analytical one-, two-, and three-component soliton solutions}

As follows from the above derivation, the special case of the integrable
model (with $a=\nu =1$ \cite{Wadati}) is physically possible, corresponding
to $c_{2}=c_{0}$ or, equivalently, $2g_{0}=-g_{2}>0$. The condition $%
2g_{0}=-g_{2}>0$ can be satisfied by imposing condition%
\begin{equation}
a_{\bot }=3ca_{0}a_{2}/\left( 2a_{0}+a_{2}\right)  \label{eq_7.6}
\end{equation}%
on the scattering lengths, provided that $a_{0}a_{2}\left(
a_{2}-a_{0}\right) >0$ holds.

However, it is also possible to find simple exact soliton solutions of both
polar and FM types in the general (non-integrable) case $a\neq \nu $.
Several soliton species and results for their stability are presented below,
following Ref. \cite{6.1}.

\subsubsection{Single-component FM solitons}

A one-component FM soliton is given by a straightforward single-component
solution (assuming $a+\nu >0$):%
\begin{equation}
\phi _{-1}=\phi _{0}=0,\text{ }\phi _{+1}=\sqrt{-\frac{2\mu }{\nu +1}}\frac{%
\exp \left( -i\mu t\right) }{\cosh \left( \sqrt{-\mu }x\right) },
\label{eq_7.7}
\end{equation}%
where chemical potential $\mu <0$ is an intrinsic parameter of the soliton
family. While expression (\ref{eq_7.7}) corresponds to the zero-velocity
soliton, moving ones can be generated from it in an obvious way by means of
the Galilean transformation. Note that condition $a+\nu >0$ implies $a>1$ or
$a>-1$ in the cases of the, respectively, repulsive ($\nu =-1$) or
attractive ($\nu =+1$) spin-independent interaction. Norm (\ref{eq_7.4}) and
energy (\ref{eq_7.3}) of this soliton are%
\begin{equation}
N=\frac{4\sqrt{-\mu }}{\nu +a}\text{, }H=-\frac{\left( \nu +a\right) ^{2}}{48%
}N^{3}.  \label{eq_7.8}
\end{equation}%
Soliton (\ref{eq_7.7}) is stable against small perturbations, as the
linearization of system (\ref{eq_7.2}) about this solution demonstrates that
the BdG equations for small perturbations are decoupled. Then, because the
standard soliton of the single NLS equation is always stable, solution (\ref%
{eq_7.7}) cannot be unstable against small perturbations of $\phi _{+1}$.
Further, the decoupled BdG equations for small perturbations $\phi _{-1}$
and $\phi _{0}$ of the other fields are
\begin{subequations}
\begin{eqnarray}
\frac{\partial \phi _{-1}}{\partial t} &=&-\frac{\partial ^{2}\phi _{-1}}{%
\partial x^{2}}-\left( \nu -a\right) \left\vert \phi _{+1}\right\vert
^{2}\phi _{-1},  \label{eq_7.9} \\
\frac{\partial \phi _{0}}{\partial t} &=&-\frac{\partial ^{2}\phi _{0}}{%
\partial x^{2}}-\left( \nu +a\right) \left\vert \phi _{+1}\right\vert
^{2}\phi _{0},  \label{eq_7.10}
\end{eqnarray}%
and it is well known that such linear equations, with $\left\vert \phi
_{+1}\right\vert ^{2}$ corresponding to the unperturbed soliton (\ref{eq_7.7}%
), give rise to no instabilities either.

\begin{figure}[tbp]
\centerline{\includegraphics[scale=0.68]{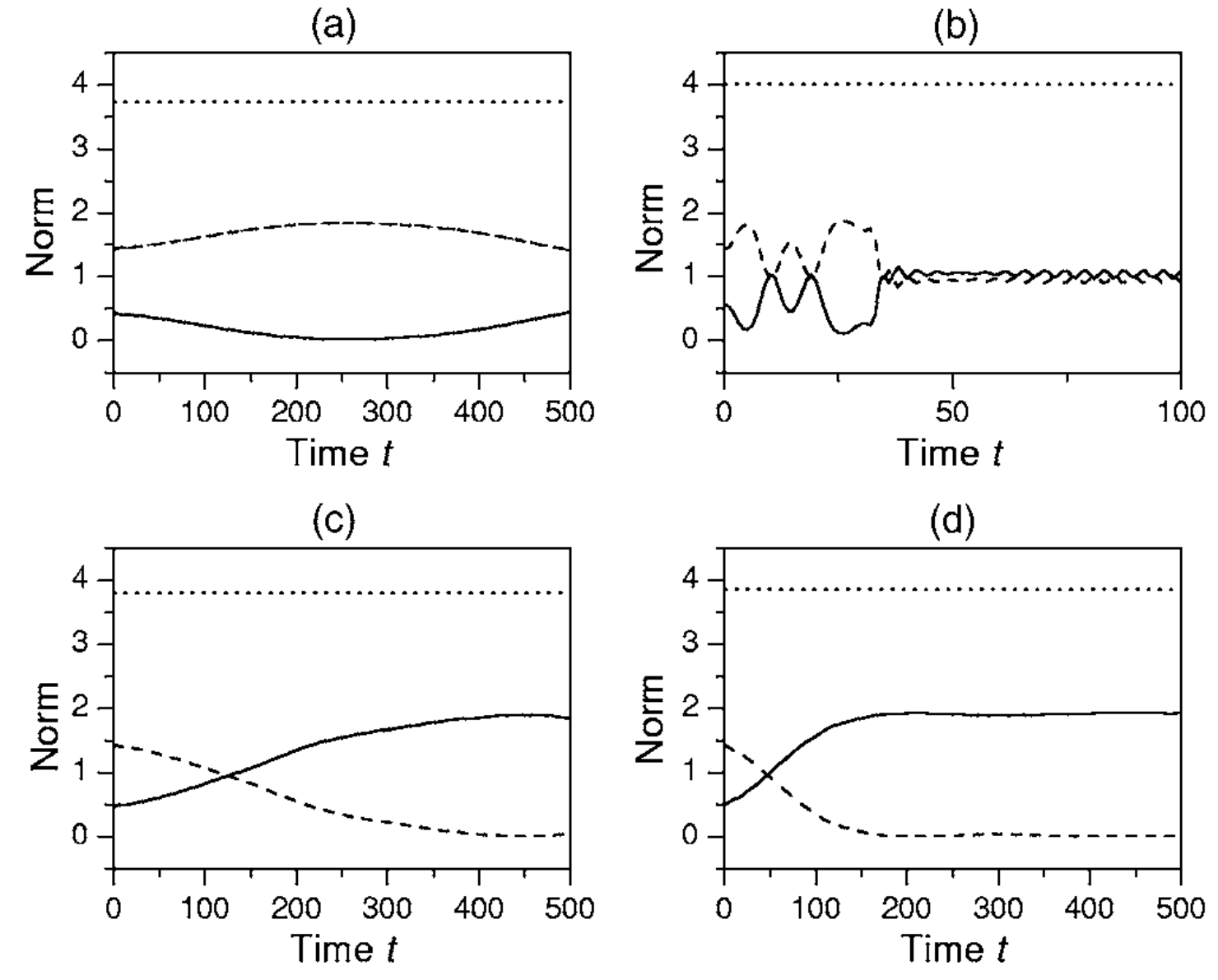}}
\caption{The evolution of the three-component polar soliton (\protect\ref%
{eq_7.20}) under the action of a small random perturbation initially added
to the $\protect\phi _{0}$ component. Parameters are $\protect\nu =1$, $%
\protect\mu =-0.8$, and $\protect\epsilon =2/\protect\sqrt{5}$ and (a) $%
a=0.5 $; (b) $a=1.5$; (c) $a=-0.5$; (d) $a=-1.5$. The meaning of the solid,
dashed, and dotted curves is the same as in Fig. \protect\ref{fig035}. The
results are reproduced from Ref. \protect\cite{6.1}.}
\label{fig036}
\end{figure}

\subsubsection{Single-component polar solitons}

The simplest polar soliton, which has only the $\phi _{0}$ component
different from zero, can be found for $\nu =+1$,
\end{subequations}
\begin{equation}
\phi _{0}(x,t)=\sqrt{-\mu }\frac{\exp \left( -i\mu t\right) }{\cosh \left(
\sqrt{-\mu }x\right) },~\phi _{\pm 1}(x,t)=0.  \label{eq_7.11}
\end{equation}%
The stability problem for this soliton is more involved, as the unperturbed
field $\phi _{0}$ gives rise to a \textit{coupled} system of BdG equations
for infinitesimal perturbations $\phi _{\pm 1}$ of the other fields:%
\begin{equation}
\left( i\frac{\partial }{\partial t}+\mu \right) \chi _{\pm 1}=-\frac{%
\partial ^{2}\chi _{\pm 1}}{\partial x^{2}}+\frac{2\mu }{\cosh ^{2}\left(
\sqrt{-\mu }x\right) }\left[ \left( 1+a\right) \chi _{\pm 1}+a\chi _{\mp
1}^{\ast }\right] ,  \label{eq_7.12}
\end{equation}%
where we have substituted $\nu =+1$ and redefined the perturbation,%
\begin{equation}
\phi _{\pm 1}(x,t)\equiv \chi _{\pm 1}(x,t)\exp \left( -i\mu t\right) .
\label{eq_7.13}
\end{equation}%
The last term in system (\ref{eq_7.12}) gives rise to parametric gain, which
may be a source of instability. This instability can be easily understood in
qualitative terms if one replaces system (\ref{eq_7.12}) by simplified
equations, which disregard the $x$ dependence, and replace the wave form $%
\mathrm{sech}^{2}\left( \sqrt{-\mu }x\right) $ by its value at the soliton's
center, $x=0$:%
\begin{equation}
i\frac{d\chi _{\pm 1}}{dt}=\mu \left[ \left( 1+2a\right) \chi _{\pm
1}+2a\chi _{\mp 1}^{\ast }\right] .  \label{eq_7.14}
\end{equation}%
An elementary consideration shows that the zero solution of linear equations
(\ref{eq_7.14}) is unstable through a double eigenvalue in the region of $%
a<-1/4$.

The stability of the single-component polar soliton (\ref{eq_7.11}) was
checked by means of direct simulations of Eqs. (\ref{eq_7.2}), adding a
small random perturbation in components $\phi _{+1}$ and $\phi _{-1}$, the
value of the perturbation being distributed uniformly between $0$ and $0.03$
(the same random perturbation is used in simulations of the stability of
other solitons; see below). The result is that the soliton (\ref{eq_7.11})
is, indeed, always unstable, as shown in Fig. \ref{fig034}. In particular,
panel (d) in the figure displays the time evolution of%
\begin{equation}
N_{0}=\int_{-\infty }^{+\infty }\left\vert \phi _{0}(x,t)\right\vert ^{2}dx,%
\text{ \ }N_{\pm 1}=\int_{-\infty }^{+\infty }\left\vert \phi _{\pm
1}(x,t)\right\vert ^{2}dx.  \label{eq_7.15}
\end{equation}

\begin{figure}[h]
\centerline{\includegraphics[scale=0.65]{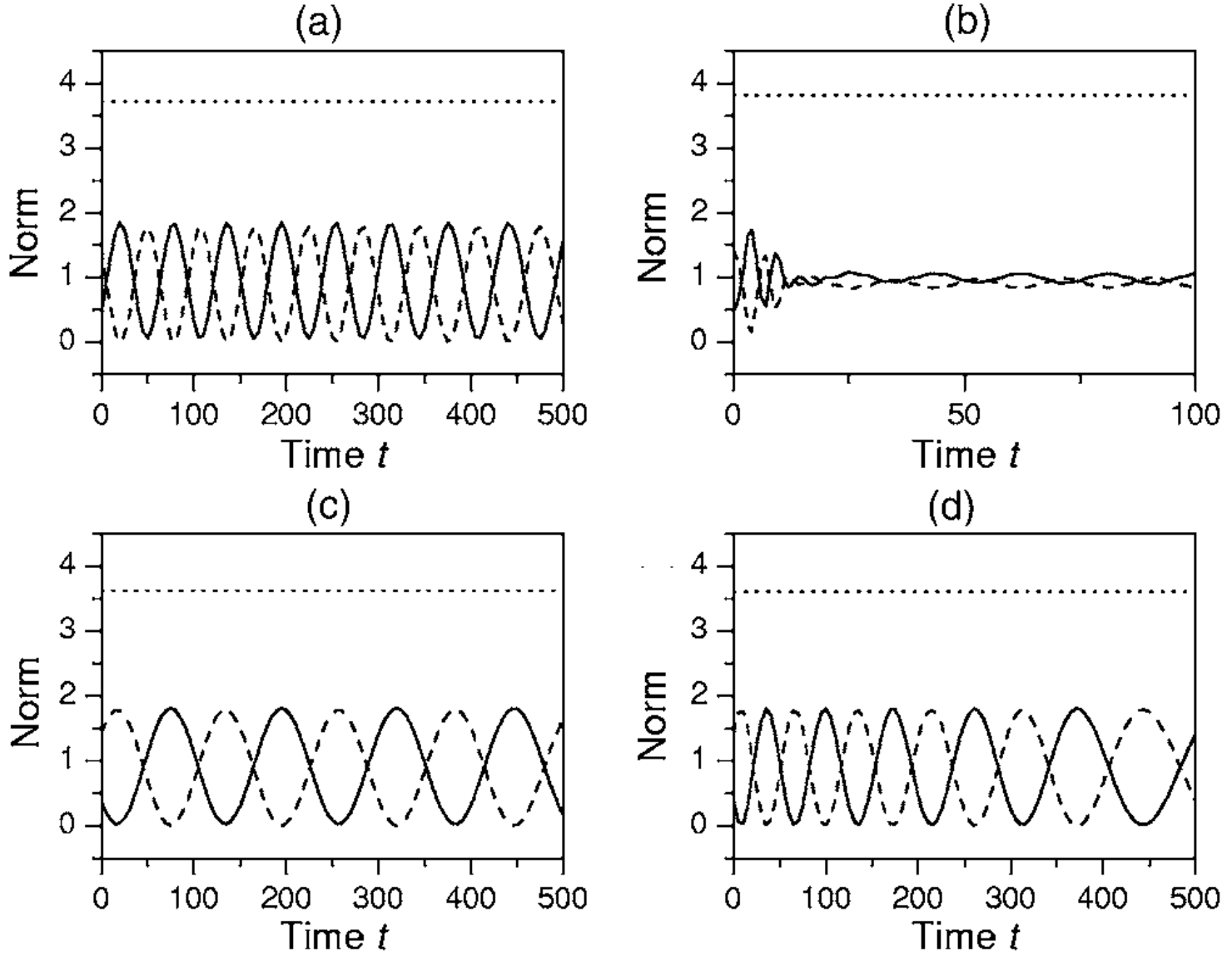}}
\caption{The same as in Fig. \protect\ref{fig035}, but for the
three-component polar soliton (\protect\ref{eq_7.21}). Parameters are $%
\protect\nu =1$, $\protect\mu =-0.8$, $\protect\epsilon =2/\protect\sqrt{5}$%
, and $a=0.5$ in (a), $a=1.5$ in (b), $a=-0.5$ in (c), and $a=-1.5$ in (d).
The meaning of the solid, dashed, and dotted curves is the same as in Figs.
\protect\ref{fig035} and \protect\ref{fig036}. The results are reproduced
from Ref. \protect\cite{6.1}.}
\label{fig037}
\end{figure}

\subsubsection{Two-component polar solitons}

In the same case as considered above, $\nu =+1$, a two-component polar
soliton can be easily found too:%
\begin{equation}
\phi _{0}(x,t)=0,\text{ }\phi _{+1}(x,t)=\pm \phi _{-1}(x,t)=\sqrt{-\mu }%
\frac{\exp \left( -i\mu t\right) }{\cosh \left( \sqrt{-\mu }x\right) }.
\label{eq_7.16}
\end{equation}%
In this case, instability is possible due to the parametric gain revealed by
the BdG equations for small perturbations. Indeed, the corresponding
equation for the perturbation in the $\phi _{0}$ component decouples from
the other equations and takes the form of%
\begin{equation}
\left( i\frac{\partial }{\partial t}+\mu \right) \chi _{0}=-\frac{\partial
^{2}\chi _{0}}{\partial x^{2}}+\frac{2\mu }{\cosh ^{2}\left( \sqrt{-\mu }%
x\right) }\left[ \left( 1+a\right) \chi _{0}+a\chi _{0}^{\ast }\right] ,
\label{eq_7.17}
\end{equation}%
where the perturbation was redefined the same way as in Eq. (\ref{eq_7.13}),
$\phi _{0}(x,t)\equiv \chi _{0}(x,t)\exp \left( -i\mu t\right) $. Similar to
the case of Eq. (\ref{eq_7.12}), the origin of the instability may be
understood, replacing Eq. (\ref{eq_7.17}) by its simplified version that
disregards the $x$ dependence, replacing the wave form $\mathrm{sech}%
^{2}\left( \sqrt{-\mu }x\right) $ by its value at the central point of the
soliton, $x=0$:%
\begin{equation}
i\frac{d\chi _{0}}{dt}=\mu \left[ \left( 1+2a\right) \chi _{0}\pm 2a\chi
_{0}^{\ast }\right] .  \label{eq_7.18}
\end{equation}%
Elementary consideration demonstrates that the zero solution of Eq. (\ref%
{eq_7.18}) (for either sign $\pm $) is unstable in exactly the same region
as in the case of system (\ref{eq_7.14}), \textit{viz}., $a<-1/4$. While
simplified equation (\ref{eq_7.18}) illustrates the qualitative mechanism of
the parametric instability, the actual stability border may be different
from $a=-1/4$. The stability of the soliton (\ref{eq_7.16}) was tested by
direct simulations of Eqs. (\ref{eq_7.2}) with a small uniformly distributed
random perturbation added to the $\phi _{0}$ component. The result shows
that the soliton is unstable in the region of $\left\vert a\right\vert \geq
1 $, as shown in Fig. \ref{fig035}, and it is stable at $\left\vert
a\right\vert <1$ (not shown here).

The change of the stability of this soliton at $a=+1$ can be explained
analytically. Indeed, splitting the perturbation into real and imaginary
parts, $\chi _{0}\left( x,t\right) \equiv $ $\chi \left( _{1}x,t\right) $ $%
+i\chi _{2}\left( x,t\right) $, and looking for a perturbation eigenmode as $%
\chi _{1,2}\left( x,t\right) =\exp \left( \sigma t\right) U_{1,2}(x)$ with
instability growth rate $\sigma $, one arrives at a system of real equations,%
\begin{eqnarray}
-\sigma U_{2} &=&\left( -\frac{d^{2}}{dx^{2}}-\mu +\frac{2\mu \left(
1+2a\right) }{\cosh ^{2}\left( \sqrt{-\mu }x\right) }\right) U_{1},  \notag
\\
&&  \label{eq_7.19} \\
-\sigma U_{1} &=&\left( -\frac{d^{2}}{dx^{2}}-\mu +\frac{2\mu }{\cosh
^{2}\left( \sqrt{-\mu }x\right) }\right) U_{2}.  \notag
\end{eqnarray}%
The simplest possibility for the onset of instability of the soliton is the
passage of eigenvalue $\sigma $ through zero. At $\sigma =0$, system (\ref%
{eq_7.19}) decouples, making each equation explicitly solvable and producing
the zero eigenvalue at $a=a_{n}\equiv n\left( n+3\right) /4$, $n=0,$ $1,$ $%
2,...$ . The vanishing of $\sigma $ at $a_{0}=0$ corresponds not to
destabilization of the soliton, but to the fact that system (\ref{eq_7.19})
becomes symmetric at this point, while the zero crossings at other points
indeed imply stability changes. In particular, the critical point $a_{1}=1$
explains the destabilization of soliton (\ref{eq_7.16}) at $a=1$ as observed
in the simulations, the corresponding eigenfunctions being
\begin{equation}
U_{1}(x)=\frac{\sinh \left( \sqrt{-\mu }x\right) }{\cosh ^{2}\left( \sqrt{%
-\mu }x\right) },U_{2}(x)=\frac{1}{\cosh \left( \sqrt{-\mu }x\right) }.
\label{two-comp-polar}
\end{equation}%
Critical points corresponding to $n>1$ imply additional destabilizations
through the emergence of new unstable eigenmodes of the already unstable
soliton. On the other hand, the destabilization of the soliton at $a=-1$,
also observed in the simulations, may be explained by a bifurcation which
changes a pair of eigenvalues $\sigma $ from purely imaginary to complex
ones, developing an unstable positive real part.

\begin{figure}[h]
\centerline{\includegraphics[scale=0.65]{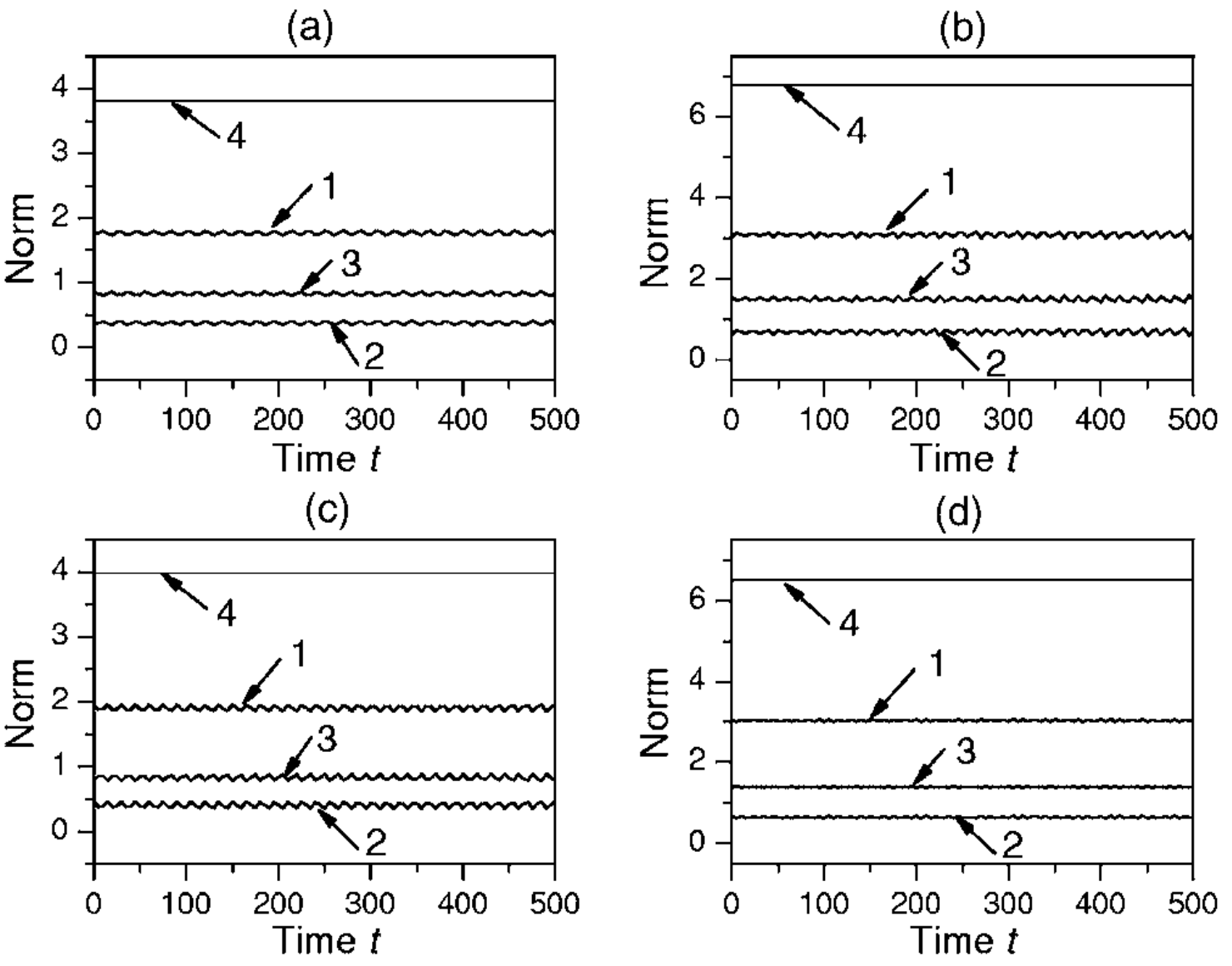}}
\caption{The evolution of the three-component polar soliton (\protect\ref%
{eq_7.22}) with initially added random perturbations. Parameters are $%
\protect\nu =1$, $\protect\mu _{+1}=-1.21$, $\protect\mu _{-1}=-0.25$, and $%
a=0.5$ in (a,c) or $a=-0.5$ in (b,d). In cases (a) and (b), the
random perturbation was added to the $\protect\phi _{0}$
component, and in cases (c) and (d) it was added to components
$\protect\phi _{\pm 1}$. The curves labelled by $1$, $2$, $3$, and
$4$ represent, respectively, the evolution of
norms $N_{+1}$, $N_{-1}$, $N_{0}$, and $N$ defined as per Eqs. (\protect\ref%
{eq_7.15}) and (\protect\ref{eq_7.4}). The results are reproduced from Ref.
\protect\cite{6.1}. }
\label{fig038}
\end{figure}

\subsubsection{Three-component polar solitons}

In the case of $\nu =+1$, three-component solitons of the polar type can be
also found. One of them is%
\begin{eqnarray}
\phi _{0} &=&\sqrt{1-\epsilon ^{2}}\sqrt{-\mu }\frac{\exp \left( -i\mu
t\right) }{\cosh \left( \sqrt{-\mu }x\right) },  \notag \\
&&  \label{eq_7.20} \\
\phi _{+1} &=&-\phi _{-1}=\pm \epsilon \sqrt{-\mu }\frac{\exp \left( -i\mu
t\right) }{\cosh \left( \sqrt{-\mu }x\right) },  \notag
\end{eqnarray}%
where $\epsilon $ is an arbitrary parameter taking values $-1<\epsilon <+1$,
and the opposite signs of $\phi _{+1}$ and $\phi _{-1}$ is an essential
ingredient of the solution. As well as the one- and two-component polar
solitons displayed above, see Eqs. (\ref{eq_7.16}) and (\ref{two-comp-polar}%
), the solution does not explicitly depend on parameter $a$ in Eq. (\ref%
{eq_7.2}).

There is another three-component polar solution similar to the above one,
i.e., containing an arbitrary parameter and independent of $a$, but with
identical signs of the $\phi _{\pm 1}$ components and a phase shift of $\pi
/2$ in the $\phi _{0}$ component. This solution is%
\begin{eqnarray}
\phi _{0} &=&i\sqrt{1-\epsilon ^{2}}\sqrt{-\mu }\frac{\exp \left( -i\mu
t\right) }{\cosh \left( \sqrt{-\mu }x\right) },  \notag \\
&&  \label{eq_7.21} \\
\phi _{+1} &=&\phi _{-1}=\pm \epsilon \sqrt{-\mu }\frac{\exp \left( -i\mu
t\right) }{\cosh \left( \sqrt{-\mu }x\right) }.  \notag
\end{eqnarray}%
Furthermore, there is a species of three-component polar solitons that
explicitly depends on $a$:%
\begin{eqnarray}
\phi _{0} &=&i\left( \mu _{+1}\mu _{-1}\right) ^{1/4}\frac{\exp \left( -i\mu
t\right) }{\cosh \left( \sqrt{-\mu }x\right) },  \notag \\
&&  \label{eq_7.22} \\
\phi _{\pm 1} &=&\sqrt{-\mu _{\pm 1}}\frac{\exp \left( -i\mu t\right) }{%
\cosh \left( \sqrt{-\mu }x\right) },  \notag
\end{eqnarray}%
where $\mu _{\pm 1}$ are two arbitrary negative parameters, and the chemical
potential is $\mu =-\left( \nu +a\right) \left( \sqrt{-\mu _{+1}}+\sqrt{-\mu
_{-1}}\right) ^{2}/2$, which implies condition $\nu +a>0$ (note that this
solution admits $\nu =-1$, i.e., repulsive spin-independent interaction).
Thus, each species of the three-component polar soliton contains two
arbitrary parameters: either $\mu $ and $\epsilon $ in solutions (\ref%
{eq_7.20}) and (\ref{eq_7.21}), or $\mu _{\pm 1}$ in solution (\ref{eq_7.22}%
).

The stability of all the species of these solitons was tested in direct
simulations \cite{6.1}. In the case of solutions (\ref{eq_7.20}) and (\ref%
{eq_7.21}), it was enough to seed a small random perturbation only in the $%
\phi _{0}$ component, to observe that both these types are unstable, as
shown in Figs. \ref{fig036} and \ref{fig037}.

On the contrary, the three-component polar soliton (\ref{eq_7.22}) is
completely stable. This point was checked in detail by seeding small random
perturbations in all the components. The result, illustrated by Fig. \ref%
{fig038}, is that the perturbation induces only small oscillations of the
amplitudes of each component of the soliton.

\subsubsection{Multistability of the solitons}

The above results demonstrate remarkable multistability in the system, as
the FM soliton (\ref{eq_7.7}), two-component polar solitons (\ref{eq_7.16})
in the regions of $-1<a<+1$ , and the three-component one (\ref{eq_7.22})
may all be stable in the same parameter region. Then, to identify which
solitons are \textquotedblleft more stable\textquotedblright\ and
\textquotedblleft less stable,\textquotedblright\ one may fix norm (\ref%
{eq_7.4}) and compare respective values of the Hamiltonian (\ref{eq_7.3})
for these solutions, as the ground-state solution should correspond to a
minimum of $H$ at given $N$.

The substitution of the solutions in Eq. (\ref{eq_7.3}) reveals another
remarkable fact: all the solutions which do not explicitly depend on $a$,
i.e., the one-, two-, and three-component polar solitons (\ref{eq_7.11}), (%
\ref{eq_7.16}), (\ref{eq_7.20}), and (\ref{eq_7.21}), produce identical
relations between $N$, $\mu $, and $H$:%
\begin{equation}
N=4\sqrt{-\mu },\text{ }H=-\frac{1}{48}N^{3}.  \label{eq_7.23}
\end{equation}%
The $N(\mu )$ and $H(N)$ relations are different for the single-component FM
soliton (\ref{eq_7.7}), see Eq. (\ref{eq_7.8}). Finally, for the stable
three-component polar soliton (\ref{eq_7.22}), the relations between $\mu $,
$N$, and $H$ take exactly the same form as Eq. (\ref{eq_7.8}) for the FM
soliton.

The comparison of expressions (\ref{eq_7.23}) and (\ref{eq_7.8}) leads to a
simple conclusion: the one-component FM soliton (\ref{eq_7.7}) and the
stable three-component polar one (\ref{eq_7.22}) simultaneously provide the
minimum of energy in the case of $\nu =+1$ and $a>0$, when both the
spin-independent and spin-exchange interactions between atoms are
attractive. In these cases, each of these two species (\ref{eq_7.7}) and (%
\ref{eq_7.22}) plays the role of the ground state in its own class of the
solitons, \textit{viz}., FM and polar ones, respectively. Simultaneously,
the two-component soliton (\ref{eq_7.16}) is also stable in the region of $%
0<a<1$, as shown above, but it corresponds to higher energy, hence it
represents a metastable state in this region.

\begin{figure}[h]
\centerline{\includegraphics[scale=0.9]{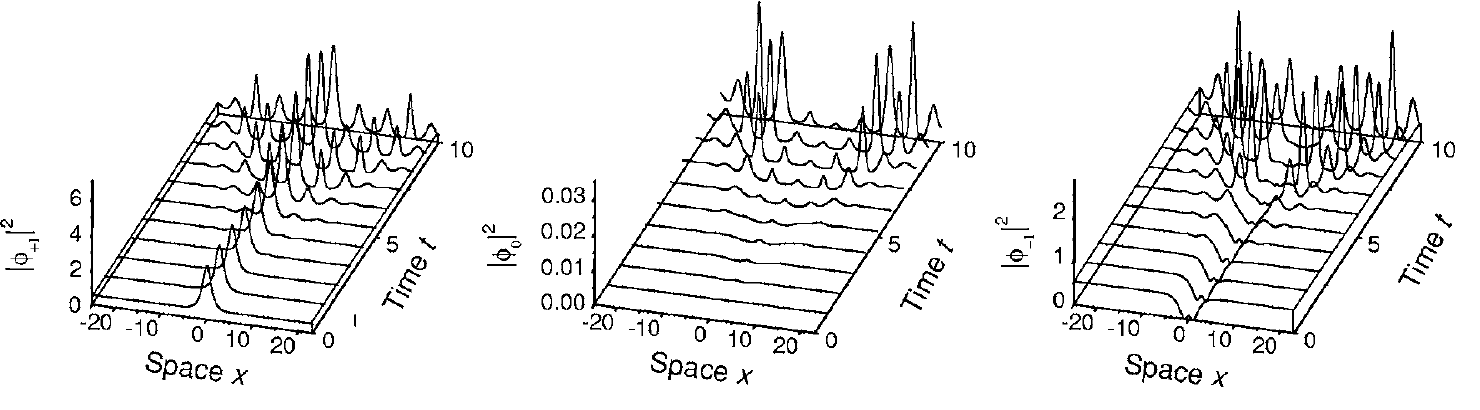}}
\caption{The evolution of the two-component polar soliton on the CW
background, given by Eq. (\protect\ref{eq_7.24}), under the action of
initial random perturbations added to the $\protect\phi _{0}$ and $\protect%
\phi _{\pm 1}$ components. Parameters are $\protect\nu =+1$, $q=-1/2$, and $%
\protect\mu =-1$. The results rep4oduced from Ref. \protect\cite{6.1}}
\label{fig039}
\end{figure}

The FM soliton (\ref{eq_7.7}) and three-component polar soliton (\ref%
{eq_7.22}) also exist in the case of $\nu =-1$ and $a>1$, when $\nu +a$ is
positive. In this case too, these two soliton species provide for the energy
minimum, simply because the other solitons, (\ref{eq_7.11}) , (\ref{eq_7.16}%
) , (\ref{eq_7.20}) , and (\ref{eq_7.21}) do not exist at $\nu =-1$.

For $\nu =+1$ and $-1<a<0$, i.e., attractive spin-independent and repulsive
spin-exchange interactions in Eqs. (\ref{eq_7.5}), Hamiltonian (\ref{eq_7.23}%
) is smaller than the competing one (\ref{eq_7.8}), which means that the
two-component polar soliton (\ref{eq_7.16}) plays the role of the ground
state in this case, as only it, among all the polar solitons whose
Hamiltonian is given by expression (\ref{eq_7.23}), is dynamically stable in
direct simulations (for $-1<a<+1$, see above). As the FM soliton (\ref%
{eq_7.7}) and three-component polar one (\ref{eq_7.22}) also exist and are
stable in this region, but correspond to greater energy, they represent
metastable states here. Finally, for $\nu $ $=+1$ and $a<-1$, there are no
stable solitons.

Dependence $N(\mu )$ for each solution family provides an additional
characteristic of the soliton stability. Indeed, the well-known known
Vakhitov-Kolokolov criterion \cite{6.4,4.11} states that a necessary
stability condition for the soliton family supported by a self-attractive
nonlinearity is $dM/d\mu <0$. It guarantees that the soliton cannot be
unstable against perturbations with real eigenvalues, but does not say
anything about oscillatory perturbations modes appertaining to complex
eigenvalues. Obviously, both relations, given by Eqs. (\ref{eq_7.8}) and (%
\ref{eq_7.23}), satisfy the criterion, i.e., solitons may be unstable only
against perturbations that grow in time with oscillations. Indeed, numerical
results which illustrate the evolution of unstable solitons, see Figs. \ref%
{fig034}-\ref{fig037}, clearly suggest that the instability, if any, is
oscillatory.

\subsubsection{Finite-background solitons}

\begin{figure}[h]
\centerline{\includegraphics[scale=0.9]{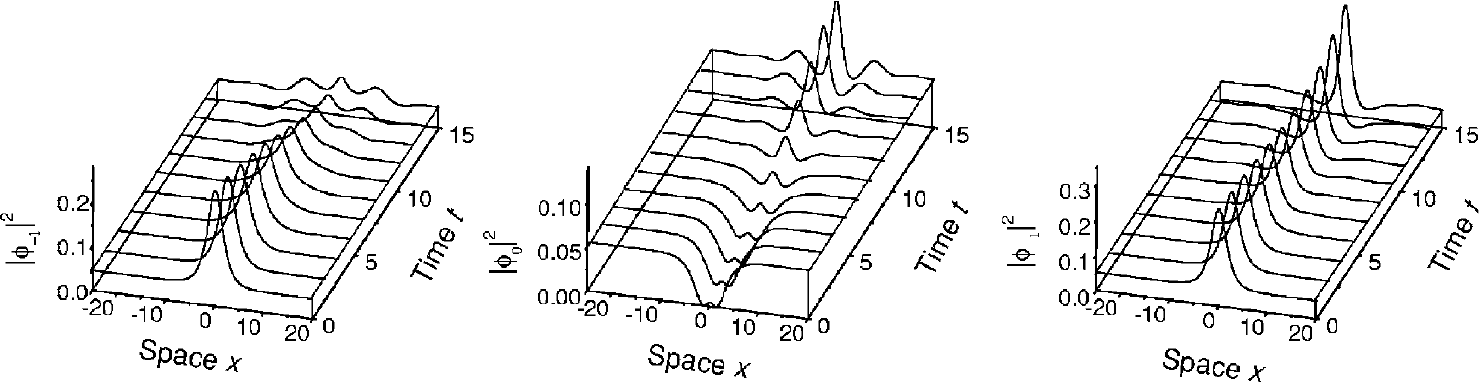}}
\caption{The same as in Fig. \protect\ref{fig039}, but for the
finite-background soliton (\protect\ref{eq_7.25}). Parameters are $\protect%
\nu =a=1$ and $\protect\mu =-0.36$. The results are reproduced from Ref.
\protect\cite{6.1}.}
\label{fig040}
\end{figure}

In special cases, it is possible to find analytical solutions for solitons
sitting on a CW background. Namely, for $\nu =1$ and $a=-1/2$, one can find
a two-component polar soliton with a CW background attached to it, in the
following form:%
\begin{eqnarray}
\phi _{0} &=&0,  \notag \\
\phi _{+1} &=&\sqrt{-\mu }\left[ \frac{1}{\sqrt{2}}\pm \frac{1}{\cosh \left(
\sqrt{-\mu }x\right) }\right] \exp \left( -i\mu t\right) ,  \label{eq_7.24}
\\
\phi _{-1} &=&\sqrt{-\mu }\left[ \frac{1}{\sqrt{2}}\mp \frac{1}{\cosh \left(
\sqrt{-\mu }x\right) }\right] \exp \left( -i\mu t\right) .  \notag
\end{eqnarray}%
For $\nu =a=1$, it is also possible to find a three-component polar solution
with the background,%
\begin{eqnarray}
\phi _{+1} &=&\phi _{-1}=\frac{1}{2}\sqrt{-\mu }\left[ \frac{1}{\sqrt{2}}\pm
\frac{1}{\cosh \left( \sqrt{-\mu }x\right) }\right] \exp \left( -i\mu
t\right) ,  \notag \\
&&  \label{eq_7.25} \\
\phi _{0} &=&\frac{1}{2}\sqrt{-\mu }\left[ \frac{1}{\sqrt{2}}\mp \frac{1}{%
\cosh \left( \sqrt{-\mu }x\right) }\right] \exp \left( -i\mu t\right) .
\notag
\end{eqnarray}

The stability of solutions (\ref{eq_7.24}) and (\ref{eq_7.25}) was tested by
simultaneously perturbing the $\phi _{0}$ and $\phi _{\pm 1}$ components by
small uniformly distributed random perturbations. Figures \ref{fig039} and %
\ref{fig040} show that both solutions are unstable, although the character
of the instability is different: in the former case, the soliton's core
seems stable, while the CW background appears to be modulationally unstable.
In the latter case, which is shown in Fig. \ref{fig040}, the background is
modulationally stable, but the core of the soliton (\ref{eq_7.25}) clearly
features oscillatory instability.

\begin{figure}[h]
\centerline{\includegraphics[scale=0.9]{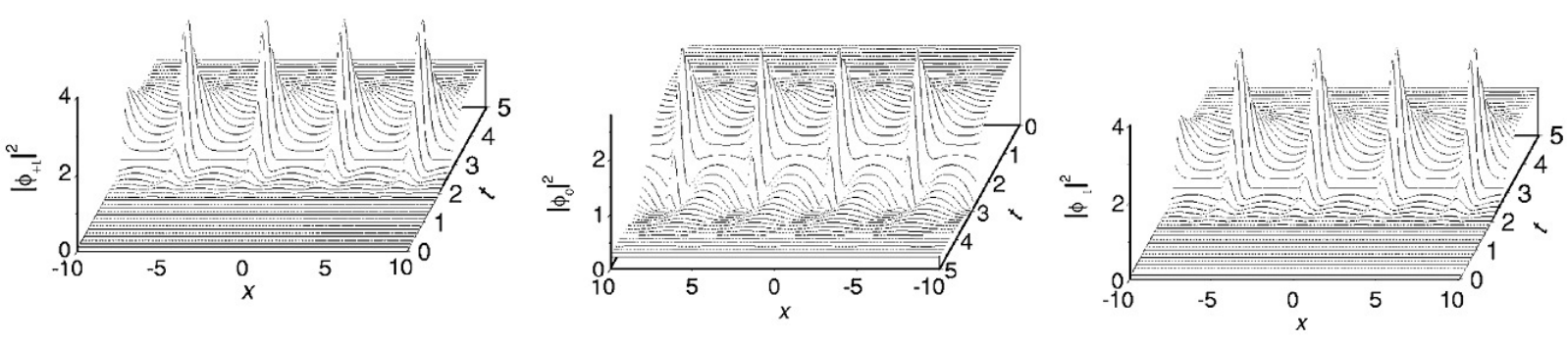}}
\caption{The nonlinear development of the modulation instability, as per
exact solution (\protect\ref{eq_7.35}), under conditions $k=2\protect\eta $
and $\protect\alpha _{c}^{2}+\protect\beta _{c}^{2}>\protect\xi ^{2}$.
Parameters are $k=0.6$, $\protect\eta =0.3$, $\protect\xi =1$, $\protect%
\alpha =\protect\beta =$ $\protect\gamma =\exp \left( -8\right) $, $\protect%
\alpha _{c}=1.2$, and $\protect\beta _{c}=0$. The results are reproduced
from Ref. \protect\cite{6.1}.}
\label{fig041}
\end{figure}

MI of the CW background in solution (\ref{eq_7.24}) is obvious, as, even
without exciting the $\phi _{0}$ field, i.e., in the framework of the
reduced equations (\ref{eq_7.5}), with $\nu +a=1/2$ and $\nu -a=3/2$, any CW
solution is subject to MI. As concerns the excitation of the $\phi _{0}$
field, an elementary consideration shows that the CW part of solution (\ref%
{eq_7.24}) exactly corresponds to the threshold of the parametric
instability driven by term $2a\phi _{+1}\phi $ $_{-1}$ $\phi _{0}^{\ast }$
in the last equation of system (\ref{eq_7.2}). Actually, Fig. \ref{fig039}
demonstrates that the parametric instability sets in, which is explained by
a conjecture that the initial development of the above-mentioned MI, that
does not involve the $\phi _{0}$ field, drives the perturbed system across
the threshold of the parametric instability.

\subsection{The Darboux transform (DT)\ and nonlinear development of MI}

The integrable case, with $\nu =a=1$, which corresponds to the attractive
interactions, makes it possible to develop deeper analysis of MI. As
mentioned above, the spinor BEC obeys the integrability condition if a
special but physically possible constraint (\ref{eq_7.6}) is imposed on the
scattering lengths of collisions between atoms. Then, Eqs. (\ref{eq_7.2})
can be rewritten as a $2\times 2$ matrix NLS equation
\begin{equation}
i\frac{\partial \mathbf{Q}}{\partial t}+\frac{\partial ^{2}\mathbf{Q}}{%
\partial x^{2}}+2\mathbf{QQ}^{\dag }\mathbf{Q=0,}\text{ \ \ }\mathbf{Q\equiv
}\left(
\begin{array}{cc}
\phi _{+1} & \phi _{0} \\
\phi _{0} & \phi _{-1}%
\end{array}%
\right) ,  \label{eq_7.26}
\end{equation}%
which is a completely integrable system \cite{Wadati}. DT for Eqs. (\ref%
{eq_7.26}) can be derived from the respective Lax pair, which is \cite%
{6.5,Wadati}

\begin{equation}
\frac{\partial \mathbf{\Psi }}{\partial x}=\mathbf{U\Psi ,}\frac{\partial
\mathbf{\Psi }}{\partial t}=\mathbf{V\Psi .}  \label{eq_7.27}
\end{equation}%
Here $\mathbf{U}=\lambda \mathbf{J}+\mathbf{P}$ and $\mathbf{V}=2i\lambda
^{2}\mathbf{J}+2i\lambda \mathbf{P}+i\mathbf{W}$, with
\begin{subequations}
\begin{eqnarray}
\mathbf{J} &=&\left(
\begin{array}{cc}
\mathbf{I} & \mathbf{0} \\
\mathbf{0} & -\mathbf{I}%
\end{array}%
\right) ,\text{ }\mathbf{P}=\left(
\begin{array}{cc}
\mathbf{0} & \mathbf{Q} \\
-\mathbf{Q}^{\dag } & \mathbf{0}%
\end{array}%
\right) ,  \label{eq_7.28} \\
\mathbf{W} &=&\left(
\begin{array}{cc}
\mathbf{QQ}^{\dag } & \frac{\partial }{\partial x}\mathbf{Q} \\
\frac{\partial }{\partial x}\mathbf{Q}^{\dag } & -\mathbf{Q}^{\dag }\mathbf{Q%
}%
\end{array}%
\right) ,  \label{eq_7.29}
\end{eqnarray}%
where $\mathbf{I}$ and $\mathbf{0}$ denote the unit and zero matrices, $%
\mathbf{\Psi =}\left( \mathbf{\Psi }_{1},\mathbf{\Psi }_{2}\right) ^{\mathrm{%
T}}$ is the matrix eigenfunction corresponding to $\lambda $, $\mathbf{\Psi }%
_{1}$ and $\mathbf{\Psi }_{2}$ are $2\times 2$ matrices, and $\lambda $ is
the spectral parameter. Accordingly, Eq. (\ref{eq_7.26}) is tantamount to
the compatibility condition of the overdetermined linear system (\ref%
{eq_7.27}), $\partial \mathbf{U}/\partial r-\partial \mathbf{V}/\partial x+%
\mathbf{UV}-\mathbf{VU}=0$.

Based on the Lax pair (\ref{eq_7.27}), one introduces a transformation in
the form of
\end{subequations}
\begin{equation}
\widetilde{\mathbf{\Psi }}=\left( \lambda -\mathbf{S}\right) \mathbf{\Psi ,~S%
}=\mathbf{D\Lambda D}^{-1},~\mathbf{\Lambda =}\left(
\begin{array}{cc}
\lambda _{1}\mathbf{I} & \mathbf{0} \\
\mathbf{0} & \lambda _{1}\mathbf{I}%
\end{array}%
\right) ,  \label{eq_7.30}
\end{equation}%
where $\mathbf{D}$ is a nonsingular matrix satisfying equation $\partial
\mathbf{D/}\partial x=\mathbf{JD\Lambda }+\mathbf{PD}$. Further, letting
\begin{equation}
\frac{\partial \widetilde{\mathbf{\Psi }}}{\partial x}=\widetilde{\mathbf{%
U\Psi }},\text{ }\widetilde{\mathbf{U}}=\lambda \mathbf{J}+\mathbf{P}_{1}%
\text{,\ }\mathbf{P}_{1}\equiv \left(
\begin{array}{cc}
\mathbf{0} & \mathbf{Q}_{1} \\
\mathbf{-Q}_{1}^{\dag } & \mathbf{0}%
\end{array}%
\right) ,  \label{eq_7.31}
\end{equation}%
one obtains%
\begin{equation}
\mathbf{P}_{1}=\mathbf{P}+\mathbf{JS}-\mathbf{SJ}.  \label{eq_7.32}
\end{equation}%
One can also verify the following involution property of the linear
equations written above: if $\mathbf{\Psi =}\left( \mathbf{\Psi }_{1},%
\mathbf{\Psi }_{2}\right) ^{\mathrm{T}}$ is an eigenfunction corresponding
to $\lambda $, where $\mathbf{\Psi }_{j}$ are $2\times 2$ matrices, then $%
\left( -\mathbf{\Psi }_{2}^{\ast },\mathbf{\Psi }_{1}^{\ast }\right) ^{%
\mathrm{T}}$ is an eigenfunction corresponding to $-\lambda ^{\ast }$. Thus
one can take $\mathbf{D}$ as
\begin{equation}
\mathbf{D=}\left(
\begin{array}{cc}
\mathbf{\Psi }_{1} & -\mathbf{\Psi }_{2}^{\ast } \\
\mathbf{\Psi }_{2} & \mathbf{\Psi }_{1}^{\ast }%
\end{array}%
\right) ,\text{ }\mathbf{\Lambda =}\left(
\begin{array}{cc}
\lambda \mathbf{I} & \mathbf{0} \\
\mathbf{0} & -\lambda ^{\ast }\mathbf{I}%
\end{array}%
\right) ,
\end{equation}%
to obtain%
\begin{equation}
\mathbf{S}=\lambda \left(
\begin{array}{cc}
\mathbf{I} & \mathbf{0} \\
\mathbf{0} & \mathbf{I}%
\end{array}%
\right) +\left( \lambda +\lambda ^{\ast }\right) \left(
\begin{array}{cc}
-\mathbf{S}_{11} & \mathbf{S}_{12} \\
\mathbf{S}_{21} & -\mathbf{S}_{22}%
\end{array}%
\right) ,
\end{equation}%
where the matrix elements of $\mathbf{S}$ are given by
\begin{eqnarray}
\mathbf{S}_{11} &=&\left( \mathbf{I}+\mathbf{\Psi }_{1}\mathbf{\Psi }%
_{2}^{-1}\mathbf{\Psi }_{1}^{\ast }\mathbf{\Psi }_{2}^{\ast ^{-1}}\right)
^{-1},\text{ \ }\mathbf{S}_{12}=\left( \mathbf{\Psi }_{2}\mathbf{\Psi }%
_{1}^{-1}+\mathbf{\Psi }_{1}^{\ast }\mathbf{\Psi }_{2}^{\ast ^{-1}}\right)
^{-1}, \\
\mathbf{S}_{21} &=&\left( \mathbf{\Psi }_{2}^{\ast }\mathbf{\Psi }_{1}^{\ast
^{-1}}+\mathbf{\Psi }_{1}\mathbf{\Psi }_{2}^{-1}\right) ^{-1},\text{ }%
\mathbf{S}_{22}=\left( \mathbf{I}+\mathbf{\Psi }_{2}\mathbf{\Psi }_{1}^{-1}%
\mathbf{\Psi }_{2}^{\ast }\mathbf{\Psi }_{1}^{\ast ^{-1}}\right) ^{-1}.
\end{eqnarray}%
Finally, DT for Eq. (\ref{eq_7.26}) follows from Eq. (\ref{eq_7.32}), taking
the form of%
\begin{equation}
\mathbf{Q}_{1}=\mathbf{Q+}2\left( \lambda +\lambda ^{\ast }\right) \mathbf{S}%
_{12}.  \label{eq_7.33}
\end{equation}%
From Eqs. (\ref{eq_7.33}) and (\ref{eq_7.27}) it can be deduced that Eq. (%
\ref{eq_7.33}) generates a new solution $\mathbf{Q}_{1}$ for Eq. (\ref%
{eq_7.36}), once seed solution $\mathbf{Q}$ is known. In particular, a
one-soliton solution can be generated if the seed is a trivial zero state.
Next, taking $\mathbf{Q}_{1}$ as the new seed solution, one can derive from
Eq. (\ref{eq_7.33}) the corresponding two-soliton solution. The procedure
can be continued to generate multisoliton solutions.

In what follows, solutions to Eq. (\ref{eq_7.26}) are considered under
nonzero boundary conditions. The simplest among them is the CW solution with
constant densities:%
\begin{equation}
\mathbf{Q}_{c}=-\mathbf{A}_{c}\exp \left[ i\varphi _{c}\right] \text{, \ \ }%
\mathbf{A}_{c}\equiv \left(
\begin{array}{cc}
\beta _{c} & \alpha _{c} \\
\alpha _{c} & -\beta _{c}%
\end{array}%
\right) ,\text{ \ }\varphi _{c}\equiv kx+\left[ 2\left( \alpha
_{c}^{2}+\beta _{c}^{2}\right) -k^{2}\right] t,  \label{eq_7.34}
\end{equation}%
where $\alpha _{c}$ and $\beta _{c}$ are real constants and $k$ is a wave
number. In this solution, the constant densities of components $\phi _{\pm
1} $ are equal, while their signs are opposite.

Applying the above DT to the CW solution $\mathbf{Q}_{c}$, solving Eqs. (\ref%
{eq_7.27}) for this case, and employing Eq. (\ref{eq_7.33}), one obtains a
new family of solutions of Eq. (\ref{eq_7.26}), in the form of
\begin{equation}
\mathbf{Q}_{1}=\left[ \mathbf{A}_{c}+4\xi \left( \mathbf{I}+\mathbf{AA}%
^{\ast }\right) ^{-1}\mathbf{A}\right] \exp \left( i\varphi _{c}\right) .
\label{eq_7.35}
\end{equation}%
Here the following definitions are used:
\begin{subequations}
\begin{eqnarray}
\mathbf{A} &=&\left[ \mathbf{\Pi }\exp \left( \theta -i\varphi \right)
+\kappa ^{-1}\mathbf{A}_{c}\right] \left[ \kappa ^{-1}\mathbf{A}_{c}\mathbf{%
\Pi }\exp \left( \theta -i\varphi \right) +\mathbf{I}\right] ^{-1},
\label{eq_7.36} \\
\theta &=&M_{I}x+\left[ 2\xi M_{R}-\left( k+2\eta \right) M_{I}\right] t,
\label{eq_7.37} \\
\varphi &=&M_{R}x-\left[ 2\xi M_{I}+\left( k+2\eta \right) M_{R}\right] ,
\label{eq_7.38} \\
M &=&\sqrt{\left( k+2i\lambda \right) ^{2}+4\left( \alpha _{c}^{2}+\beta
_{c}^{2}\right) }=M_{R}+iM_{I},  \label{eq_7.39}
\end{eqnarray}%
where $\kappa \equiv \frac{1}{2}\left( ik-2\lambda +iM\right) ,$ $\lambda
=\xi +i\eta $ is the spectral parameter, and $\mathbf{\Pi =}\left(
\begin{array}{cc}
\beta & \alpha \\
\alpha & \gamma%
\end{array}%
\right) $ is an arbitrary complex symmetric matrix. Solution (\ref{eq_7.35})
reduces back to CW (\ref{eq_7.34}) when $\xi =0$. Note that the
three-component polar soliton (\ref{eq_7.25}), considered in the previous
subsection, is not a special example of solution (\ref{eq_7.35}), because
the background fields in components $\phi _{\pm 1}$ in solution (\ref%
{eq_7.25}) have identical signs.

In particular, with zero background, $\mathbf{A}_{c}=\mathbf{O}$, Eq. (\ref%
{eq_7.35}) yields a soliton solution:
\end{subequations}
\begin{equation}
\mathbf{Q}_{1}=\frac{4\xi \left[ \mathbf{\Pi }_{1}\exp \left( -\theta
_{1}\right) +\left( \mathbf{\sigma }^{y}\mathbf{\Pi }_{1}^{\ast }\mathbf{%
\sigma }^{y}\right) \exp \left( \theta _{s}\right) \det \mathbf{\Pi }_{1}%
\right] }{\exp \left( -2\theta _{s}\right) +1+\exp \left( 2\theta
_{s}\right) \left\vert \det \mathbf{\Pi }_{1}\right\vert ^{2}}\exp \left(
i\varphi _{s}\right) ,  \label{eq_7.40}
\end{equation}%
where $\theta _{s}=2\xi \left( x-4\eta t\right) -\theta _{0}$, $\varphi
_{s}=2\eta x+4\left( \xi ^{2}-\eta ^{2}\right) t$, $\theta _{0}$ is an
arbitrary real constant which determines the initial position of the
soliton, $\mathbf{\sigma }^{y}$ is the Pauli matrix, and $\mathbf{\Pi }_{1}$
is the polarization matrix \cite{6.6},%
\begin{equation}
\mathbf{\Pi }_{1}=\left( 2\left\vert \alpha \right\vert ^{2}+\left\vert
\beta \right\vert ^{2}+\left\vert \gamma \right\vert ^{2}\right) ^{-1/2}%
\mathbf{\Pi \equiv }\left(
\begin{array}{cc}
\beta _{1} & \alpha _{1} \\
\alpha _{1} & \gamma _{1}%
\end{array}%
\right) .
\end{equation}

\begin{figure}[tbp]
\centerline{\includegraphics[scale=0.6]{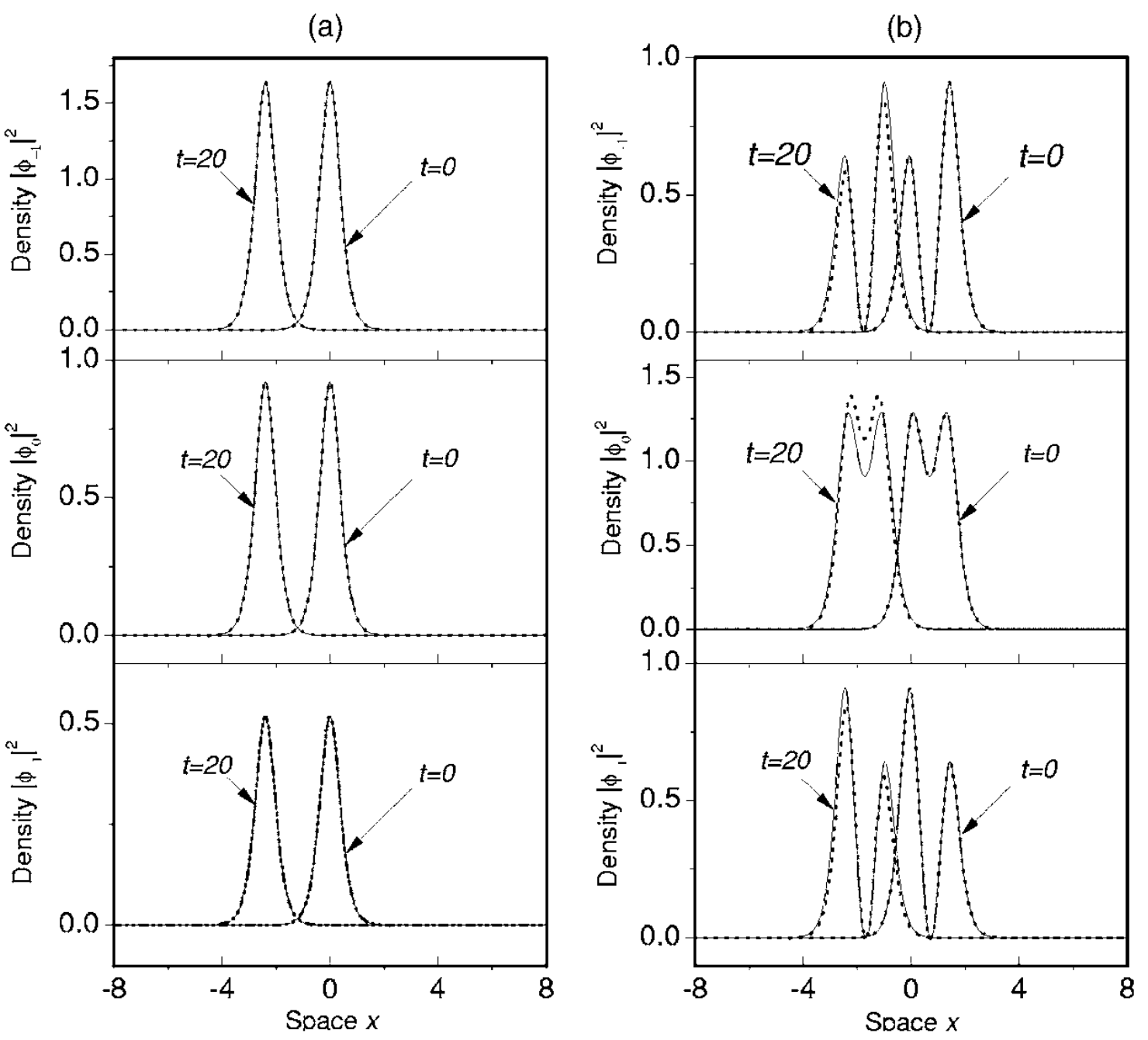}}
\caption{The evolution of the density distribution in the soliton, as
obtained from solution (\protect\ref{eq_7.35}) by adding a random
time-dependent perturbation of the nonlinear-coupling constant $a$, with the
perturbation amplitude $\pm 5\%$. Parameters are $\protect\nu =a=1$ (as
concerns the unperturbed value of $a$), $\protect\xi =1$, $\protect\eta %
=-0.03$, and $\protect\theta _{0}=0$. (a) The FM soliton, with $\protect%
\alpha _{1}=0.48$, $\protect\beta _{1}$ and $\protect\gamma _{1}$ determined
by conditions $\protect\alpha _{1}^{2}=\protect\beta _{1}\protect\gamma _{1}$
and $2\left\vert \protect\alpha _{1}\right\vert ^{2}+\left\vert \protect%
\beta _{1}\right\vert ^{2}+\left\vert \protect\gamma _{1}\right\vert ^{2}=1$%
; (b) the polar soliton, with $\protect\alpha _{1}=0.53$ and $\protect\beta %
_{1}=0.43$, $\protect\gamma _{1}$ being determined by condition $2\left\vert
\protect\alpha _{1}\right\vert ^{2}+\left\vert \protect\beta _{1}\right\vert
^{2}+\left\vert \protect\gamma _{1}\right\vert ^{2}=1$. Solid and dotted
curves show, respectively, the exact solutions and the perturbed ones,
produced by numerical simulations. The results are reproduced from Ref.
\protect\cite{6.1}.}
\label{fig042}
\end{figure}

Solitons (\ref{eq_7.40}) are tantamount to ones found in Ref. \cite{6.6},
where they were classified into the above-mentioned types, \textit{viz}.,
the FM and polar ones. Indeed, one can see from expression (\ref{eq_7.40})
that, when $\det \mathbf{\Pi }_{1}=0$, soliton (\ref{eq_7.40}) represents a
FM state. However, when $\det \mathbf{\Pi }_{1}\neq 0$, solution (\ref%
{eq_7.40}) may have two peaks, corresponding to a polar state.

Another relevant example of exact solutions (\ref{eq_7.35}) describes the
onset and nonlinear development of MI of CW states. It can be obtained by
noting that, under conditions $k=2\eta $ and $\alpha _{c}^{2}$M $\beta
_{c}^{2}$M$\xi ^{2}$, Eq. (\ref{eq_7.39}) yields $M_{I}=0$, hence there is
no dependence on $x$ in expression (\ref{eq_7.37}), and the solution
features no localization. Instead, it is periodic in $x$, through the $x$
dependence of $\varphi $, as given by Eq. (\ref{eq_7.38}); in this
connection, note that%
\begin{equation}
M_{R}^{2}=4\left( \alpha _{c}^{2}+\beta _{c}^{2}\right) -4\xi ^{2}
\label{eq_7.41}
\end{equation}%
does not vanish when $M_{I}=0$. This spatially periodic state represents a
particular mode of the nonlinear development of MI, in the exact form. To
look at it in detail, one can take a special case with $\mathbf{\Pi }%
=\epsilon \mathbf{E}$, where $\mathbf{E}$ is a matrix with all elements
equal to $1$, and $\epsilon $ is a small amplitude of the initial
perturbation added to the CW background, with the aim to initiate the onset
of MI. Indeed, the linearization of the initial profile of solution (\ref%
{eq_7.35}) with respect to $\epsilon $ yields%
\begin{equation}
\mathbf{Q}_{1}(x,0)\approx \left[ \mathbf{A}_{c}\rho -\epsilon \chi _{1}%
\mathbf{E}\cos \left( M_{R}x\right) -\epsilon \chi _{2}\mathbf{\sigma }^{z}%
\mathbf{\sigma }^{x}\mathbf{A}_{c}\exp \left( iM_{R}x\right) \right] \exp
\left( ikx\right) ,  \label{eq_7.42}
\end{equation}%
where $\rho \equiv 1-\xi \left( 2\xi +iM_{R}\right) /\left( \alpha
_{c}^{2}+\beta _{c}^{2}\right) $ with $\left\vert \rho \right\vert =1,$ $%
\chi _{1}\equiv \xi M_{R}\left( 2i\xi -M_{R}\right) /\left( \alpha
_{c}^{2}+\beta _{c}^{2}\right) $, $\chi _{2}\equiv \beta _{c}\chi
_{1}/\left( \alpha _{c}^{2}+\beta _{c}^{2}\right) ,$ $M_{R}=2\sqrt{\alpha
_{c}^{2}+\beta _{c}^{2}-\xi ^{2}}$, and $\mathbf{\sigma }^{x,z}$ are the
Pauli matrices. Clearly, the first term in Eq. (\ref{eq_7.42}) represents
the CW background, while the second and third ones are small spatially
modulated perturbations.

Comparing exact solution (\ref{eq_7.35}), obtained in this special case,
with results of direct numerical simulations of Eqs. (\ref{eq_7.32}) with
the initial condition (\ref{eq_7.42}), one can verify that the numerical
solution is very close to the analytical one, both displaying the
development of the MI initiated by the small modulational perturbation in
expression (\ref{eq_7.42}) (not shown here). Figure \ref{fig041} displays
the solution in terms of densities of the three components, as provided by
exact result (\ref{eq_7.35}). From this picture one concludes that atoms are
periodically transferred, in the spinor BEC, from the spin state $m_{F}=0$
into ones $m_{F}=\pm 1$ and vice versa.

As said above, exact solution (\ref{eq_7.35}) of Eqs. (\ref{eq_7.2}) is only
valid under the special condition $a=1$ (i.e., $2g_{0}=-g_{2}$). In the
general case ($a\neq 1$), the onset of the MI can be analyzed directly from
Eqs. (\ref{eq_7.2}) linearized for small perturbations \cite{6.7}. To this
end, take the CW solution (\ref{eq_7.34}) with $k=0$, which can be fixed by
means of the Galilean transformation, and consider its perturbed version
\begin{equation}
\widetilde{\mathbf{Q}}=\left( \mathbf{A}_{c}+\mathbf{B}\right) \exp \left[
2i\nu \left( \alpha _{c}^{2}+\beta _{c}^{2}\right) t\right] ,\text{ \ }%
\mathbf{B}=\left(
\begin{array}{cc}
b_{+1}(x,t) & b_{0}(x,t) \\
b_{0}(x,t) & b_{-1}(x,t)%
\end{array}%
\right) ,
\end{equation}%
where $b_{\pm 1}$ and $b_{0}$ are weak perturbations. Substituting this
expression in Eqs. (\ref{eq_7.2}), linearizing them, and looking for
perturbation eigenmodes in the usual form,%
\begin{equation}
b_{j}(x,t)=b_{jR}\cos \left( Kx-\omega t\right) +ib_{jI}\sin \left(
Kx-\omega t\right) ,\text{ \ \ }j=+1,-1,0,
\end{equation}%
where $b_{jR,I}$ are real amplitudes, the BdG equations give rise to two
branches of the dispersion relation between wave number $K$ and frequency $%
\omega $ of the perturbation,
\begin{subequations}
\begin{eqnarray}
\omega ^{2} &=&K^{2}\left( K^{2}-4a\alpha _{c}^{2}-4a\beta _{c}^{2}\right) ,
\label{eq_7.43} \\
\omega ^{2} &=&K^{2}\left( K^{2}-4\nu \alpha _{c}^{2}-4\nu \beta
_{c}^{2}\right) .  \label{eq_7.44}
\end{eqnarray}

The MI sets in when $\omega ^{2}$ given by either expression ceases to be
real and positive. Obviously, this happens if, at least, one condition
\end{subequations}
\begin{equation}
K^{2}<4a\left( \alpha _{c}^{2}+\beta _{c}^{2}\right) ,\text{ }K^{2}<4\nu
\left( \alpha _{c}^{2}+\beta _{c}^{2}\right)  \label{eq_7.45}
\end{equation}%
holds. Note that these conditions agree with the above exact result, which
corresponds to $a=\nu =1$. Indeed, in that case one may identify $K\equiv
M_{R}$, according to Eq. (\ref{eq_7.42}), and then Eq. (\ref{eq_7.41})
entails constraint $M_{R}^{2}<4\left( \alpha _{c}^{2}+\beta _{c}^{2}\right) $%
, which is identical to conditions (\ref{eq_7.45}). Note that the MI may
occur in the case of the repulsive spin-independent interactions in Eqs. (%
\ref{eq_7.2}), i.e., for $\nu =-1$, provided that the nonlinear-coupling
constant $a$, which accounts for the strength of the spin-dependent
interaction, is positive.

The stability of the soliton solutions (\ref{eq_7.40}) was explored too, of
both the polar and FM types, against finite random variations of the
coupling constant $a$ in time. The issue is relevant, as the exact solutions
are only available for $a=1$; hence, it is necessary to understand if the
solitons survive random deviations of $a$ from this value corresponding to
the integrability. Physically, this situation may correspond to a case when
the scattering length, which determines the nonlinear coefficient $a$ in
Eqs. (\ref{eq_7.2}), follows random variations of an external magnetic or
laser field which affects the scattering length through FR. The evolution of
the density profiles in the thus perturbed solitons is displayed in Fig. \ref%
{fig042}. This figure demonstrates that the solitons of both the FM and
polar types are robust against random changes of $a$, once they are
dynamically stable as exact solutions to the integrable version of the
model. Note also that the FM soliton seems more robust than its polar
counterpart.

\subsection{Conclusion of the section}

Soliton states have been presented in the model based on the coupled GP
equations describing the dynamics of spinor BEC with atomic spin $F=1$. In
the general nonintegrable version of the model, several types of elementary
exact soliton solutions are reported, which include one, two, or three
components, and represent either polar or FM solitons. Their stability is
checked by direct simulations and, in some cases, in an exact analytical
form, based on BdG equations for small perturbations. The global stability
of the solitons is analyzed by comparing the respective values of the energy
for a fixed norm. As the main result, stable ground-state solitons of the
polar and FM types have been found. Metastable solitons coexisting with the
ground-state ones are possible too, in certain parameter regions.

In the special case of the integrable spinor-BEC model \cite{Wadati}, DT has
been introduced and applied to derive a family of exact solutions on top of
the CW background. This family includes a solution that explicitly displays
full nonlinear evolution of the MI of the CW state, seeded by a small
spatially periodic perturbation. Robustness of one-soliton solutions of both
the FM and polar types against finite perturbations of the coupling constant
(deviations from the integrability) has been verified by means of direct
simulations.

\section{Soliton dynamics in one-dimensional spin-orbit-coupled BEC}

This section deals with motion of matter-wave solitons in 1D
spin-orbit-coupled (SOC) BEC. It is demonstrated that the spin dynamics of
solitons in this system is governed by a nonlinear Bloch equation, which
affects the orbital motion of solitons, leading to SOC effects in the motion
of macroscopic quantum objects. The solitons perform oscillations with a
frequency determined by the interplay of SOC, Raman coupling, and intrinsic
nonlinearity \cite{7.2}. Analytical predictions are corroborated by
numerical simulations of the underlying system of GP equations.

Results collected in this section are based on Ref. \cite{7.2}.

\subsection{The model}

In the presence of SOC, the dynamics of quasi-1D BEC, elongated in the
direction of $z$, is modeled by the GP equation for the pseudo-spinor
mean-filed wave function,
\begin{equation}
i\frac{\partial }{\partial t}\left(
\begin{array}{c}
\psi _{\uparrow } \\
\psi _{\downarrow }%
\end{array}%
\right) =\widehat{h}_{0}\left(
\begin{array}{c}
\psi _{\uparrow } \\
\psi _{\downarrow }%
\end{array}%
\right) +\left(
\begin{array}{cc}
g_{\uparrow \uparrow }\left\vert \psi _{\uparrow }\right\vert ^{2} &
g_{\uparrow \downarrow }\left\vert \psi _{\downarrow }\right\vert ^{2} \\
g_{\downarrow \uparrow }\left\vert \psi _{\uparrow }\right\vert ^{2} &
g_{\downarrow \downarrow }\left\vert \psi _{\downarrow }\right\vert ^{2}%
\end{array}%
\right) \left(
\begin{array}{c}
\psi _{\uparrow } \\
\psi _{\downarrow }%
\end{array}%
\right) ,  \label{eq_8.1}
\end{equation}%
where $\psi _{\sigma }$ are pseudospin components, with $\sigma
=\uparrow ,\downarrow $ labelling the spin states. These can
represent, for instance, the hyperfine states $|1,-1\rangle $ and
$|1,0\rangle $ of $^{87}$\textrm{Rb} atoms
\cite{7.1,SOC-review1,SOC-review2}. Here,
\begin{equation}
\widehat{h}_{0}=-\frac{1}{2}\frac{\partial ^{2}}{\partial z^{2}}%
+V(z)+i\lambda \frac{\partial \sigma _{z}}{\partial z}+\Omega \sigma
_{x}+\delta \sigma _{z}  \label{eq_8.2}
\end{equation}%
is the single-particle Hamiltonian which includes the Raman-induced SOC with
frequency $\Omega $ and strength $\lambda $, while $\delta $ is the Zeeman
detuning. Further, $V(z)=\gamma ^{2}z^{2}/2$ is the 1D HO trapping potential
with aspect ratio $\gamma \equiv \omega _{z}/\omega _{\perp }$, where $%
\omega _{z}$ and $\omega _{\perp }$ are the trapping frequencies along the
longitudinal and transverse directions, respectively. The frequencies and
lengths are measured here in units $\omega _{\perp }$ and $a_{\perp }=\sqrt{%
\hslash /\left( m\omega _{\perp }\right) }$, respectively, and the SOC
strength is $\lambda =k_{L}a_{\perp }$, with $k_{L}$ being the momentum
transfer. Note that, as the scattering lengths characterizing the inter- and
intra-species atomic interactions are very close in the experiment, it is
reasonable to assume $SU(2)$-symmetric spin interactions, with all strengths
$g_{\sigma \sigma ^{\prime }}$\ taking a single value, $g$. To focus on SOC
effects in the dynamics of solitons, the free space is considered first,
while effects of the external trap are discussed afterwards. Note that the
SOC term $\sim \lambda $ in Eq. (\ref{eq_8.2}) breaks the Galilean
invariance of the system of GP equations (\ref{eq_8.1}) in the free space
(although the total momentum remains a dynamical invariant). The latter fact
makes mobility of solitons in the SOC system a nontrivial issue, which is a
particular motivation for studying them.

For $\lambda =\Omega =0$, the system reduces to the normal binary BEC
without SOC. In this case, Eq. (\ref{eq_8.1}) with $g_{\sigma \sigma
^{\prime }}\equiv g$ is the integrable Manakov's system \cite{4.14}. In
particular, it produces bright-bright solitons $\psi _{\sigma }=\eta
\epsilon _{\sigma }\left[ \sqrt{-g}\cosh \left( \eta z\right) \right]
^{-1}\exp \left( i\eta ^{2}t/2\right) $ for the attractive sign of the
nonlinearity, $g<0$, where $\eta ^{-1}$ is the soliton's width, and $%
|\epsilon _{\uparrow }|^{2}+|\epsilon _{\downarrow }|^{2}=-g/(2\eta )$. Such
exact soliton solutions may be naturally used as an initial wave function,
while SOC is switched on. Note that the effective single-particle
Hamiltonian (\ref{eq_8.2}) can be rewritten in the frame transformed by the
local pseudospin rotation with angle $\vartheta =2\lambda z$ about the $z$
axis. The transformation adds phase factors $\exp \left( \pm i\lambda
z\right) $ to the two components of the wave function.

\subsection{Approximate analytical results (variational approximation)}

\begin{figure}[tbp]
\centerline{\includegraphics[scale=0.8]{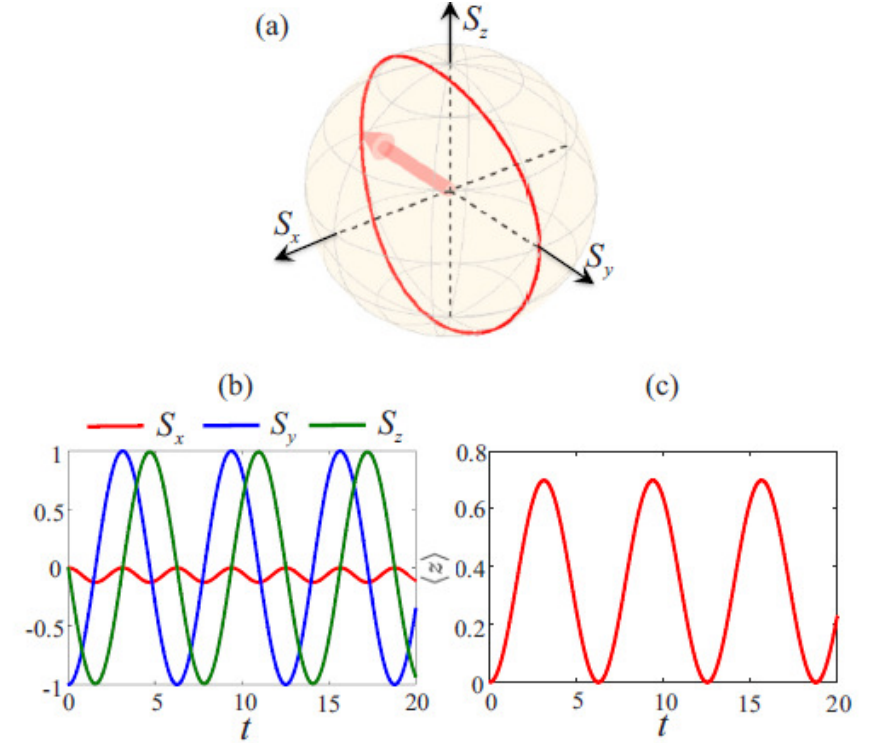}}
\caption{The track of the motion of the spin density on the Bloch sphere
(a), and the corresponding evolution of the spin components (b) and
center-of-mass coordinate (c) of the soliton for the initially balanced
state, with $\protect\theta (t=0)=\protect\pi /4$, and initial phase
difference $\protect\varphi _{-}(t=0)=\protect\pi /4$, see Eq. (\protect\ref%
{eq_8.3}). Other parameters are $\Omega =0.5$, $\protect\lambda =0.5\protect%
\sqrt{\Omega }$, $\protect\delta =0$, and $g=-10$. $z$ and $t$ being
measured in units of $a_{\perp }$ and $\protect\omega _{\perp }^{-1}$. The
results are reproduced from Ref. \protect\cite{7.2}.}
\label{fig043}
\end{figure}

The full GP system (\ref{eq_8.1}) is not integrable. Therefore, it is
relevant to employ VA for the analysis of the soliton dynamics \cite{3.5,7.2}%
, making use of the Lagrangian
\begin{equation}
L=\int_{-\infty }^{+\infty }\left\{ \frac{i}{2}\sum_{\sigma =\uparrow
,\downarrow }\left[ \psi _{\sigma }^{\ast }(\psi _{\sigma })_{t}-\psi
_{\sigma }(\psi _{\sigma }^{\ast })_{t}\right] -\mathcal{H}\right\} dz,
\end{equation}%
where $\mathcal{H}$ is the Hamiltonian density of system (\ref{eq_8.1}). In
the case of the attractive nonlinearity, $g<0$, the following variational
ansatz for bright-bright solitons, with the total norm fixed to be $1$, was
introduced in Ref. \cite{7.2}:%
\begin{equation}
\left(
\begin{array}{c}
\psi _{\uparrow } \\
\psi _{\downarrow }%
\end{array}%
\right) =\sqrt{\frac{\eta }{2}}\left(
\begin{array}{c}
\sin \theta \cosh ^{-1}\left( \eta z+\xi \right) \exp \left[ i\left(
k_{\uparrow }z+\varphi _{\uparrow }\right) \right] \\
\cos \theta \cosh ^{-1}\left( \eta z+\xi \right) \exp \left[ i\left(
k_{\downarrow }z+\varphi _{\downarrow }\right) \right]%
\end{array}%
\right) ,  \label{eq_8.3}
\end{equation}%
where $\theta $, $\eta $, $\xi $, $k_{\sigma }$, $\varphi _{\sigma }$ are
time-dependent variational parameters. Here, $\theta $ determines the
population imbalance between the pseudospin components, $\eta ^{-1}$ defines
their common width, $k_{\sigma }$ is the wave number, and $\varphi _{\sigma
} $ the phase.

Inserting ansatz (\ref{eq_8.3}) into the Lagrangian and performing the
integration, one obtains the effective Lagrangian,
\begin{eqnarray}
L &=&\frac{\xi }{\eta }\frac{dk_{+}}{dt}-\frac{\xi }{\eta }\cos \left(
2\theta \right) \frac{dk_{-}}{dt}-\frac{d\varphi _{+}}{dt}+\cos \left(
2\theta \right) \frac{d\varphi _{-}}{dt}  \notag \\
&&-\frac{1}{2}\left[ k_{+}^{2}-2k_{+}k_{-}\cos \left( 2\theta \right)
+k_{-}^{2}\right] -\frac{1}{6}\eta ^{2}-\frac{1}{6}g\eta  \notag \\
&&-\frac{\Omega \pi k_{-}\sin \left( 2\theta \right) \cos \left( 2\varphi
_{-}-2k_{-}\xi /\eta \right) }{\eta \sinh \left( \pi k_{-}/\eta \right) }
\label{eq_8.4} \\
&&+\delta \cos \left( 2\theta \right) -\lambda \left[ k_{+}\cos \left(
2\theta \right) -k_{-}\right] ,  \notag
\end{eqnarray}%
where $k_{\pm }\equiv (1/2)(k_{\uparrow }\pm k_{\downarrow })$ and $\varphi
_{\pm }\equiv (1/2)(\varphi _{\uparrow }\pm \varphi _{\downarrow })$. The
evolution of the variational parameters is governed by the corresponding
Euler-Lagrangian equations. These equations produce simple results, $\eta
\approx -g/2$ and $k_{-}\approx \lambda $, in the case of a weak SOC, $\pi
\lambda \ll \eta $. Indeed, in the absence of SOC, relation $\eta =-g/2$
holds for the normalized wave function, indicating that the width of the
solitons is determined by the nonlinearity, and $k_{-}$ remains equal to the
initial relative momentum $\lambda $ between the components of the soliton.
As a result, one arrive at a reduced system of dynamical equations, in which
$\eta $ and $k_{-}$ are considered as frozen quantities:
\begin{subequations}
\begin{eqnarray}
\frac{d}{dt}k_{+} &=&2\widetilde{\lambda }\sin \left( 2\theta \right) \sin
\phi ,  \label{eq_8.5} \\
\frac{d}{dt}\phi &=&-2\widetilde{\Omega }\cot \left( 2\theta \right) \cos
\phi +2\lambda k_{+}-2\delta ,  \label{eq_8.6} \\
\frac{d}{dt}\theta &=&-\widetilde{\Omega }\sin \phi ,  \label{eq_8.7} \\
\frac{d}{dt}\left\langle z\right\rangle &=&k_{+}.  \label{eq_8.8}
\end{eqnarray}%
Here, $\langle z\rangle =\int_{-\infty }^{+\infty }z(|\psi _{\uparrow
}|^{2}+|\psi _{\downarrow }|^{2})dz\equiv -\xi /\eta $ is the center-of-mass
coordinate, $\widetilde{\Omega }\equiv \pi \lambda /[\eta \sinh \left( \pi
\lambda /\eta \right) ]$, and $\phi \equiv 2\varphi _{-}+2k_{-}\langle
z\rangle $ is the phase difference between the two components of the
soliton. Thus, Eqs. (\ref{eq_8.5})-(\ref{eq_8.8}) account for the dynamical
coupling of the center-of-mass momentum $k_{+}$, phase difference $\phi $,
population imbalance $\theta $, and center-of-mass coordinate $\langle
z\rangle $.

Next, introduce a normalized complex-valued spinor $\chi =(\chi _{\uparrow
},\chi _{\downarrow })$ proportional to the two-component wave function, as $%
\psi _{\sigma }=\sqrt{\rho (z,t)}\chi _{\sigma }$, where $\rho \equiv |\psi
_{\uparrow }|^{2}+|\psi _{\downarrow }|^{2}$ is the total density, hence $%
\chi $ is subject to normalization $|\chi _{\uparrow }|^{2}+|\chi
_{\downarrow }|^{2}=1$. Then, one defines spin density $S=\chi ^{T}\sigma
\chi $, where $\sigma \equiv \{\sigma _{x},\sigma _{y},\sigma _{z}\}$ is the
vector set of the Pauli matrices, and one has
\end{subequations}
\begin{equation}
\{S_{x},S_{y},S_{z}\}=\{\sin \left( 2\theta \right) \cos \left( 2\lambda
z+2\varphi _{-}\right) ,-\sin \left( 2\theta \right) \sin \left( 2\lambda
z+2\varphi _{-}\right) ,-\cos \left( 2\theta \right) \}
\end{equation}%
for the ansatz given by Eq. (\ref{eq_8.3}). Focusing the consideration on
the center-of-mass motion of the soliton, by setting $z=\langle z\rangle
=-\xi /\eta $, the evolution of the atomic spin at the soliton's
center-of-mass is governed by the following equations:
\begin{subequations}
\begin{eqnarray}
\frac{d}{dt}S_{x} &=&2(\lambda c_{1}-\delta )S_{y}-2\lambda ^{2}S_{z}S_{y},
\label{eq_8.9} \\
\frac{d}{dt}S_{y} &=&-2\widetilde{\Omega }S_{z}-2(\lambda c_{1}-\delta
)S_{x}+2\lambda ^{2}S_{z}S_{x},  \label{eq_8.10} \\
\frac{d}{dt}S_{z} &=&2\widetilde{\Omega }S_{y},  \label{eq_8.11}
\end{eqnarray}%
where $c_{1}\equiv \lambda S_{z,0}$, with $S_{\alpha ,0}$ ($\alpha =x,y,z$)
being initial values of the components. These equations of motion for the
soliton's spin can be rewritten as
\end{subequations}
\begin{equation}
\frac{d}{dt}\mathbf{S}=\mathbf{S}\times \mathbf{B}_{\mathrm{eff}},\text{ \ \
}\mathbf{B}_{\mathrm{eff}}=\{2\widetilde{\Omega },0,2\lambda
^{2}S_{z}-2(\lambda c_{1}-\delta )\}.  \label{eq_8.12}
\end{equation}%
This is the Bloch equation for the spin precession under the action of the
effective magnetic field, $\mathbf{B}_{\mathrm{eff}}$. The macroscopic SOC
for the soliton as a quantum body is exhibited by the effect of the
evolution of the spin on the soliton's longitudinal momentum, resulting in
coupled nonlinear dynamics of the soliton's spin and position.

To construct solutions of the nonlinear Bloch equation, one first integrates
Eq. (\ref{eq_8.9}), dividing it by Eq. (\ref{eq_8.11}). This yields $%
S_{x}=c_{2}-(\lambda ^{2}/2)\widetilde{\Omega }^{-1}S_{z}^{2}+(c_{1}\lambda
-\delta )\widetilde{\Omega }^{-1}S_{z}$, where $c_{2}\equiv S_{x,0}+(\delta
-c_{1}\lambda )\widetilde{\Omega }^{-1}S_{z,0}+(\lambda ^{2}/2)\widetilde{%
\Omega }^{-1}S_{z,0}^{2}$ is a constant determined by the initial
conditions. Next, the analysis focuses on the case of
\begin{equation}
\delta =c_{1}\lambda ,  \label{delta}
\end{equation}%
which implies a particular relation between the strengths of the Zeeman
splitting and SOC, making the analysis more explicit. By differentiating Eq.
(\ref{eq_8.11}), one arrives at a standard equation for anharmonic
oscillations for the single spin component $S_{z}$,%
\begin{equation}
\frac{d^{2}S_{z}}{dt^{2}}+\Xi S_{z}+2\lambda ^{4}S_{z}^{3}=0,
\label{eq_8.13}
\end{equation}%
where $\Xi \equiv 4\widetilde{\Omega }(\widetilde{\Omega }-c_{2}\lambda
^{2}) $ may be positive or negative. Equation (\ref{eq_8.13}) has a usual
solution,
\begin{equation}
S_{z}(t)=\frac{\sqrt{1-\Xi \tau ^{2}}}{2\lambda ^{2}\tau }\mathrm{cn}\left(
t/\tau ,k\right) ,\text{ }k^{2}=\frac{1}{2}\left( 1-\Xi \tau ^{2}\right) ,
\label{eq_8.14}
\end{equation}%
where $\mathrm{cn}$ is the Jacobi's cosine with modulus $k$, and $\tau $ is
an arbitrary parameter taking values $\tau <1/\sqrt{\left\vert \Xi
\right\vert }$. In the case of $\Xi >0$, the linearized version of Eq. (\ref%
{eq_8.13}), which corresponds to $\tau \rightarrow 1/\sqrt{\left\vert \Xi
\right\vert }$ in Eq. (\ref{eq_8.14}), gives rise to free Rabi oscillations
with frequency $\sqrt{\left\vert \Xi \right\vert }$. In the general case,
the frequency given by solution (\ref{eq_8.14}), $\omega _{\mathrm{osc}}=\pi
/[2\tau K(k)]$, where $K(k)$ is the complete elliptic integral, exceeds $%
\sqrt{\left\vert \Xi \right\vert }$ due to the nonlinear shift. Note also
that the nonlinearity may give rise to oscillations in the case of $\Xi <0$,
when the Rabi oscillations are impossible.

\begin{figure}[tbp]
\centerline{\includegraphics[scale=0.8]{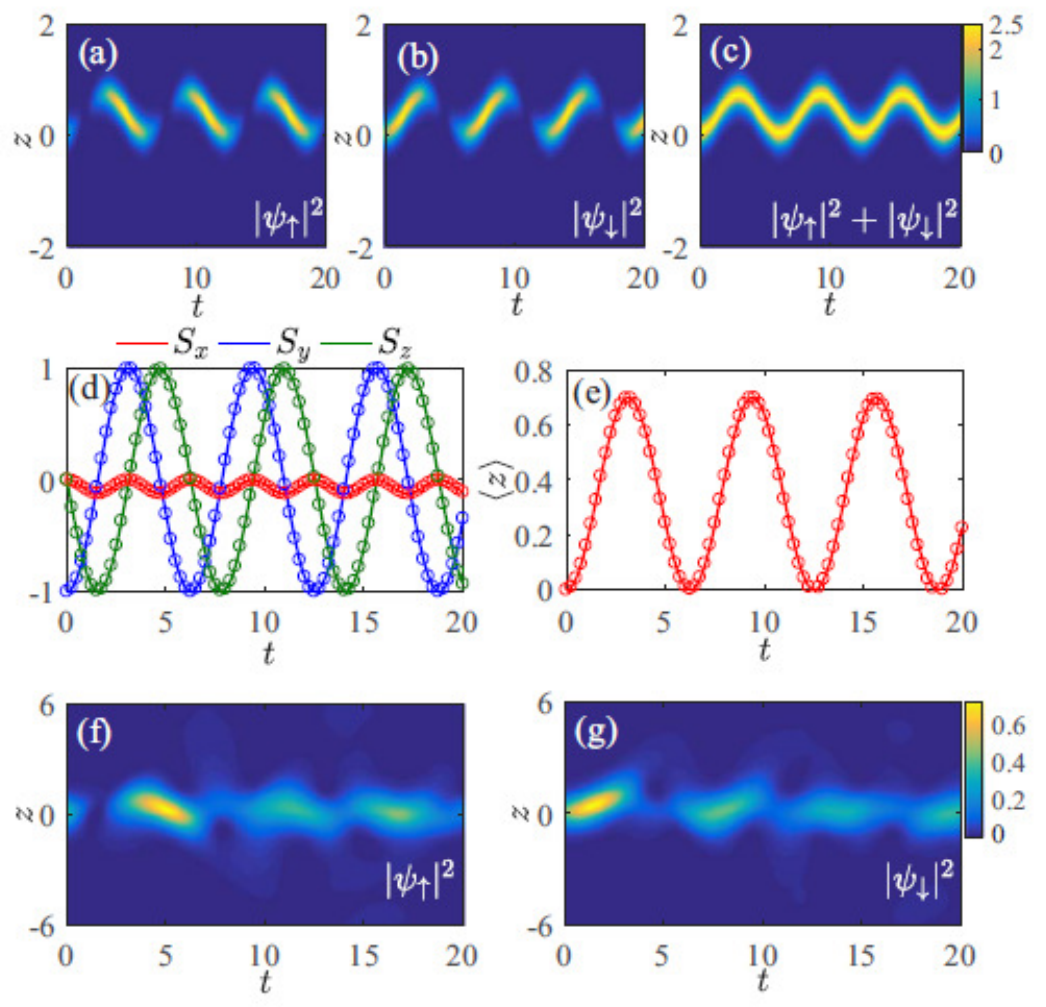}}
\caption{(a-c) The evolution of the density in the two components of the
soliton, produced by simulations of Eq. (\protect\ref{eq_8.1}) for the
initially balanced state with $\protect\theta (t=0)=\protect\pi /4$ and $%
\protect\varphi _{-}(t=0)=\protect\pi /4$. The parameters are $\Omega =0.5$,
$\protect\lambda =0.5\protect\sqrt{\Omega }$, $\protect\delta =0$, and $%
g=-10 $. The spin dynamics and center-of-mass motion of the soliton,
generated by these simulations (circles), and by VA based on Eqs. (\protect
\ref{eq_8.9})-(\protect\ref{eq_8.11}) and (\protect\ref{eq_8.15}) (solid
lines), are depicted in panels (d) and (e). Panels (f) and (g) display decay
of the soliton in the case of a weaker atomic interaction, $g=-3$. The
results are reproduced from Ref. \protect\cite{7.2},}
\label{fig044}
\end{figure}

Further, in the spin representation, Eq. (\ref{eq_8.5}) can be written as $%
dk_{+}/dt=-2\lambda \widetilde{\Omega }S_{y}$, which accounts for the
feedback of the evolution of the spin on the center-of-mass momentum (this
effect makes the SOC system essentially different from others considered in
the present review). This leads to the following equation of motion for the
center-of-mass coordinate:%
\begin{equation}
\frac{d^{2}\langle z\rangle }{dt^{2}}=-2\lambda \widetilde{\Omega }S_{y}.
\label{eq_8.15}
\end{equation}%
In other words, if one considers the soliton as a macroscopic quantum body
carrying the intrinsic angular momentum, Eq. (\ref{eq_8.15}) represents the
driving force, exerted by the intrinsic momentum and acting on the linear
momentum. This is, literally, a macroscopic realization of SOC. It is
relevant to stress that the mechanism of the soliton's motion is definitely
different from the conventional collective dipole oscillations of BEC in the
HO trap. Indeed, in the present case the SOC-induced force which drives the
motion of the soliton is the spin precession, rather than the force defined
by the external potential.

To illustrate the soliton dynamics in detail, Fig. \ref{fig044} displays
numerical solutions of Eqs. (\ref{eq_8.9})-(\ref{eq_8.11}) and (\ref{eq_8.15}%
), with initially balanced populations in the two components, which
correspond to $\theta (t=0)=\pi /4$ and $\varphi _{-}(t=0)=\pi /4$. First,
Fig. \ref{fig043}(a) shows that the soliton's spin moves along a closed
orbit on the Bloch sphere. Accordingly, perfect periodic oscillations of the
spin, coupled to the periodic motion of the soliton's central coordinate,
are observed in Figs. \ref{fig043}(b) and (c), respectively. In particular,
for weaker SOC, the center-of-mass oscillations can be approximated by $%
\langle z\rangle \simeq \left( \lambda /\widetilde{\Omega }\right) \sin
^{2}\left( \widetilde{\Omega }t\right) $ with amplitude $\lambda /\widetilde{%
\Omega }$ and period $\pi /\widetilde{\Omega }$, which depend on the
strengths of SOC, Raman coupling, and intrinsic BEC\ nonlinearity.

\subsection{Numerical results}

To test the predictions of VA (i.e., effectively quasi-particle
approximation for the motion of the soliton), the system of
coupled GP equations (\ref{eq_8.1}) was solved numerically in Ref.
\cite{7.2}, with the input in the form of the bright-bright
soliton,
\begin{equation}
\psi _{\uparrow }=\sqrt{\eta /2}(\sin \theta _{0})\cosh ^{-1}\left( \eta
z\right) \exp \left[ i(\lambda z+\varphi _{\uparrow ,0})\right]  \label{up}
\end{equation}%
\begin{equation}
\psi _{\downarrow }=\sqrt{\eta /2}\left( \cos \theta _{0}\right) \cosh
^{-1}\left( \eta z\right) \exp \left[ i(-\lambda z+\varphi _{\downarrow ,0})%
\right] .  \label{down}
\end{equation}%
In Figs. \ref{fig044}(a-c), the density of the soliton's components exhibits
robust periodic oscillations, and the soliton maintains its initial
hyperbolic-secant profile in the course of many oscillation cycles. Note
that, although the initial momenta of the two components are opposite, the
soliton does not split. Figures \ref{fig044}(d) and (e) show that results of
VA agree very well with the direct GP simulations. For weaker nonlinearity,
the simulations show that solitons decay under the action of SOC for $\pi
\lambda /\eta >(\pi \lambda /\eta )_{c}\approx 0.4$, as shown in Figs. \ref%
{fig044}(f) and (g). Furthermore, when condition (\ref{delta}) does not
hold, it is found that, besides the SOC-driven oscillations, the soliton
exhibits additional linear motion, as can be found, in particular, by
simulations for the initially polarized state with $\theta (t=0)=\pi /2$
(not shown here in detail).

\begin{figure}[tbp]
\centerline{\includegraphics[scale=0.8]{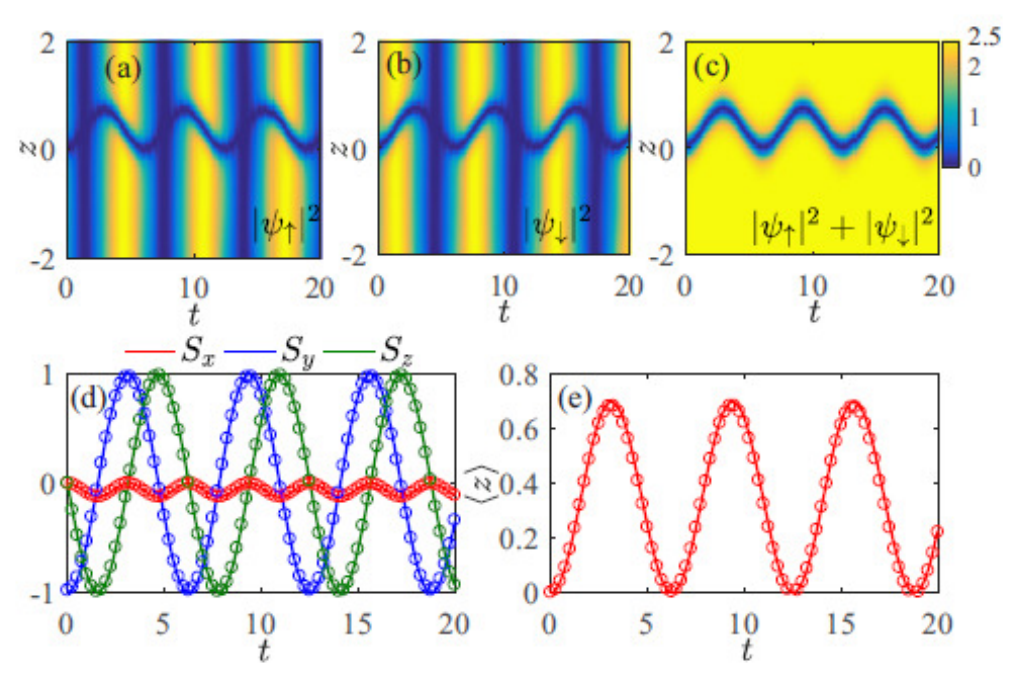}}
\caption{The same as in panels (a)-(e) of Fig. \protect\ref{fig044}, but for
the dark soliton, in the case of $g=10$. The results are reproduced from
Ref. \protect\cite{7.2}.}
\label{fig045}
\end{figure}

Next the case of the repulsive BEC nonlinearity is addressed, which is more
relevant to current experiments with SOC condensates \cite{7.1,7.3}. In this
case, the following variational ansatz can be used for dark-dark solitons:%
\begin{equation}
\left(
\begin{array}{c}
\psi _{\uparrow } \\
\psi _{\downarrow }%
\end{array}%
\right) =\sqrt{\frac{\eta }{2}}\left(
\begin{array}{c}
\left( \sin \theta \right) \tanh \left( \eta z+\xi \right) \exp \left(
k_{\uparrow }z+\varphi _{\uparrow }\right) \\
\left( \cos \theta \right) \tanh \left( \eta z+\xi \right) \exp \left(
k_{\downarrow }z+\varphi _{\downarrow }\right)%
\end{array}%
\right) ,  \label{eq_8.16}
\end{equation}%
cf. Eqs. (\ref{up}) and (\ref{down}). Inserting ansatz (\ref{eq_8.16}) into
the Lagrangian, one needs to renormalize the integrals, so as to exclude
divergent contributions of the nonvanishing background. The analysis yields
the same Euler-Lagrange equations as Eqs. (\ref{eq_8.6})-(\ref{eq_8.8}), but
with $\eta =g/2$ for repulsive $g>0$. The corresponding direct simulations
of the GP system were performed in Ref. \cite{7.2} with initial conditions
corresponding to the dark-dark soliton,
\begin{eqnarray}
\psi _{\uparrow } &=&\sqrt{\eta /2}(\sin \theta _{0}\sin )\tanh \left( \eta
z\right) \exp \left[ i(\lambda z+\varphi _{\uparrow ,0})\right] , \\
\psi _{\downarrow } &=&\sqrt{\eta /2}(\cos \theta _{0})\tanh \left( \eta
z\right) \exp \left[ i(-\lambda z+\varphi _{\downarrow ,0})\right] .
\end{eqnarray}%
The results are displayed in Fig. \ref{fig045}, where the dark-dark soliton
performs oscillations similar to those displayed above in the case of the
bright-bright soliton. Note that the two components of the background
alternately disappear and revive, as seen in Figs. \ref{fig045}(a) and (b).

\begin{figure}[tbp]
\centerline{\includegraphics[scale=0.9]{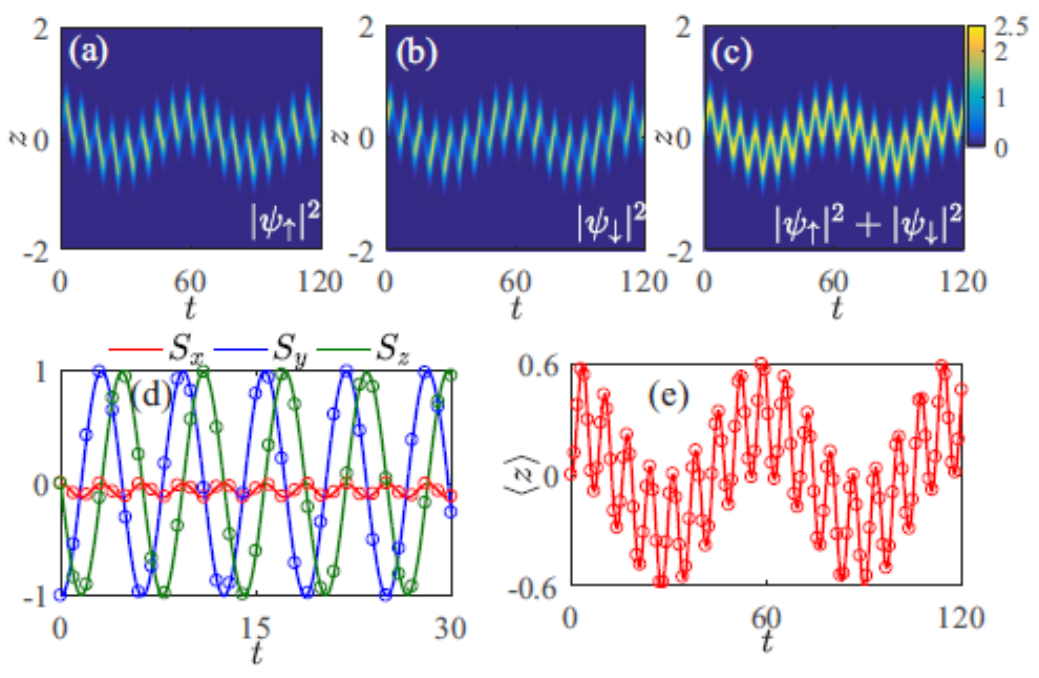}}
\caption{Dynamics of bright-bright solitons in the trapping potential, see
Eqs. (\protect\ref{kxi}) and (\protect\ref{gamma}), with $\protect\gamma %
=0.1 $, the other parameters being the same as in Fig. \protect\ref{fig044}.
The results are reproduced from Ref. (\protect\cite{7.2}).}
\label{fig046}
\end{figure}

In the experiment, the condensate is trapped in an HO potential, $%
V(z)=\gamma ^{2}z^{2}/2$, which affects the motion of solitons. In this
case, the Euler-Lagrange equations for variational parameters $k_{+}$ and $%
\xi $ are modified to
\begin{equation}
\frac{d}{dt}k_{+}=-2\lambda \widetilde{\Omega }S_{y}+\frac{\gamma ^{2}}{\eta
}\xi ,~\frac{d}{dt}\xi =-k_{+}\eta ,  \label{kxi}
\end{equation}%
with $\eta $ satisfying equation $4\eta ^{4}+2g\eta ^{3}=\pi ^{2}\gamma ^{2}$%
, cf. Eqs. (\ref{eq_8.5})-(\ref{eq_8.8}). Accordingly, the center-of-mass
motion acquires an additional trap-induced term,
\begin{equation}
\frac{d^{2}}{dt^{2}}\langle z\rangle +\gamma ^{2}\langle z\rangle =-2\lambda
\widetilde{\Omega }S_{y},  \label{gamma}
\end{equation}%
which gives rise to additional collective oscillations with period $\simeq
2\pi /\gamma $, as shown in Fig. \ref{fig046}. For a weak trap, the period
of the oscillations and the characteristic length scale are much larger than
the period and amplitude of the SOC-driven oscillations, therefore the
SOC-driven oscillations of the soliton are not conspicuously affected by the
trap. Similar results are also produced by the simulations for a dark
soliton in the HO trap (oscillations of a trapped single-component dark
soliton, in the absence of SOC, were studied in work \cite{Busch}).

Finally, it is relevant to briefly discuss conditions for the experimental
realization of the predicted dynamical regimes. The Raman-induced SOC has
been realized for BEC in the $^{87}$\textrm{Rb} gas with repulsive atomic
interactions. Typically, an elongated condensate of $\sim 10^{4}$ atoms is
trapped in a weak HO trap with frequencies $\omega _{\perp }=2\pi \times 85$
$\mathrm{Hz}$ and $\omega _{z}=2\pi \times 5.9$ $\mathrm{Hz}$, and the ratio
of the scattering lengths is $a_{\uparrow \uparrow }:a_{\uparrow \downarrow
}:a_{\downarrow \downarrow }=1:1$ $:1.005$. In this case, a stationary
dark-dark soliton with width $\sim 0.5$ $\mathrm{\mu m}$ can be created at
the center of the condensate by means of PIT \cite{imprint}. Then two $804.1$
\textrm{nm} Raman lasers with intersection angle $20^{\circ }$ are used to
induce weak SOC. With the above parameters, the period and amplitude of the
SOC-driven oscillations of the dark-dark soliton are about $11$ \textrm{ms}
and $2$ $\mathrm{\mu m}$, respectively, which may be further adjusted by
varying strengths of the SOC, Raman coupling, and atomic interaction. These
estimates clearly suggest that the SOC-driven oscillations can be
experimentally observed under available experimental conditions.

\subsection{Conclusion of the section}

In conclusion, it has been shown in this section that the interplay of SOC,
Raman coupling, and intrinsic nonlinearity in quasi-1D BEC may realize the
mechanism of \emph{macroscopic} SOC, in the form of a mechanical motion of
bright and dark solitons, considered as macroscopic quantum bodies. The
soliton's angular momentum (spin) evolves according to the Bloch equation
under the action of an effective magnetic field, and induces a force
affecting the motion of the soliton's central coordinate. The results have
been obtained by means of VA and numerical simulations, which demonstrate
very good agreement. These findings suggest directions for experimental
studies of the motion of matter-wave solitons under the action of SOC.

\section{Conclusion of the article}

Theoretical and experimental work focused on solitons and other nonlinear
modes in BEC and similar physical settings (in particular, nonlinear optics)
subject to the temporal, spatial, and spatiotemporal \textquotedblleft
management", has grown into a vast research area. In terms of the theory,
the central theme is analytical and numerical investigation of the dynamics
of solitons and other modes in the framework of the underlying models
(chiefly, NLS and GP equations, as well as multicomponent systems of coupled
equations), in which coefficients in front of specific linear and nonlinear
terms are made functions of time and/or spatial coordinates. Well-known
examples are the dispersion management in optics and nonlinearity management
in BEC. In this context, basic subjects of the theoretical studies are the
existence and stability of solitons and extended nonlinear states in this
class of models. In particular, an important direction is search for
nontrivial models which may be transformed into well-known integrable
equations with constant coefficients, such as the 1D cubic NLS equation and
the Manakov's system. Then, by means of the inverse transformation, one can
produce a vast set of exact stable solutions in the original models, using,
as inputs, well-known exact solutions of the integrable equations. Search
for robust solitons in nonintegrable systems is a still more challenging
direction, which has also yielded many nontrivial results. These findings
are summarized in the present review. Also included are theoretical results
obtained in some other models, which, strictly speaking, do not include
spatiotemporal modulation, but the methods and findings elaborated for such
models are closely related to those produced in models which explicitly
include the spatiotemporal \textquotedblleft management". Particular
results, which are collected in the article, are summarized in detail in
concluding subsections included in Sections 2-8.

The most essential point which should be stressed to conclude the review is
that experimental findings lag far beyond the theoretical predictions.
Nevertheless, the remarkable progress in the development of experimental
methods in realms of matter waves in BEC, as well as in nonlinear optics,
suggest that essential development of the experimental studies in the areas
outlined in the present review may be expected. In particular, many
theoretical predictions, presented in this review, provide straightforward
clues for the experiment.

As concerns further development of the theory, there are many interesting
problems to be tackled. In particular, this may be the investigation of 2D
and 3D solitons and similar objects, especially solitary vortices and more
complex self-trapped modes with an intrinsic topological structure, as well
as studies of interactions between two or several solitons, and formation of
multi-soliton bound states, in one- and multidimensional settings alike. The
study of multidimensional models is most interesting in cases when they
cannot be reduced, by means of exact or approximate factorization, to 1D
counterparts. As concerns interactions between solitons, they may yield most
nontrivial results in models which are not tantamount to integrable
equations, hence results of the interactions may be strongly different from
familiar elastic collisions.

\section*{Acknowledgments}

E. Kengne's work has been supported by the Initiative of the
President of the Chinese Academy of Sciences for Visiting
Scientists (PIFI) under Grants No. 2020VMA0040, the National Key
R\&D Program of China under grants No. 2016YFA0301500, NSFC under
grants Nos.11434015, 61227902. W.-M Liu's work has been supported
by the National Key R\&D Program of China under grants No.
2016YFA0301500, NSFC under grants Nos.11434015, 61227902. The work
of B. A. Malomed was supported, in part, by the Israel Science
Foundation, through grant No. 1286/17, and by a fellowship
provided by PIFI.

\end{document}